\documentclass[useAMS,usenatbib]{mn2e}
\usepackage[pdftex]{graphicx}
\usepackage{subfigure,aas_macros}
\usepackage{times}

\newcommand\msun{\rm{M_{\odot}}}

\def\stacksymbols #1#2#3#4{\def\theguybelow{#2}
        \def\verticalposition{\lower#3pt}
        \def\spacingwithinsymbol{\baselineskip0pt\lineskip#4pt}
        \mathrel{\mathpalette\intermediary#1}}
\def\intermediary #1#2{\verticalposition\vbox{\spacingwithinsymbol
        \everycr={}\tabskip0pt
        \halign{$\mathsurround0pt#1\hfil##\hfil$\crcr#2\crcr
                \theguybelow\crcr}}}

\def\lta{\stacksymbols{<}{\sim}{2.5}{.2}}
\def\gta{\stacksymbols{>}{\sim}{2.5}{.2}}

\newcounter{figuresub}

\title[Raining onto black holes: accretion driven by thermal instability]{Chaotic cold accretion onto black holes}

\author[Gaspari, Ruszkowski \& Oh]{M. Gaspari$^{1}$\thanks{E-mail:
mgaspari@mpa-garching.mpg.de}, M. Ruszkowski$^{2,4}$, S. Peng Oh$^{3}$\\
$^{1}$Max Planck Institute for Astrophysics, Karl-Schwarzschild-Strasse 1, 85741 Garching, Germany\\
$^{2}$Department of Astronomy, University of Michigan, 500 Church Street, Ann Arbor, MI 48109, USA\\
$^{3}$Department of Physics, University of California, Santa Barbara, CA 93106, USA\\
$^{4}$The Michigan Center for Theoretical Physics, 3444 Randall Lab, 450 Church Street, Ann Arbor, MI 48109, USA}

\begin{document}

\pagerange{\pageref{firstpage}--\pageref{lastpage}} \pubyear{2013}
\maketitle
\label{firstpage}

\begin{abstract}
Bondi theory is often assumed to adequately describe the mode of accretion in astrophysical environments. However, the Bondi flow must be adiabatic, spherically symmetric, steady, unperturbed, with constant boundary conditions. Using 3D AMR simulations, linking the 50 kpc to the sub-pc scales over the course of 40 Myr, we systematically relax the classic assumptions in a typical galaxy hosting a supermassive black hole. In the more realistic scenario, where the hot gas is {\it cooling}, while {\it heated} and {\it stirred} on large scales, the accretion rate is boosted up to two orders of magnitude compared with the Bondi prediction. The cause is the nonlinear growth of thermal instabilities, leading to the condensation of cold clouds and filaments when $t_{\rm cool}/t_{\rm ff}\lta10$. The clouds decouple from the hot gas, `raining' onto the centre. Subsonic turbulence of just over 100 km s$^{-1}$ ($M$\,$>$\,$0.2$) induces the formation of thermal instabilities, even in the absence of heating, while in the transonic regime turbulent dissipation inhibits their growth ($t_{\rm turb}/t_{\rm cool}\lta 1$). When heating restores global thermodynamic balance, the formation of the multiphase medium is violent, and the mode of accretion is fully {\it cold} and {\it chaotic}. The recurrent {\it collisions} and tidal forces between clouds, filaments and the central clumpy torus promote angular momentum cancellation,
hence boosting accretion. On sub-pc scales the clouds are channelled to the very centre via a funnel. 
In this study, we do not inject a fixed initial angular momentum, though vorticity is later seeded by turbulence.
A good approximation to the accretion rate is the cooling rate, which can be used as subgrid model, physically reproducing the boost factor of 100 required by cosmological simulations, while accounting for the frequent fluctuations. Since our modelling is fairly general (turbulence/heating due to AGN feedback, galaxy motions, mergers, stellar evolution), chaotic cold accretion may be common in many systems, such as hot galactic halos, groups, and clusters. In this mode, the black hole can quickly react to the state of the {\it entire} host galaxy, leading to efficient self-regulated feedback and the symbiotic Magorrian relation. Chaotic accretion can generate high-velocity clouds, likely leading to strong variations in the AGN luminosity, and the deflection or mass-loading of jets.
During phases of overheating, the hot mode becomes the single channel of accretion, though strongly suppressed by turbulence. High-resolution data could determine the current mode of accretion: assuming quiescent feedback, the cold mode results in a quasi flat temperature core as opposed to the cuspy profile of the hot mode.
\end{abstract}


\section{Introduction}\label{s:intro}

Accretion onto compact objects plays central role in the evolution of astrophysical systems. A number of physical processes, such as radiative cooling,  
heating and turbulence, 
as well as the presence of magnetic fields and transport processes, could all affect the effective accretion rate onto compact objects. The complexity of the real accretion process makes it impossible to predict the accretion rate analytically. Only under very restrictive assumptions of adiabaticity, spherical symmetry, unperturbed and steady initial conditions, in the absence of magnetic fields, self-gravity, and feedback, can the 
nonlinear hydrodynamic equations
be solved analytically (\citealt{Bondi:1952}). In this idealised case, the Bernoulli equation reduces to 
\begin{equation}\label{Bernoulli}
\frac{v^2}{2}+\frac{\gamma}{\gamma-1}\frac{p_\infty}{\rho_\infty}\left[\left(\frac{\rho}{\rho_\infty}\right)^{\gamma-1}-1\right] = \frac{GM_\bullet}{r},
\end{equation}
with gas pressure $p$, density $\rho$, velocity $v$, adiabatic index $\gamma$, and mass of the accretor $M_\bullet$. 
After some algebraic manipulation, conservation of mass and Eq.~(\ref{Bernoulli}) lead to the famous Bondi accretion rate formula
\begin{equation}\label{MdotB}
\dot M_{\rm B} = \lambda\, 4\pi (GM_\bullet)^2\rho_{\infty}/c^3_{{\rm s}, \infty},
\end{equation}
where $\lambda=(1/2)^{(\gamma+1)/2(\gamma-1)}[(5-3\gamma)/4]^{-(5-3\gamma)/2(\gamma-1)}$ is a normalisation factor of order unity, and $c_{\rm s}$ is the gas sound speed (the infinity symbol denotes very large radii).

This elegant and convenient formula for mass accretion has been extensively used in the last fifty years in countless astrophysical studies. In particular, a large majority of investigations, either theoretical/numerical or observational, assume that that accretion dynamics follows the Bondi solution even when the region of influence of the accretor (the Bondi radius, $r_{\rm B}=GM_\bullet/c^2_{\rm s, \infty}$) is not resolved (e.g. \citealt{Reynolds:1996, Loewenstein:2001, Churazov:2002, Baganoff:2003, DiMatteo:2003, DiMatteo:2005, Springel:2005, Allen:2006, Croton:2006,Hopkins:2006, Rafferty:2006, Sijacki:2007, Hardcastle:2007, McCarthy:2008, Booth:2009, Cattaneo:2007, Cattaneo:2009, Yang:2012}).
However, since realistic astrophysical conditions are dramatically different from the classic scenario, it is essential to include more realistic physics and relax the Bondi assumptions. 
This is especially important in the context of accretion onto supermassive black holes, as the SMBH feedback governs the formation and evolution of galaxies, groups, and clusters throughout the cosmic time (\citealt{McNamara:2012}).

In the last decade, investigators have started to quantify the deviations from the classic Bondi accretion scenario, using numerical (e.g. \citealt{Proga:2003, Pen:2003, Krumholz:2005, Krumholz:2006, Igumenshchev:2006}) or semi-analytical approaches (e.g. \citealt{Quataert:2000, King:2006, Soker:2006, Nayakshin:2007, Pizzolato:2010, Narayan:2011, Hobbs:2012_subgrid, Mathews:2012}). The latter models tend to be based on restrictive assumptions on the flow properties (e.g. steady, spherically symmetric, isobaric, constant cooling function, free boundary conditions, simplified potential), while the former have focused on a particular physics or specific conditions, as a rotating shell or a quasar
(\citealt{Hobbs:2011,Barai:2012}). Previous simulations typically studied either the region within $r_{\rm B}$ for a few Bondi times ($t_{\rm B}=r_{\rm B}/c_{\rm s, \infty}$), or the opposite regime involving cosmological timescales without resolving the Bondi radius.
Other limitations include the choice of the geometry (1D, 2D) or the discretisation method. 
We dedicate \S\ref{s:comp} for the comparison and review of the works closer to our research.

In the present investigation, we intend to provide a coherent and unified picture of the departures from the classic and idealised Bondi accretion scenario. Using three-dimensional adaptive mesh refinement (AMR) simulations, we link the small sub-Bondi spatial scales to the large 50 kpc scales (up to roughly 10 million dynamical range), and study accretion on timescales of over 40 Myr ($\gta200$ $t_{\rm B}$). As reference case, we consider accretion onto a black hole in a common astrophysical environment -- an elliptical galaxy in a hot gaseous halo. We progressively increase the realism of the accretion process, while including physics that is simple and generalizable to a wide range of situations.
First, we start from purely adiabatic evolution in a stratified atmosphere (\S\ref{s:adi}). 
Second, we relax the assumption of adiabaticity and include radiative cooling (\S\ref{s:cool}).
Third, we drive turbulence to model the effect of weak or strong gas stirring (e.g. AGN feedback, galaxy motions, mergers, cosmic flows, stellar evolution), studying the dynamics with and without cooling (\S\ref{s:stir_cool} and \S\ref{s:stir}).
Finally, we introduce distributed heating to emulate the effects of AGN and stellar feedback (\S\ref{s:heat}). This allows us to keep the system in global -- though not local -- thermal equilibrium, as observed in most galaxies, groups, and clusters.

After taking into consideration the above realistic processes, the accretion dynamics profoundly differs from the classic case. The mode of accretion is {\it cold} and {\it chaotic}, driven by stochastically forming and moving multiphase filaments, which condense via nonlinear thermal instability (TI).  The cold clouds are complex extended structures, as opposed to infinitesimally small objects on ballistic orbits. The frequent inelastic collisions and tidal motions between clouds, filaments, and the central volatile torus, promote angular momentum cancellation in the cold phase.
The  amount of gas that is accreted trough the central region is thus boosted, up to 100 times the Bondi rate.
On sub-parsec scales (down to our highest resolution of few tens gravitational radii), the cold gas is channelled toward the BH through a funnel, maintaining a similar high level of accretion (\S\ref{s:conv}).
We show that the cooling rate is a good approximation to the accretion rate, and we suggest to employ it as fiducial subgrid prescription for accretion (and thus feedback), especially in cosmological simulations. 
In \S\ref{s:disc}, we present an in-depth discussion of the results, along with the limitations of our models, the role of additional physics, and the implications of chaotic cold accretion.

The primary focus of this work is to answer the question: do black holes accrete gas in the hot or cold mode?
In recent years, there has been a revolution in the field of galaxy formation driven by the realisation that many galaxies accrete primarily in the cold mode (e.g. \citealt{Keres:2005}), rather than in the hot mode. We address the same question in the context of BH accretion, highlighting the role of thermal instabilities and turbulence.

\section[]{Physics \& Numerics} \label{s:init}
\subsection[]{Initial Conditions}
We study accretion physics in a very common astrophysical system, i.e. an elliptical galaxy embedded in a hot atmosphere (intragroup/intracluster medium -- IGM/ICM). The initial density profile is in hydrostatic equilibrium (neglecting the black hole) with the temperature profile corresponding to that observed in the typical massive galaxy/group NGC 5044 (Fig.~\ref{pure1}, darker line; cf. \citealt{Gaspari:2011b}). 

The gravitational potential in the centre is dominated by the supermassive black hole of mass $M_\bullet=3\times10^9$ $\msun$. Schwarzschild radius is thus $R_{\rm S}\equiv 2GM_\bullet/c^2 \simeq 3\times10^{-4}$ pc, while the initial Bondi radius is $r_{\rm B}\equiv GM_\bullet/c_{\rm s, \infty}^2 \simeq 85$ pc (the sound speed is $c_{\rm s, \infty}\simeq390$ km s$^{-1}$ on kpc-scales away from the black hole), and the corresponding Bondi time is $t_{\rm B}\equiv r_{\rm B}/c_{\rm s, \infty}\simeq 210$ kyr. Adopting a large black hole mass increases the Bondi radius and thus allows us to resolve this radius better, though the timestep becomes smaller (due to higher central $T$ and $v$). In this work, we study the regime of accretion associated with the kinetic feedback, rather than the rarer quasar-like regime shaped by the strong X-ray radiation (\S\ref{s:comp}).

On kpc scales, the total potential is dominated by the deVaucouleurs profile of the cD galaxy with stellar mass $M_{\ast}\simeq 3.4\times10^{11}$ $\msun$ and effective radius $r_{\rm e} \simeq 10$ kpc. Within the central $\sim$300 pc, $M_{\ast}$ is assumed to be constant as commonly observed (\citealt{Mathews:2003}), thus inducing a flat gas density core. 
On larger scales (tens of kpc), the major contribution to total gravity $g$ comes from the dark matter potential approximated via the NFW profile with virial mass
$M_{\rm vir} \simeq 3.6\times10^{13}$ $\msun$ ($r_{\rm vir} \simeq 860$ kpc) and concentration parameter $c \simeq 9.5$, in the concordance $\Lambda$CDM universe ($H_0 \simeq 70$ km s$^{-1}$ Mpc$^{-1}$). The normalisation of the gas density is set by the gas fraction $f_{\rm gas} \simeq 0.1$ (at $r_{\rm vir}$), yielding central electron number density $n_{\rm e, 0}\simeq 0.5$ cm$^{-3}$.

The ratio of the cooling time, $t_{\rm cool}\equiv1.5\,nk_{\rm B}T/ n_{\rm e} n_{\rm i} \Lambda$ (\S\ref{s:rcool}), to the free-fall time, $t_{\rm ff}\equiv(2r/g)^{1/2}$, starts below 10 in the range $\sim 100$ pc - 8 kpc. This is typical for cool-core systems which constitute the majority of groups, clusters, and massive galaxies. In the case of our galaxy the minimum `TI-ratio' is $\sim$ 5 at 250 pc. The value of 10 corresponds to the threshold below which the growth of thermal instabilities becomes possible (\citealt{Gaspari:2012a, McCourt:2012, Sharma:2012}). This threshold can also be related to the observed entropy threshold, $K\equiv k_{\rm B}T/n^{2/3}_{\rm e}\sim30$ keV cm$^2$, below which 
substantial star formation is triggered and extended multiphase gas is observed (\citealt{Rafferty:2008, Cavagnolo:2008}).

We use a modified version of the AMR FLASH4 code (\citealt{Fryxell:2000,Lee:2009}) to integrate the equations of hydrodynamics (see \S\ref{s:adi} and \citealt{Gaspari:2011a,Gaspari:2011b} for more details).
The computational box is $\simeq52$ kpc on a side with the black hole located at the centre. We follow the evolution of the gas for 40 Myr. Such long evolutionary times ($\sim$$200\ t_{\rm B}$) in a domain that large ($\sim$$600\ r_{\rm B}$) are crucial in order to understand the role of turbulence and cooling on the formation of extended cold gas at large radii, and to follow the subsequent accretion onto the central black hole. Large dynamical range also removes any undesirable influence of the boundary conditions on our results. For the sake of completeness, we assumed 
boundaries where the gas is allowed to flow out of the computational domain, but is not allowed to enter it.
The use of three dimensions is key to study the stochastic dynamics and the instabilities, otherwise leading to spurious compression, fragmentation, and/or condensation.

In order to link the sub-pc scales to the scales of 50 kpc over long evolutionary times, our only option is to adopt concentric fixed AMR mesh. We typically employ 14 levels of refinement, reaching up to 21 levels when we study accretion on sub-pc scales. This implies an effective linear resolution of 65536$\,$-$\,$8388608 zones, i.e. a dynamical range of almost 10 million! This is one of the first studies achieving such an extended range (cf. \citealt{Levine:2008}).

Each `spherical' shell covers over 50 cells in the radial direction at each level of refinement. Typical resolution of the reference runs is $\sim$0.8 pc. In a few runs we reach the sub-pc resolution, down to $\sim$20 $R_{\rm S}$. At this point additional physics certainly becomes important and we intend to address this in future work. As shown by the convergence runs (\S\ref{s:conv}), this resolution is nevertheless sufficient to track the key features of chaotic cold accretion.

In the present study, we assume the gas has negligible initial rotational velocity.
The majority of elliptical galaxies have low $v_{\rm rot, gas}$, typically less than 10 per cent of the stellar velocity dispersion (e.g. \citealt{Caon:2000}). Local angular momentum is here seeded by turbulence and enhanced by the baroclinic instability (a clumpy torus is indeed formed). 
Under idealised conditions, for the faster rotating galaxies ($\sim0.1\ \sigma_\ast$ on the kpc scale),
a thin disc would develop near the circularisation radius, $\sim50-100$ pc. However, 
\citealt{Hobbs:2011} (\S\ref{s:heat_maps}) showed that when turbulent velocities are $\gta v_{\rm rot, gas}$, initial rotation 
is no more a hindrance for accretion. Realistic ISM/ICM turbulence, as modelled in this work, typically resides in this regime (\S\ref{s:turb}). Turbulence also represents an efficient mechanism able to diffuse angular momentum, to induce instabilities, and hence to alter the circularisation process. We intend to study more complex initial conditions in dedicated works (see also \S\ref{s:disc_phys}). In this study, we focus on the mode of accretion (cold vs. hot) and the role of thermal instability.

\subsection[]{Black Hole Sink}\label{s:sink}

After several tests, the best general and robust sink method we found is applying a vacuum region, by simply evacuating the gas that crosses the accretion sphere. For numerical stability (cf. \citealt{Ruffert:1994}), the gas density is not exactly set to zero, but is given a value of $10^{-35}$ g cm$^{-3}$ in a radius $r_{\rm a}\sim$ 4$\,$-$\,$5 zones. Within the sink region, velocities are reset to zero and the gravitational potential is softened (though the latter is not crucial\footnote{For instance, $\phi=-GM_\bullet/(r^2+s^2)^{1/2}$ with a softening parameter $s^2=(0.7\, r_{\rm a})^2\,\exp[-(r/0.7r_{\rm a})^2]$; we also tested Plummer and cubic spline softening and did not find significant differences.}).
Throughout this work, we refer to this sinked gas mass per timestep as the `black hole accretion rate', $\dot M_\bullet$.
 
The vacuum region is essential to properly evacuate the condensed cold gas, avoiding artificial overpressure bounces. 
On the other hand, the vacuum sphere defines a new (pc-scale) sonic point, 
which at first might seem artificial for black hole accretion\footnote{In general, the accretor has a finite hard surface, e.g. white dwarfs, neutron stars, planets.}.
However, a more realistic relativistic potential makes this assumption physical, even in the adiabatic case, for the following reason. Assuming the relativistic potential $\phi_{\rm PW}=-GM_\bullet/(r-R_{\rm S})$ (\citealt{Paczynski:1980}), and substituting it in the Bernoulli Eq.~(\ref{Bernoulli}), yields$^3$
\begin{equation}\label{sonic}
\hat{r}_{\rm s} = (5-3\gamma+8\hat{R}_{\rm S}+\Delta^{1/2})/8,
\end{equation}
where $\Delta=(3\gamma-5-8\hat{R_{\rm S}})^2-64\,\hat{R}_{\rm S}(1-\gamma+\hat{R}_{\rm S})$; the sonic and Schwarzschild radius are normalised to $r_{\rm B}$ (hat symbol). The key point is that the physical sonic radius resides now at a finite distance from the centre. This means that our simulated flow, adiabatic or not, always becomes supersonic near the parsec region, justifying the complete mass evacuation via the vacuum sphere.
 
Since the sonic point determines $\dot M_\bullet$, using the vacuum accretor or the relativistic potential results in a similar mild increase in the accretion rate. The non-dimensional accretion rate (Eq.~\ref{MdotB}) for the adiabatic reference case with the relativistic potential is in fact given by\footnote{From Eqs.~\ref{Bernoulli} and \ref{MdotB}, $\lambda=[g(\hat{r}=\hat{r}_{\rm s})/f(\hat{M}=1)]^{\frac{\gamma+1}{2(\gamma-1)}}$, where $f=\hat M^{4/\gamma+1}[1/2+1/(\hat M^2(\gamma-1))]$, $g=\hat r_{\rm s}^{4(\gamma-1)/\gamma+1}[1/(\hat{r}_{\rm s}-\hat{R}_{\rm S})+1/(\gamma-1)]$, and $\hat M$ the Mach number normalised to $c_\infty\,\rho_\infty^{-(\gamma-1)/2}$. The sonic radius in Eq.~(\ref{sonic}) is given by the minimum of $g(\hat r)$.} 
\begin{equation}\label{lambdaPW}
\lambda =  \hat{r}_{\rm t}^2\left(\frac{\gamma-1}{\gamma+1}\frac{2}{\hat{r}_{\rm t} - \hat{R}_{\rm S} } + \frac{2}{\gamma+1} \right)^{\frac{\gamma+1}{2(\gamma-1)}},
\end{equation}
where $\hat{r}_{\rm t}$ corresponds to the accretor surface, if $\hat{r}_{\rm a}>\hat{r}_{\rm s}$, or the sonic point in Eq.~(\ref{sonic}), if $\hat{r}_{\rm a}\leq\hat{r}_{\rm s}$. For the adopted $\gamma=5/3$,  $\lambda=[\hat{r}_{\rm t}/(2({\hat{r}_{\rm t} - \hat{R}_{\rm S}))} + 3\hat{r}_{\rm t}/4]^2$ and $\hat r _{\rm s}=\hat R_{\rm S}+(2\hat R_{\rm S}/3)^{1/2}$, instead of the classic Bondi values $\lambda=1/4$ and $\hat{r}_{\rm s}=0$. Note that the difference in the accretion rate with respect to the classic case is moderate ($\lta20$ per cent, for our typical setup),
while the key relativistic correction resides in avoiding the singular sonic radius. 
Since the flow is supersonic, all of the gas crossing the sink radius can be removed, because the internal region is causally disconnected. The more general formula given by Eq.~(\ref{lambdaPW}) is useful for detailed comparisons with the adiabatic (Bondi-like) reference case, as well as for future accretion studies.

\subsection[]{Radiative Cooling}\label{s:rcool}
The hot plasma in galaxies, groups and clusters emits radiation mainly in the X-ray band due to Bremsstrahlung at $T \gta 10^7$ K and line emission at lower temperatures. The radiative emissivity of the gas is proportional to the electron and ion number densities  $\mathcal{L}=n_{\rm{e}} n_{\rm{i}} \Lambda(T,Z)$. As in our previous works, the cooling function $\Lambda$ is modelled following the work of \citet{Sutherland:1993} and assuming metallicity of $Z\simeq1\ Z_\odot$ (\citealt{Rasmussen:2009}). The stable cold phase has a temperature floor set at $10^4$ K (which can be justified by UV background heating). We adopt for simplicity a constant\footnote{$\mu$ starts to significantly increase only below $\sim1.6\times10^4$ K.} molecular weight of $\mu\simeq0.62$.

For numerical integration of the cooling term, we use the exact cooling module described in \citet{Gaspari:2012a}. Even if not strictly required, we adopt a timestep limiter imposed by the cooling time, in order to achieve higher accuracy. We tested also a few runs with an explicit Runge-Kutta cooling solver and found comparable results, albeit the explicit solver requires smaller timesteps.

\subsection[]{Turbulence Driving}\label{s:turb}
We model continuous injection of turbulence via a spectral forcing scheme that generates statistically stationary velocity fields (\citealt{Eswaran:1988, Fisher:2008}). This scheme utilises an Ornstein-Uhlenbeck (OU) random process (coloured noise), analogous to Brownian motion in a viscous medium. The driven acceleration field is time-correlated with a zero-mean and constant root-mean-square value. The time-correlation is important for modelling realistic driving forces. In the OU process, the value of the gas acceleration at previous timestep $a_n$ decays by an exponential damping factor $f = \exp (-\,dt / \tau_{\rm d})$, where $\tau_{\rm d}$ is the correlation time. Simultaneously, a new Gaussian-distributed acceleration with variance $\sigma_a^2=\epsilon^\ast/\tau_{\rm d}$ is added in the following way:
\begin{equation}\label{OU}
a_{n+1} = f \,a_n + \sigma_a\, \sqrt{1 - f^2} \, G_n,
\end{equation}
where $G_n$ is the Gaussian random variable, $\epsilon^\ast$ is the specific energy input rate, and $a_{n+1}$ is the updated acceleration. The six phases of the stirring modes (three real and three imaginary) are evolved in Fourier space and then converted to physical space. In this approach, turbulence can be driven by stirring the gas on large scales and letting it cascade to smaller scales. This is an efficient approach as the alternative would involve executing FFTs for the entire range of scales, where the vast majority of modes would have small amplitudes. In most runs, we impose a divergence-free condition on the subsonic turbulent velocity. 
In \S\ref{s:strong1}-\ref{s:strong2} we test transonic turbulence, including the compressional modes. 

The key physical quantity of interest is the final 3D turbulent velocity dispersion, $\sigma_v$, which affects the dynamics of accretion. The driving of turbulence is intentionally kept simple as our goal is not to consider any specific stirring source, but rather keep the calculation fairly general. For example, the true source of turbulence may be galaxy motions, substructure mergers, supernovae or AGN feedback. In recent works, \citet{Gaspari:2011b, Gaspari:2012b} showed that turbulence can be at the level of $100$$\,$-$\,$$300$ km s$^{-1}$ for extended periods of time following an AGN outburst. Recent observations found very similar values of turbulence in ellipticals, groups, and clusters (\citealt{dePlaa:2012, Sanders:2013}). Notice that such a level of turbulence can preserve the observed metallicity gradient, as shown by
\citet{Rebusco:2006}.

In the reference runs the gas is stirred with $\sigma_v\sim100$$\,$-$\,$$180$ km s$^{-1}$ (Mach number $M\equiv\sigma_v/c_{\rm s}\sim0.3$$\,$-$\,$$0.4$). This is achieved by adjusting the energy per mode\footnote{Via simple dimensional analysis (\citealt{Ruszkowski:2010}), $\sigma_v\propto(N\, L\,\epsilon^\ast)^{1/3}$; the number of modes is typically $N\sim5000$.} $\epsilon^{\ast}$ and correlation time $\tau_{\rm d}$ (usually $\sim10^{-5}$ cm$^2$ s$^{-3}$ and $3.15\times10^{11}$ s, respectively). As long as different choices of these parameters result in the same velocity dispersion, the dynamics of the flow remains unaffected. We stir the gas only on large scales, $L\gta4$ kpc, letting turbulence to naturally cascade to smaller scales. In the absence of gravity the velocity field is approximately Kolmogorov-like, but in a stratified medium, as we consider here, the turbulence spectrum is slightly different. Finally, since the intensity of stirring in our fiducial models is low, the turbulent heating, which is proportional to $\sigma_v^3/L$, is negligibly small (see \S\ref{s:stir}).

\subsection[]{Global Heating}\label{s:gheat}
The final set of more realistic simulations includes heating of the gaseous atmosphere. Numerous {\it XMM} and {\it Chandra} observations of the hot gas in galaxies, groups, and clusters (e.g. \citealt{Vikhlinin:2006, Diehl:2008b, Rasmussen:2009, Sun:2009a}) have clearly demonstrated that the majority of systems have a cool core in roughly global thermal equilibrium. Despite the presence of radiative cooling, central temperatures do not drop below one third of the virial temperature. Spectroscopic data shows that the cooling rates are severely quenched (\citealt{Peterson:2006}) and the global thermal equilibrium is likely maintained at a $\sim$10 percent level. 

The main source of the hot gas heating is likely AGN feedback (\citealt{Gaspari:2011a, Gaspari:2011b, Gaspari:2012b} and references within), with other possible contributions from mergers, conduction or stellar evolution (\citealt{Brighenti:2003}). Importantly, these heating processes preserve the gentle positive temperature gradient seen in the data. We do not model AGN outflows, jets or bubbles as in previous work (\citealt{Gaspari:2012c}), but instead consider a simplified, yet general, heating model which encapsulates several phenomena. Specifically, we force the heating rate (per unit volume) to be equal to the average radiative emissivity in radial shells. In G12b, we showed that a prolonged jet feedback heating results in the system settling roughly to an equilibrium state (i.e. lack of cooling catastrophe) characterised by sustained turbulent motions (\S\ref{s:turb}). The current heating prescription achieves the same goal of keeping the atmosphere in a quasi-stable state (cf. \citealt{McCourt:2012}). As demonstrated in \citealt{Gaspari:2012a}, in a jet-heated atmosphere, thermal quasi-steady equilibrium is achieved on a timescale longer than the duration of the simulations presented here. Since now we are not interested in simulating this transition period, we want to ensure that the system settles into global thermodynamic equilibrium as quickly as possible.

The details of how the global equilibrium is achieved do not alter the accretion dynamics and our conclusions. Nevertheless, regarding the shell thickness (where the average cooling is computed), it is preferable to use 4$\,$-$\,$8 zones in order to model the stochastic spatial variation of heating in a more realistic way (Fig.~9 in \citealt{Gaspari:2012b}). 
The very dense cooling gas, $T\lta10^5$ K, can sometimes produce artificial peaks in heating in the inner tiny shells, which can be prevented by excluding those outliers from the average of emissivity. When stirring is activated, we first let the system evolve for a brief period of time in order to seed the perturbations before imposing global heating. 
As for turbulence, the heating module is fairly general: our findings should thus be valid for a range of astrophysical conditions. 

\renewcommand{\thefigure}{\arabic{figure}\alph{figuresub}}
\setcounter{figuresub}{1}

\begin{figure} 
    \begin{center}
       \subfigure{\includegraphics[scale=0.3]{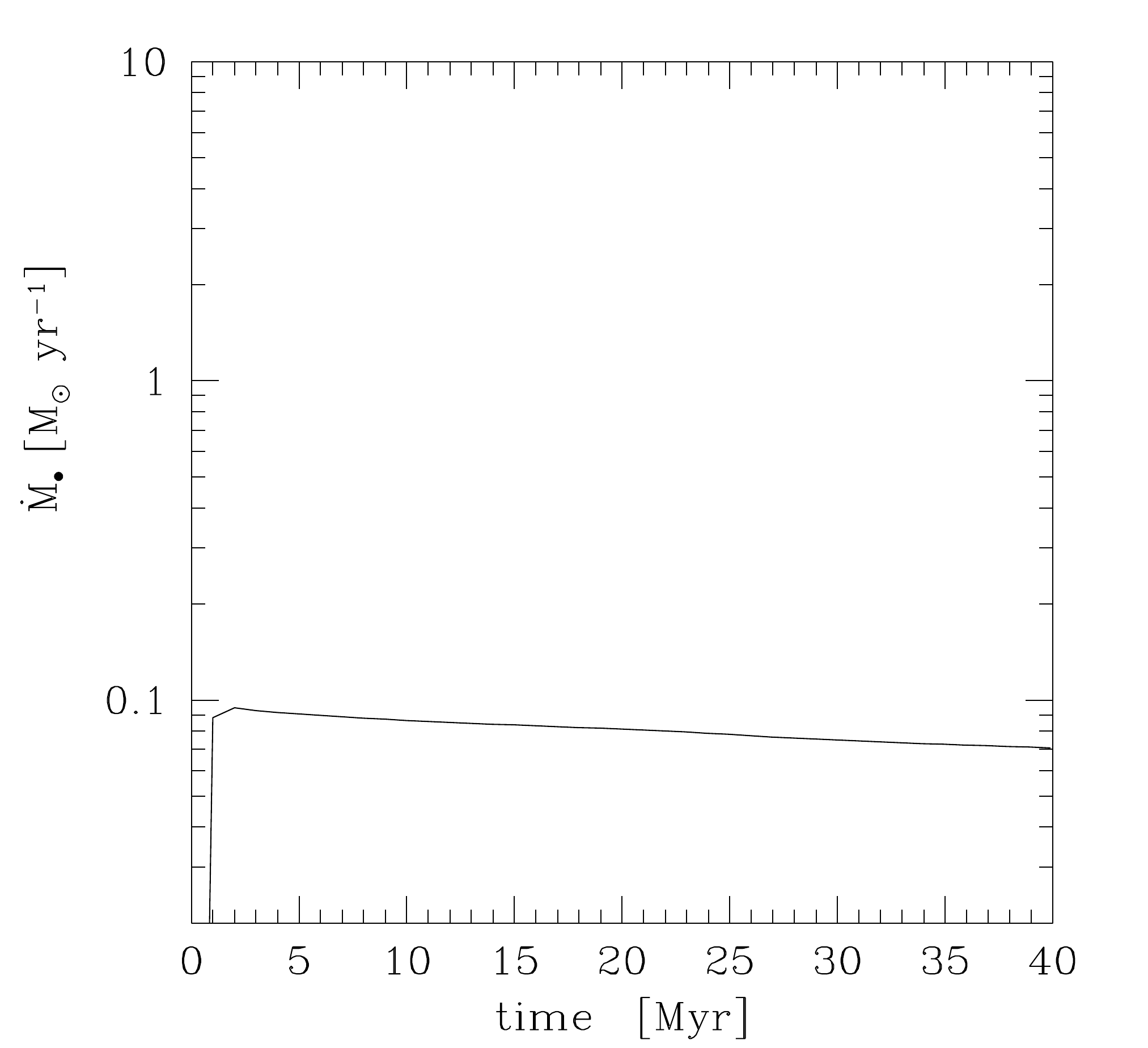}}
       \subfigure{\includegraphics[scale=0.3]{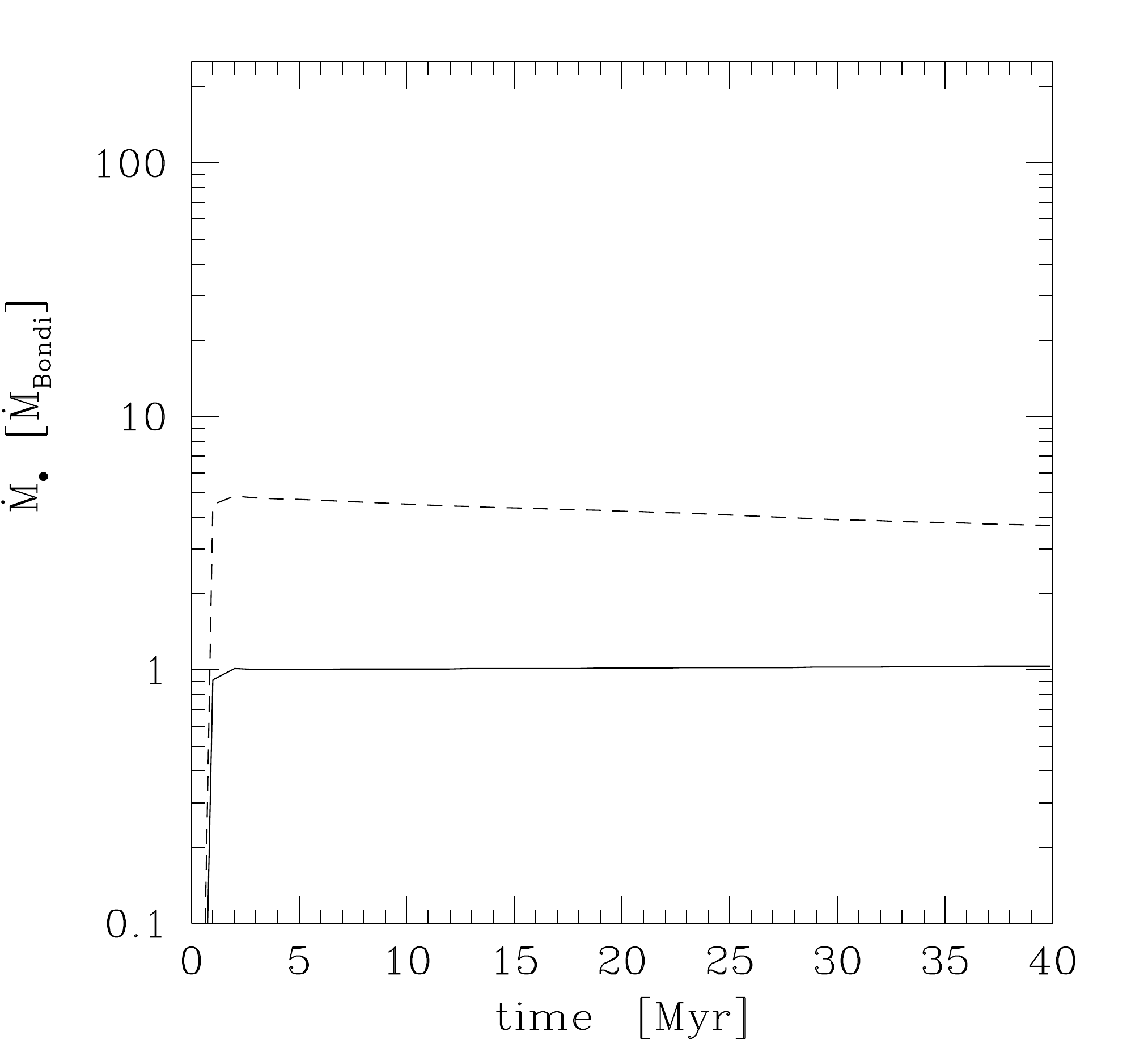}}
     \end{center}
     \caption{Adiabatic (Bondi-like) accretion: evolution of the accretion rate (1 Myr average).
      ({\it Top}) Accretion rate in physical units, $\msun$ yr$^{-1}$. The rate is slightly decreasing due to the presence of the galactic gradients. ({\it Bottom}) Black hole accretion rate normalised to the Bondi formula (\S\ref{s:sink}): the `boundary conditions' are taken at $r_{\rm B}$ (solid) or averaged over 1-2 kpc (dashed). The latter, commonly adopted procedure introduces a small bias in the accretion rate by a factor of a few (as opposed to $\sim$100 times or more that is sometimes assumed in cosmological simulations). Note the excellent match between the prediction and the numerical solution (solid line).
          \label{pure1}}
\end{figure}  

\section[]{Adiabatic accretion -- Bondi} \label{s:adi}
We start by simulating a purely adiabatic Bondi-like accretion, i.e. we do not include stirring, heating or cooling. As opposed to the classical setup and previous works, we initiate the system by using realistic astrophysical conditions, i.e. by employing the temperature and density profiles corresponding to a representative galaxy (\S\ref{s:init}). This allows us to quantify the differences between the accretion computed from the Bondi formula and the numerical simulation that are solely due to the non-zero gradients of the thermodynamic quantities on large kpc scales. 

\subsection{Accretion rate} \label{s:adi1}

\addtocounter{figure}{-1}
\addtocounter{figuresub}{1}

\begin{figure} 
    \begin{center}
       \subfigure{\includegraphics[scale=0.28]{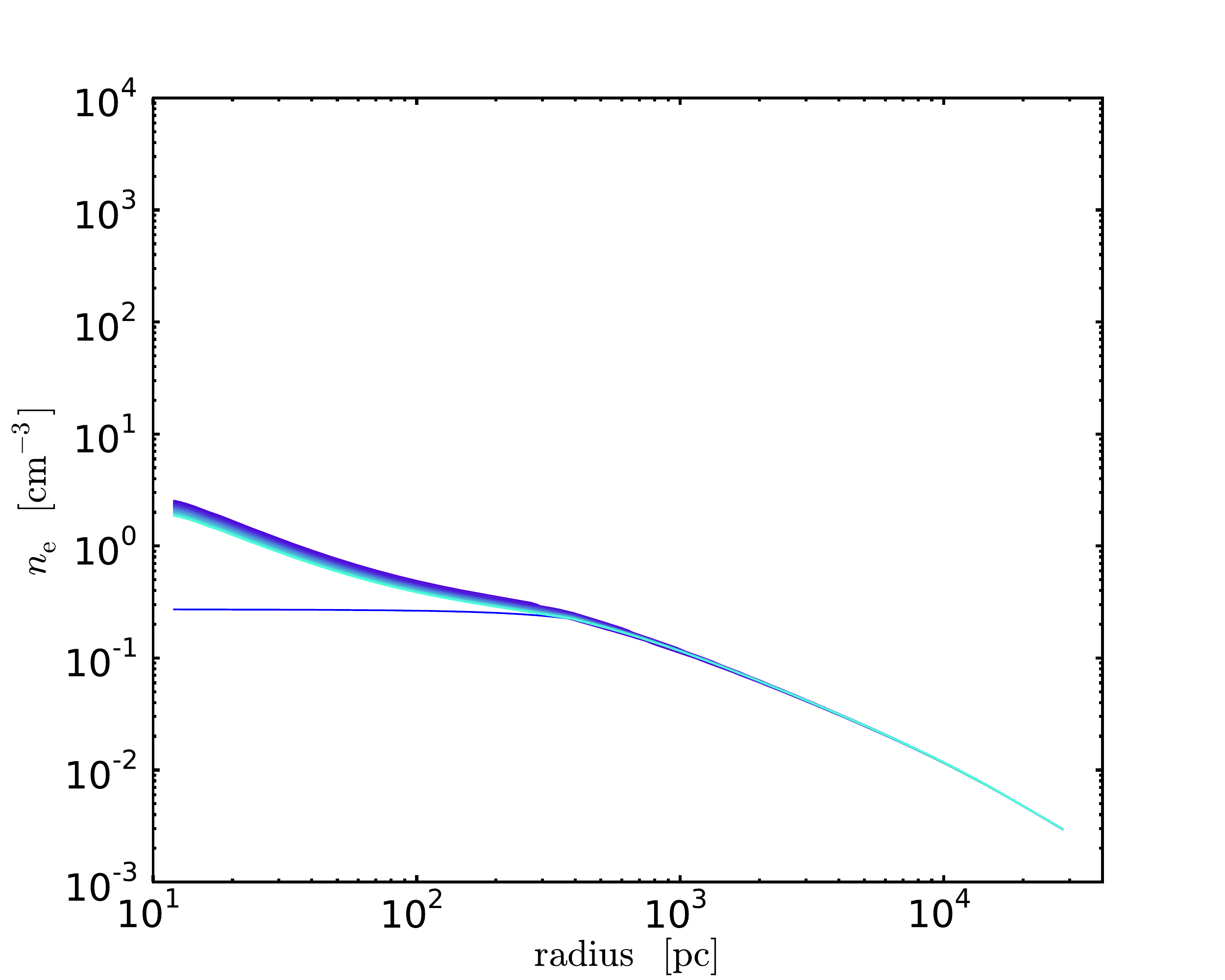}}
       \subfigure{\includegraphics[scale=0.28]{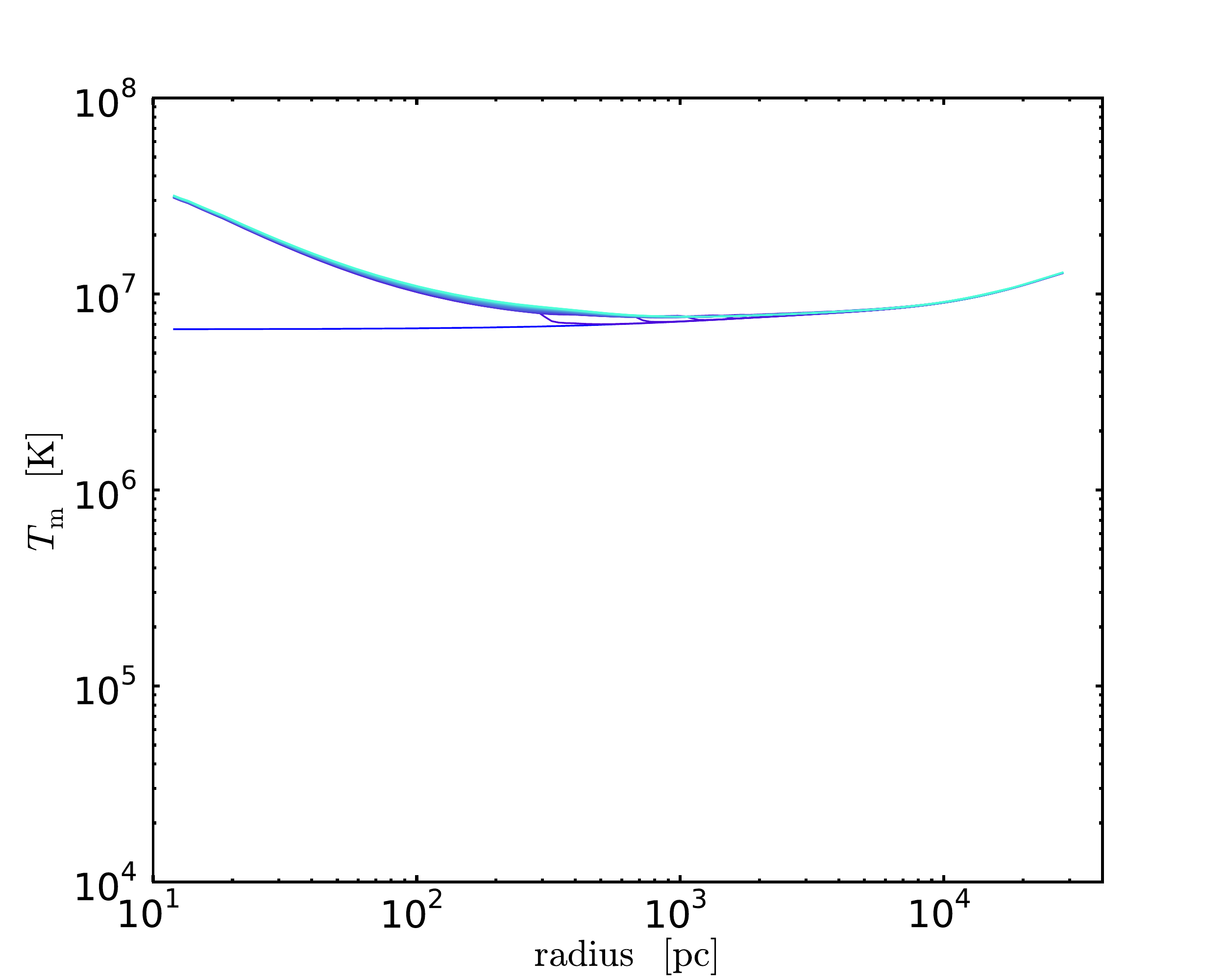}}
       \caption{Adiabatic (Bondi-like) accretion: evolution of the mass-weighted electron density ({\it top}) and temperature ({\it bottom}) radial profiles -- sampled every 1 Myr, from darker blue to cyan. Within $\sim$300 pc from the centre, the profiles are identical to the Bondi solution, and they smoothly join with the galactic gradients at large radii. The emission-weighted profiles (not shown) are very similar. Note the characteristic central cusp in temperature. \label{pure2}}
     \end{center}
\end{figure}

The reference accretion rate\footnote{When referring to the reference Bondi formula, $\dot{M}_{\rm B}$, we mean the usual Eq.~(\ref{MdotB}), with the normalisation $\lambda$ provided by Eq.~(\ref{lambdaPW}).} at the initial time is  $\dot{M}_{\rm B}\simeq 0.09\ \msun$ yr$^{-1}$. In the classic adiabatic case ($\gamma=5/3$), $\rho/c_{\rm s}^3$ is constant (in our stratified atmosphere only up to $\sim$3 $r_{\rm B}$). Therefore, the Bondi formula can be conveniently applied at $r_{\rm B}$ avoiding the complications due to the gravitational potential of the galaxy. 

As shown in Figure \ref{pure1} (bottom), the 3D numerical simulation is in excellent agreement with the analytic estimate (\S\ref{s:sink}), reaching the steady state after $\sim t_{\rm B}$: the black hole accretion rate $\dot{M}_\bullet$, through the sink sphere, is identical to $\dot{M}_{\rm B}$ (solid line). After circa 10 Myr, the large scale gradients start to affect the $\dot{M}_\bullet$ evolution and introduce a 30 percent decrement in the accretion rate ($\simeq0.07\ \msun$ yr$^{-1}$ at final time; top panel), while the difference with the instantaneous Bondi rate still remains within few per cent. 
Computing instead the reference $\dot{M}_{\rm B}$ on the kpc scale (averaging over 1$\,$-$\,$2 kpc), as commonly done in cosmological and large-scale simulations, introduces an increase in the normalised accretion rate by a factor of 3$\,$-$\,$4 (dashed line). This small bias is again caused by the declining density and slightly increasing temperature profiles, thus overestimating entropy in the Bondi formula.

\addtocounter{figure}{-1}
\addtocounter{figuresub}{1}

\begin{figure} 
    \begin{center}
     \subfigure{\includegraphics[scale=0.45]{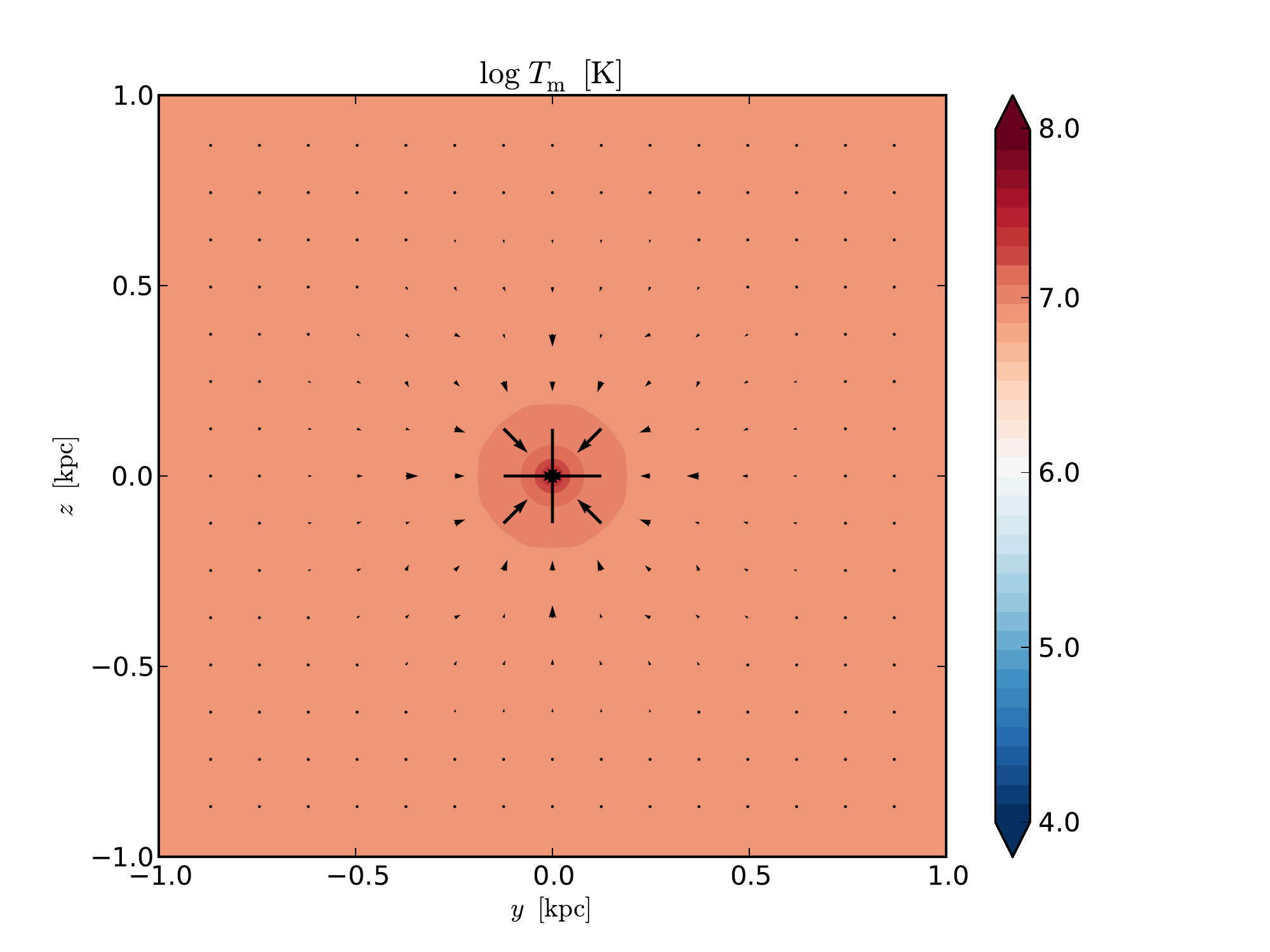}}
    \caption{Adiabatic (Bondi-like) accretion: mid-plane cross section through the mass-weighted temperature at final time (central 2 kpc$^2$); the normalisation for the velocity field is $500$ km s$^{-1}$ (the unit arrow length is 1/8 of the box width). Adiabatic accretion is smooth and steady, even in the presence of a galactic potential, and displays a significant temperature peak (even when emission-weighted). \label{pure3}}
     \end{center}
\end{figure} 

Several works adopting subgrid modelling, especially in cosmological simulations, boost the Bondi rate by an ad-hoc factor $\alpha\sim$100, mainly arguing that the entropy profiles are not resolved near $r_{\rm B}$ (\citealt{Booth:2009} for a literature review\footnote{They also note the importance of a multiphase medium, using a density-dependent boost factor, which in our cooling simulations would result similar to $\alpha\sim100$.}). However, as indicated by Fig.~\ref{pure1}, this bias is typically just a factor of a few. 
On the other hand, we will show that the onset of TI and cold accretion (\S\ref{s:cool}-\ref{s:heat}) can physically induce the boosted $\dot M_\bullet$ up to two orders of magnitude, which seems required to reproduce the observed growth of black holes and their host galaxies (\citealt{DiMatteo:2005}).

\subsection{Radial profiles \& dynamics} \label{s:adi2}

The radial profiles within 300 pc are identical to the analytic Bondi solution, with the correct power-law slopes\footnote{The limiting slopes are recovered by imposing 
the free-fall velocity $v_{\rm ff}=(2\,GM_\bullet/r)^{1/2}$ in the continuity equation $\dot{M}=4\pi r^2\rho v$, since the gas tends to become transonic at small radii.}: at very small radii the density approaches $r^{-3/2}$ and the temperature $r^{-1}$, and the slopes of both quantities flatten rapidly on larger scales, joining with the galactic/group gradients (Fig.~\ref{pure2}). The profiles are very smooth and have a low scatter, a sign that the code integrates the hydrodynamics equations in a consistent and stable way. Over 4 $r_{\rm B}$, the profiles are steady and very close to the initial conditions.

In the case of adiabatic accretion, the emission-weighted profiles (not shown) are very similar to the mass-weighted ones. This is due to the lack of cold gas. It is important to emphasise that the profile of the emission-weighted temperature $T_{\rm ew}$ within $\lta r_{\rm B}$ rises substantially. This is due to adiabatic compression and is in contrast to the pure unperturbed cooling case that exhibits only a slight decline, or the heated case where the profile is roughly flat. With future deeper and high-resolution X-ray observations, the mode of accretion can thus be unveiled through the measurements of the gas temperature profiles. 

The temperature map shown in Figure \ref{pure3} illustrates excellent stability of the simulation: even after 40 Myr (200 $t_{\rm B}$), there are only tiny oscillations and asymmetries due to the cartesian AMR grids ($\lta1$ per cent). The inflow velocity is overall subsonic (in contrast to the cooling run), approaching $M\sim1$ only near the accretor (\S\ref{s:sink}).
As we show below, this spherical (hot) accretion mode, although very convenient from the analytical point of view, is too idealised. Additional physics, such as cooling, stirring, and heating, drastically alters the whole evolution.

We note that several numerical schemes were tested (also for the non-adiabatic runs), to check the robustness of the results.  We tested different flux formulations (split, unsplit), Riemann solvers (HLLC, ROE, hybrid), data reconstruction methods (MUSCL, PPM), characteristic slope limiters (minmod, VanLeer, Toro), and key parameters (e.g. cfl number, interpolation order). They all give comparable results. Overall, we opted for the unsplit formulation with PPM plus hybrid solver, which is in principle more accurate.

\addtocounter{figuresub}{-2}

\begin{figure} 
      \begin{center}
      \subfigure{\includegraphics[scale=0.3]{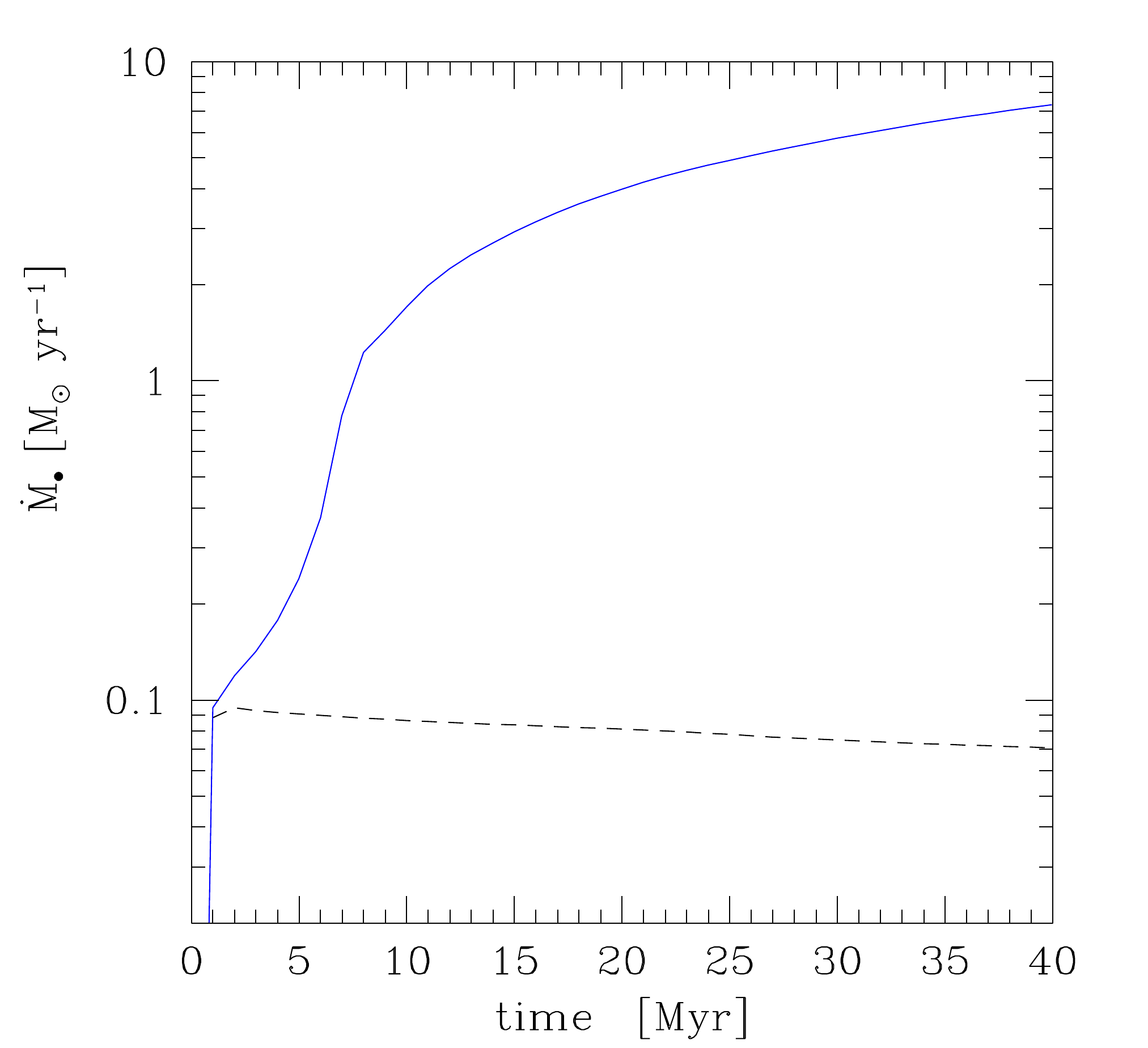}}
      \subfigure{\includegraphics[scale=0.3]{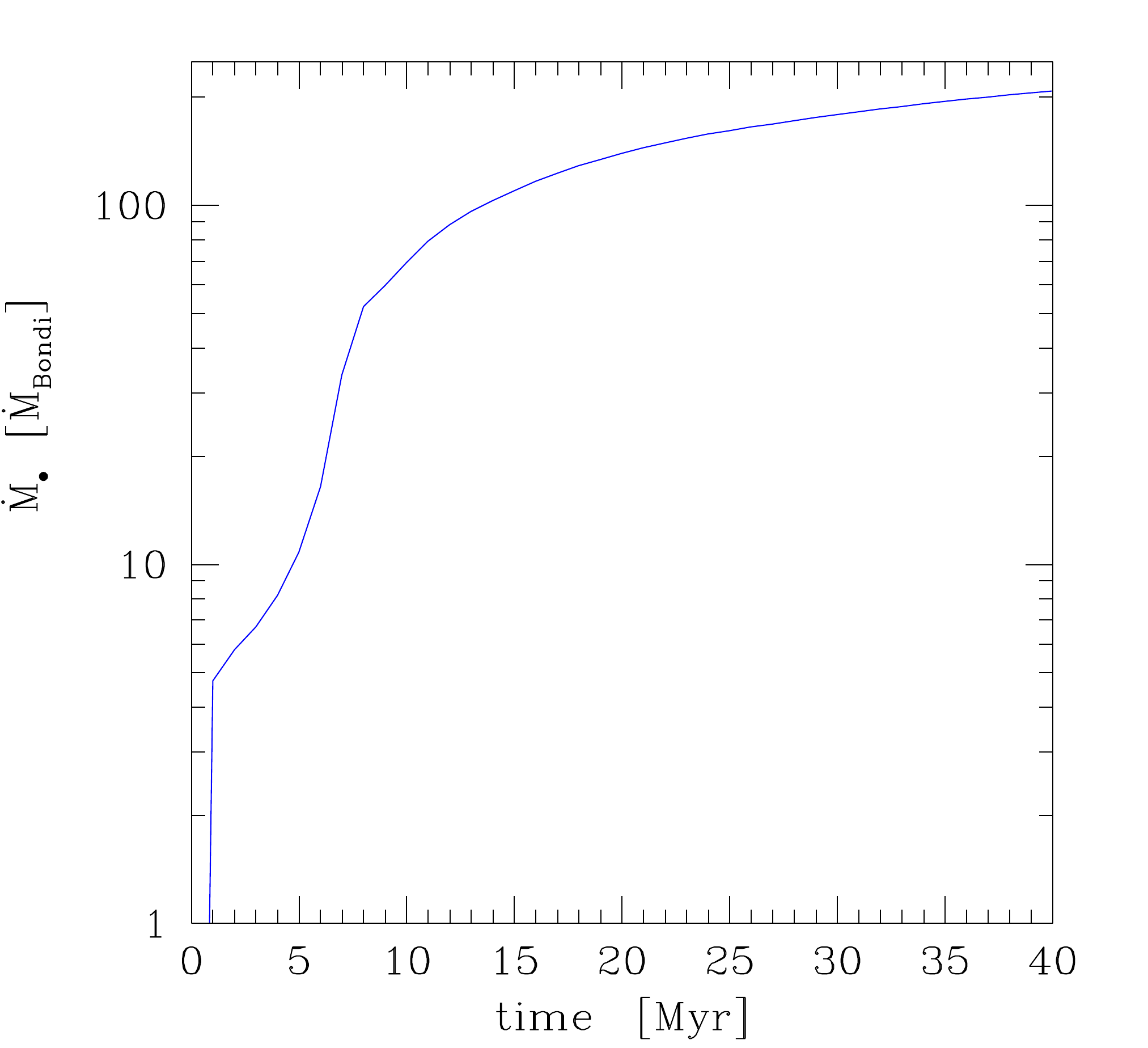}}
      \subfigure{\includegraphics[scale=0.3]{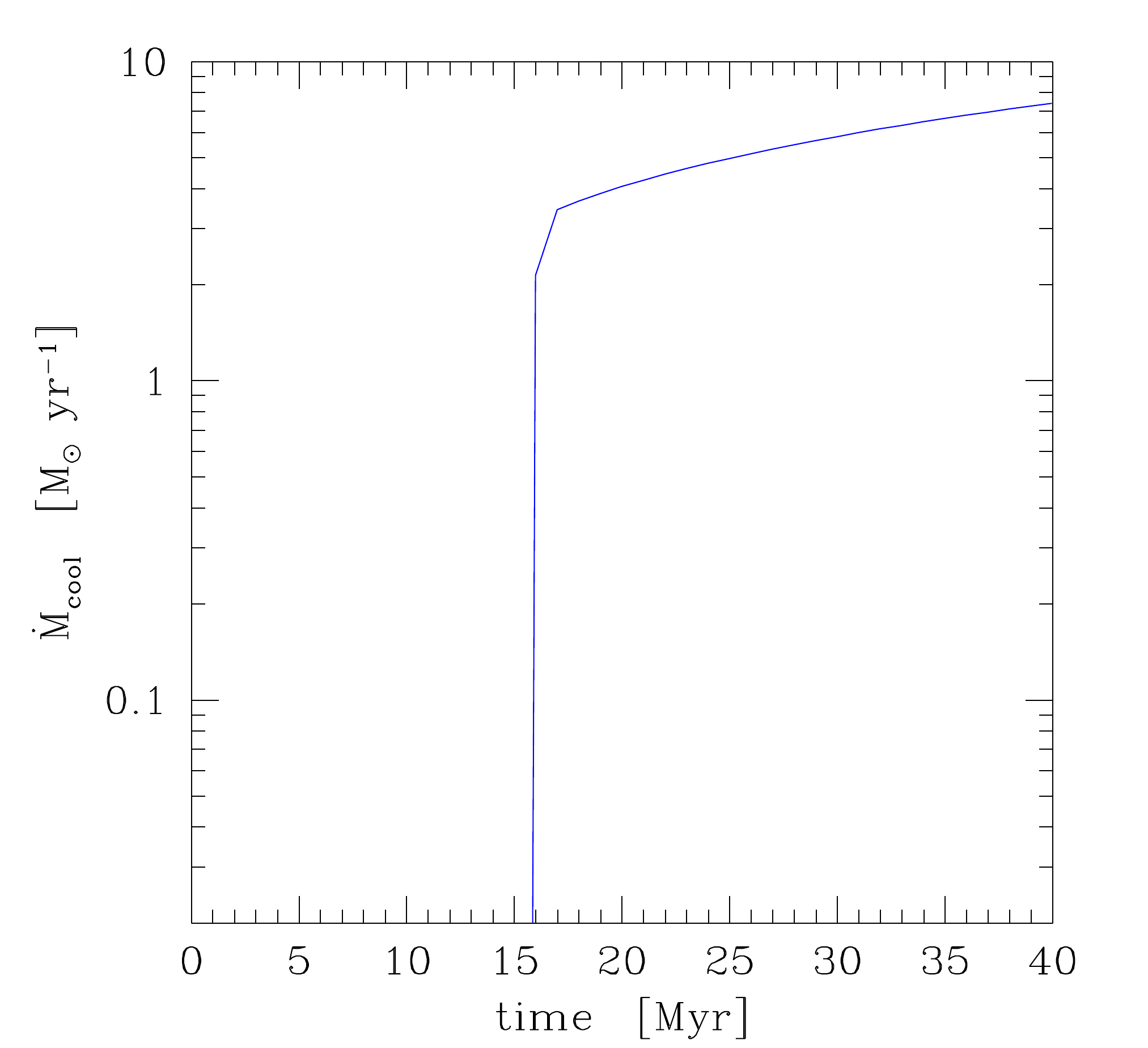}}
      \end{center}
      \caption{Accretion with cooling: evolution of the accretion rate (physical and normalised to the Bondi rate on kpc scale -- {\it top} and {\it middle} panel) and the cooling rate ({\it bottom}; we track only the very cold gas, with $T$ below $\sim$$10^5$ K). The dashed line corresponds to the previous adiabatic run. Note the dramatic boost in the accretion rate, by over two orders of magnitude with respect to the Bondi estimate, and the tight link with the cooling rate.  \label{cool1}}   
      \end{figure}    
      
\addtocounter{figure}{-1}
\addtocounter{figuresub}{1}      
      
\begin{figure} 
      \begin{center}
      \subfigure{\includegraphics[scale=0.28]{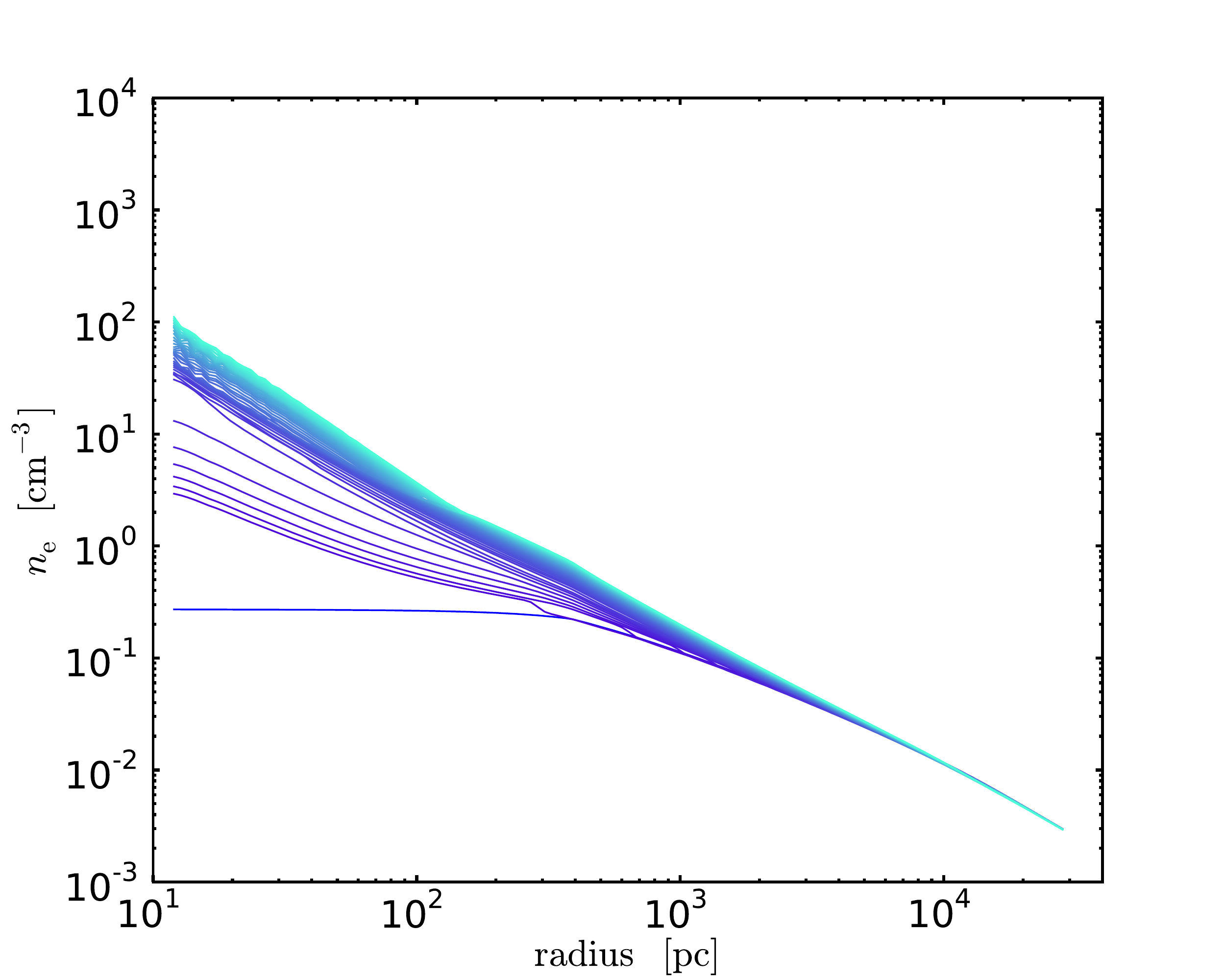}}
      \subfigure{\includegraphics[scale=0.28]{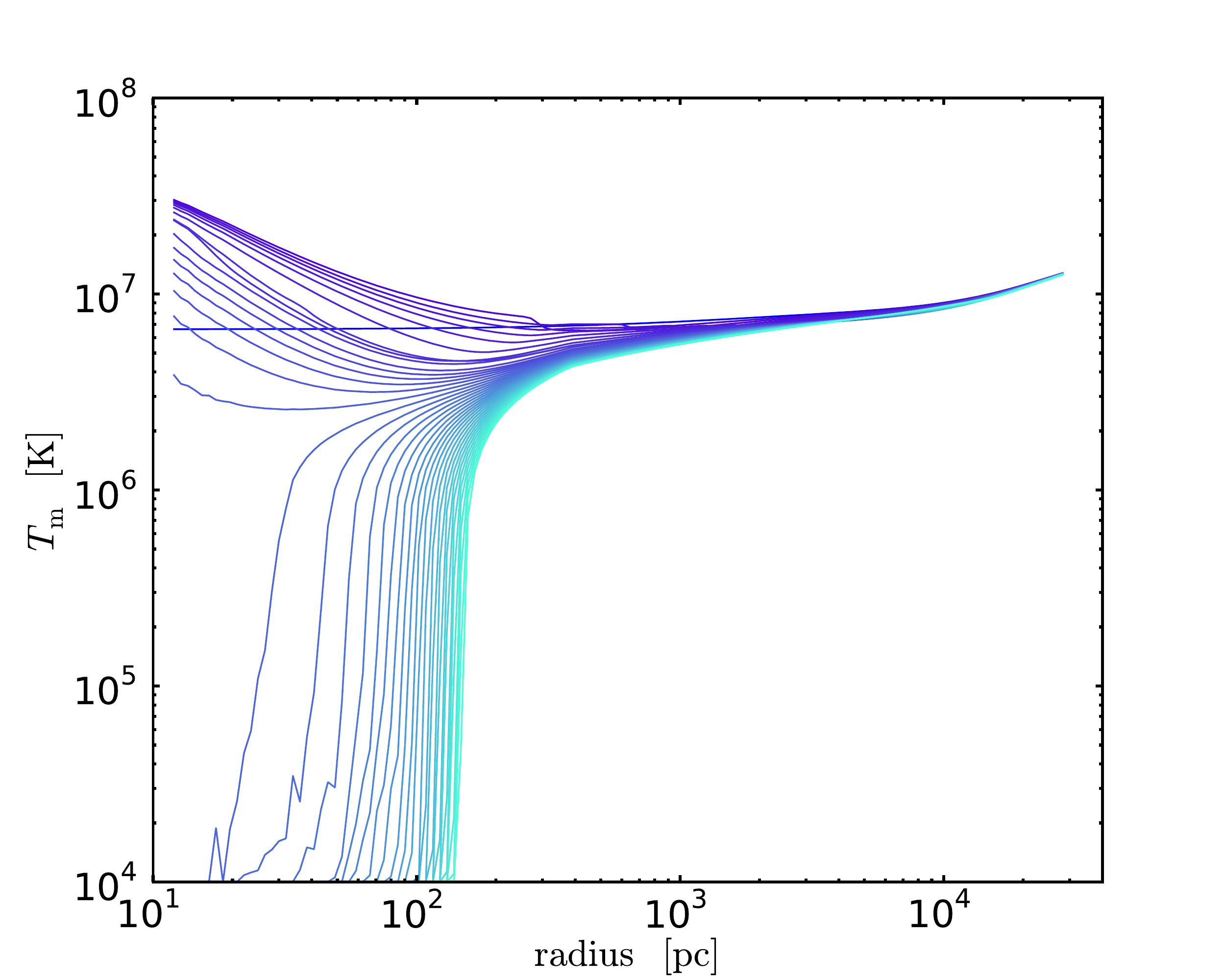}}
      \subfigure{\includegraphics[scale=0.28]{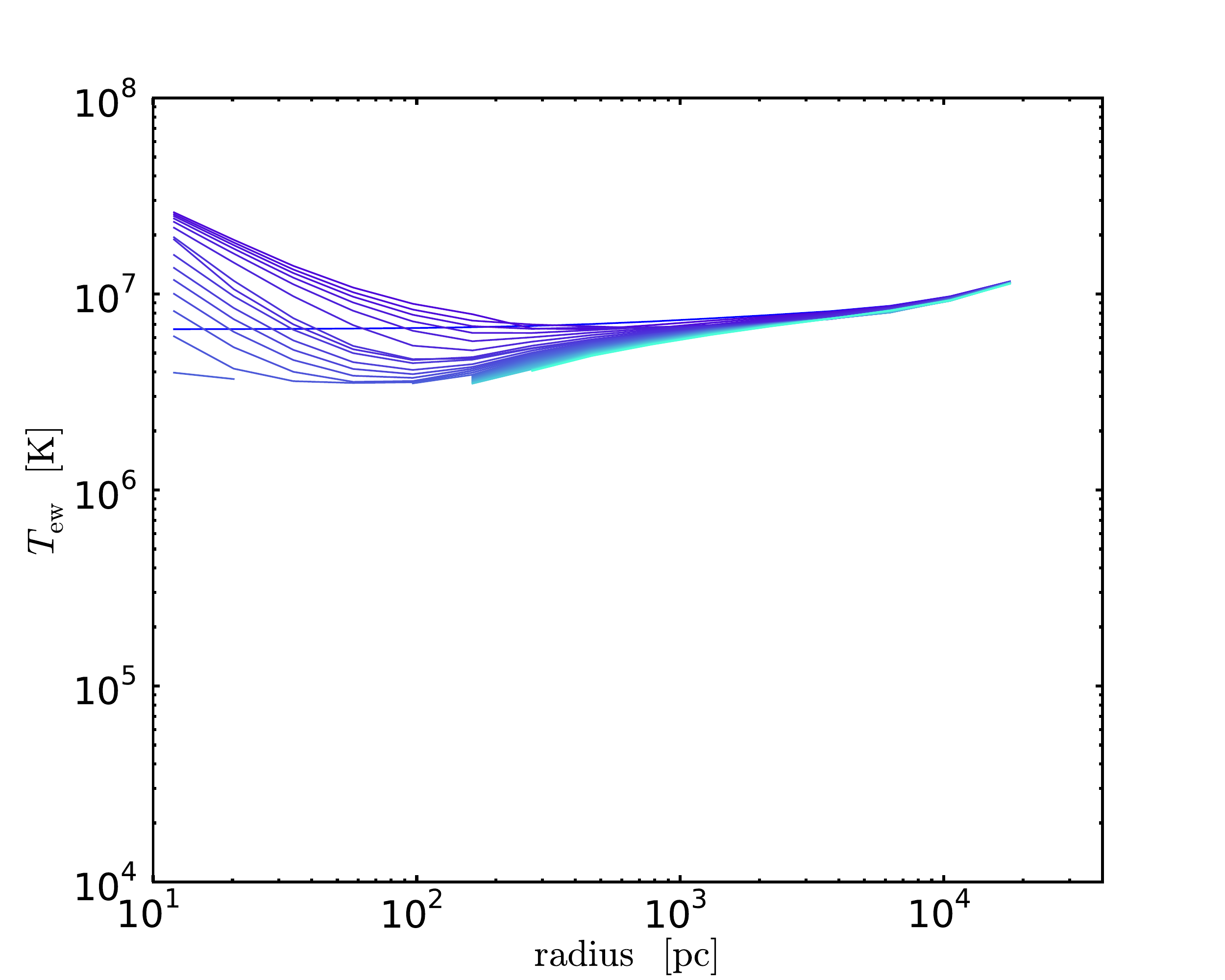}} 
     \end{center}
      \caption{Accretion with cooling: 3D mass- and emission-weighted radial profiles of density and temperature (cf. Fig.~\ref{pure2}). The $T_{\rm ew}$ profile has an X-ray cut of 0.3 keV and is computed in larger radial bins (to emulate a {\it Chandra} observation). It is evident that the massive condensation of cold gas, out of the hot phase, is entering the supersonic regime within few $r_{\rm B}$.  
       \label{cool2}}      
\end{figure} 

\section[]{Accretion with cooling}  \label{s:cool}
Radiative cooling radically changes the evolution of accretion flows. A fully realistic environment would imply a rough balance between cooling and heating (\S\ref{s:heat}). Nevertheless, it is very instructive to study the accretion flow when adiabaticity is violated and cooling prevails. The evolution may represent a phase in which the feedback heating is somehow delayed (as in the Phoenix cluster; \citealt{McDonald_Phoenix_Nat}) or still in the embryonic stage.

\subsection[]{Accretion \& cooling rate} \label{s:cool_mdot}
The accretion rate (Figure \ref{cool1}) exponentially increases until 7 Myr and then saturates around 7.5 $\msun$ yr$^{-1}$ at 40 Myr ($\sim$$5\ t_{\rm cool}$). The cooling rate follows a very similar evolution (bottom panel; the rapid transition is due to the fact that we are tracking only the very cold gas,  $T\lta$$10^5$ K). The key result is the drastic increase of the accretion rate by $\gta 100$ times (from 0.07 to 7 $\msun$ yr$^{-1}$). If we tried to predict $\dot{M_\bullet}$ via the Bondi formula, we would be wrong by over 200 times (middle panel). As noted in \S\ref{s:adi1}, even computing $\rho_\infty/c^3_{\rm s, \infty}$ on different scales would change the normalised accretion rate just by a factor of a few (see physical $\dot{M}_\bullet$). This simulation demonstrates that the Bondi prescription provides incorrect estimates for the accretion rate when cooling is present. Further modifications will be induced by stirring and heating (\S\ref{s:stir_cool}-\ref{s:heat}), but the break of adiabaticity plays the major role.

It could be tempting to modify the Bondi formula in order to retrieve an estimate for the accretion rate.
However, the latter is suited only for accretion in a single-phase symmetric and unperturbed medium. In the current scenario, the cold gas decouples from the hot flow, and enters the quasi free-fall regime (though there is still some friction with the hot medium). Therefore, we can infer that $\dot{M}_\bullet\sim M_{\rm cold}/t_{\rm ff}$ (but not $M_{\rm gas}/t_{\rm ff}$, as argued by \citealt{Hobbs:2012_subgrid} -- \S\ref{s:comp} -- especially in the presence of TI discussed below). Since the amount of the cold gas mass depends on the cooling time, which is the relevant timescale for achieving approximate steady state, our estimate for the accretion rate reduces to $\dot{M}_\bullet\sim M_{\rm gas}/t_{\rm cool}$, which is the cooling rate. Figure \ref{cool1} shows this tight relationship, which will still hold in the presence of stirring and thermal instabilities.
This argument (elaborated in more detail in the following Sections) provides a basis for the subgrid modelling based on cold feedback, which has been previously shown to be extremely efficient in self-regulating the thermodynamical evolution of galaxies, groups, and clusters (\citealt{Gaspari:2012c}; \S\ref{s:disc}). 

\addtocounter{figure}{-1}
\addtocounter{figuresub}{1}

\begin{figure*} 
     \subfigure{\includegraphics[scale=0.432]{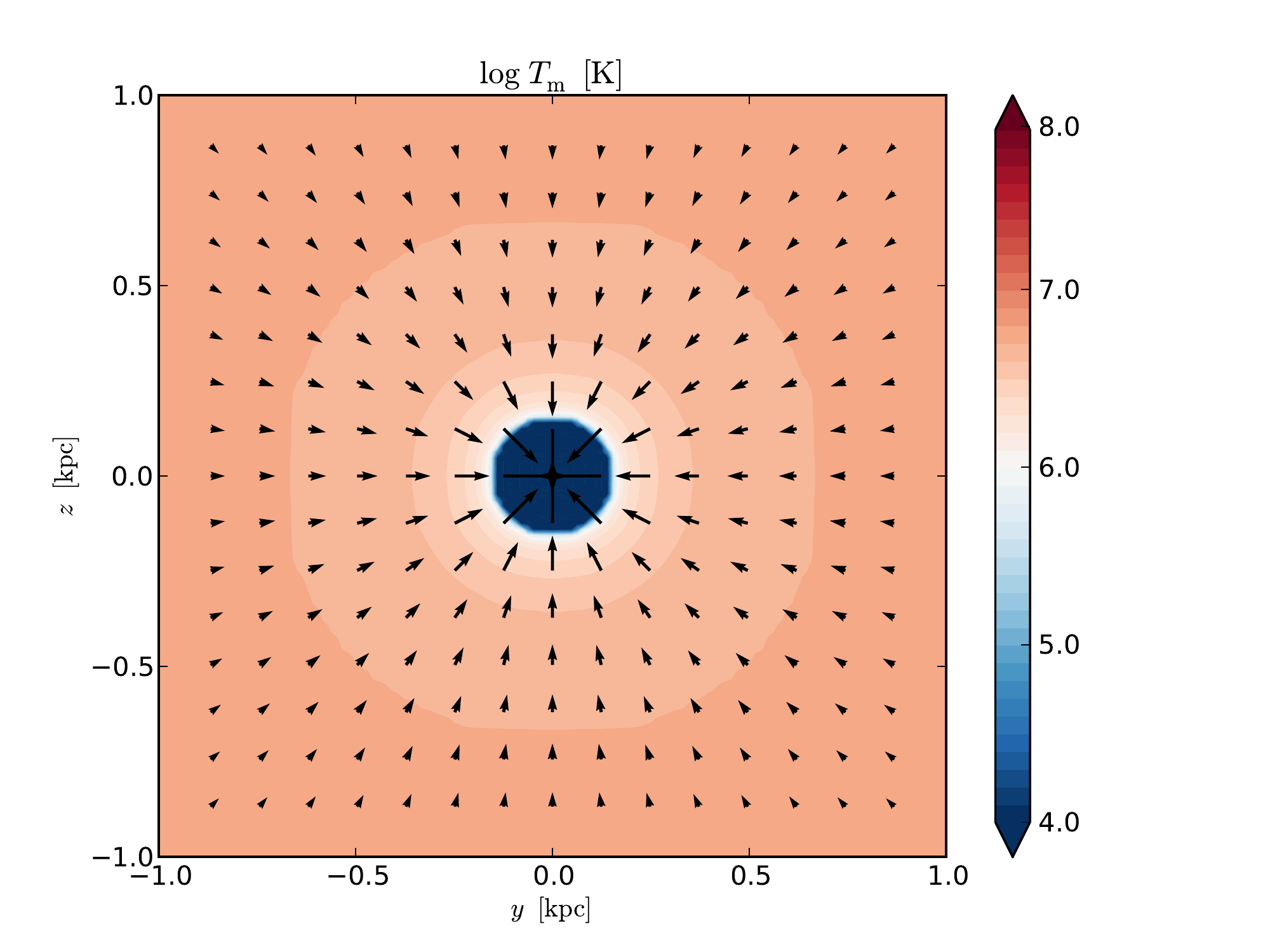}}
     \subfigure{\includegraphics[scale=0.432]{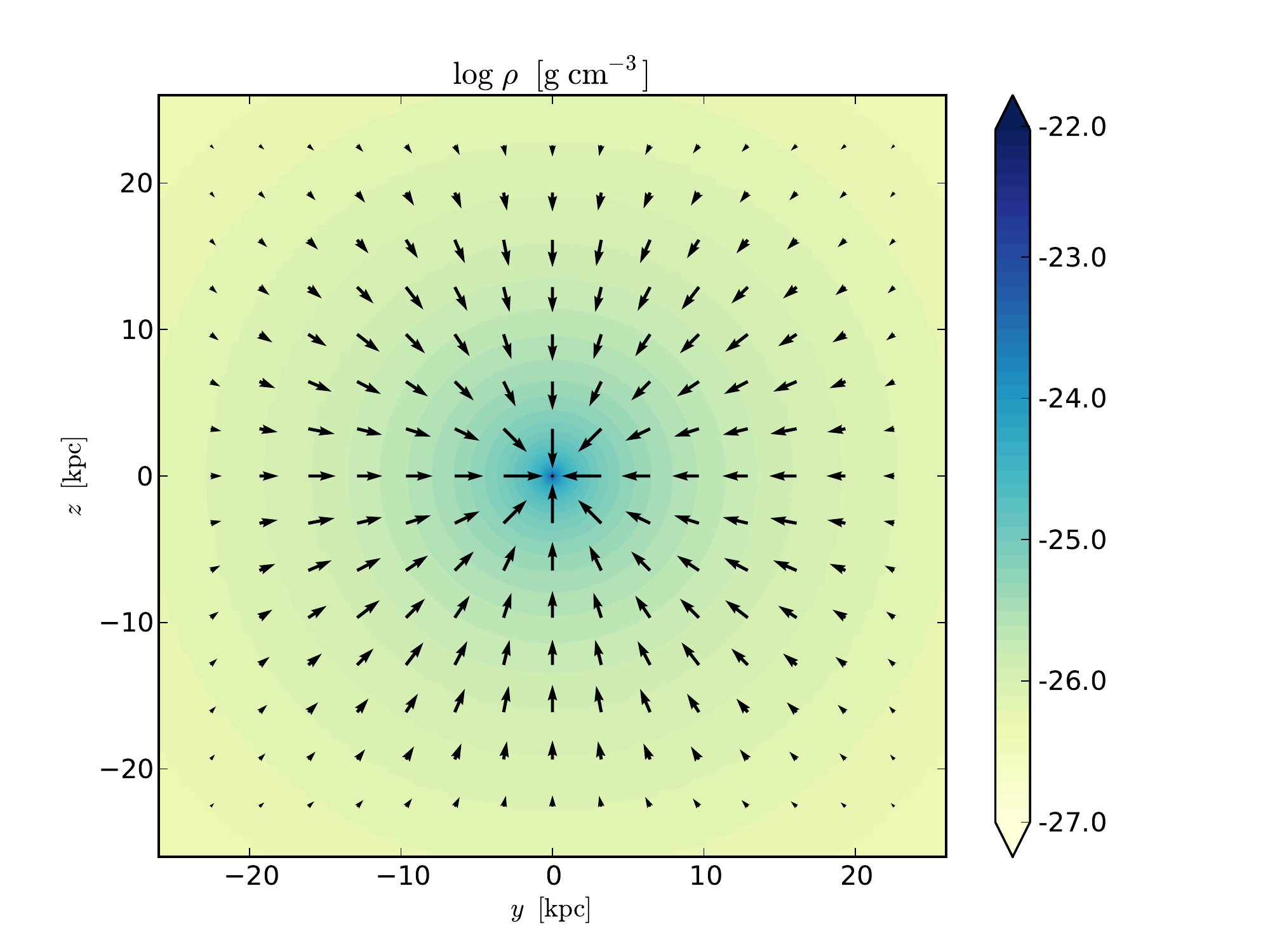}} 
     \caption{Accretion with cooling: mid-plane cross sections through $T_{\rm m}$ (central 2 kpc$^2$) and density (52 kpc$^2$) after 40 Myr; the velocity field normalisation is $6000$ km s$^{-1}$ (1/8 of the box width). The cooling gas becomes supersonic near few $r_{\rm B}$ due to the monolithic and symmetric condensation, which dramatically boosts $\dot M_\bullet$. The inflow continues to increase over 10s kpc due to the loss of pressure support.
         \label{cool3}}
    \end{figure*}  

\subsection[]{Radial profiles} \label{s:cool_prof}
The presence of cooling drastically changes the evolution of radial profiles (Figure \ref{cool2}). The central flat density profile, $n_{\rm e}$, is erased in less than 1 Myr as in the adiabatic run. However, the pile-up of gas is now almost two orders of magnitude larger, reaching 100 cm$^{-3}$ at 10 pc. In addition, the density slope is approximately $r^{-1.5}$ within two Bondi radii ($\sim$150 pc; rather than only very near the accretor), and it smoothly joins with the galactic gradient with $\sim r^{-1.3}$. A negative slope of 3/2 is typical of an accreting gas in free-fall (\S\ref{s:adi2}). In fact, the cooling gas quickly loses pressure and can be thus considered to be in free-fall within few $r_{\rm B}$.

The mass-weighted temperature profile (middle panel) clearly shows that the whole core has condensed out of the hot phase and reached a stable floor at $10^4$ K. This is illustrated by the significant drop in the emission-weighted profile within $2\ r_{\rm B}$ (bottom panel) due to the zero sensitivity of X-ray detectors below 0.3 keV.

One key result concerns compressional heating. Since $t_{\rm B}$ is much lower than the central initial $t_{\rm cool}$ ($\sim$30$\times$), it is tempting to study accretion models neglecting cooling (e.g. \citealt{Narayan:2011}). However, as the temperature profiles show, adiabatic compression
can delay cooling only for $\sim3$ Myr. Even if the ratio of $t_{\rm cool}/t_{\rm ff}$ initially increases toward smaller radii ($\propto r^{-1/2}$, assuming a transonic inflow with $T\propto r^{-1}$), the dense warm gas near two Bondi radii starts to cool substantially creating an exponentially colder dip in temperature. After one cooling time ($\sim$8 Myr), the {\it conditions near one Bondi radius are completely altered} by radiative cooling, and the temperature in the inner parsec region starts to gradually collapse\footnote{Note that very low central resolution, as in cosmological simulations, will create
an artificial inversion of the temperature gradient ($\nabla T<0$), in contrast to the findings of Fig.~\ref{cool2}.}
down to $10^4$ K. At this point, the Bondi solution becomes unrealistic. Adiabatic compression can not stop cooling, it can only somewhat delay the collapse. A quasi adiabatic interstellar medium, even within a few parsec from the centre, represents a short transient phase in the history of accretion. Interestingly, as we explain in the following Sections, the same conclusions hold when the spatially distributed heating and turbulence are present. These processes also lead to a globally thermally stable atmosphere with local thermal instability and the formation of multiphase gas clouds.

On the larger galactic scales ($r\sim1$$\,$-$\,$10 kpc), adiabatic compression due to the cD potential seems to be able to partially inhibit the weak cooling (the $T$ decrement is no more than 40 per cent), at least for 40 Myr. However, the central temperature dip continues to expand and a massive cooling flow would develop in a few 100 Myr 
(with average cooling rate $\sim$20 $\msun$ yr$^{-1}$). 
Note that the smoothly varying emission-weighted $T_{\rm ew}$ profile (with X-ray cut 0.3 keV; not projected) paradoxically conceals the development of the massive cooling flow, which is instead exposed by the high cooling (and star formation) rates.

\subsection[]{Dynamics}
The temperature slice in Figure \ref{cool3} (left panel) shows complete condensation of the galactic core into the cold $10^4$ K phase. The collapse is symmetric and monolithic and no thermal instabilities are present. This giant cold core/sphere continues to slowly increase in size. This is accompanied by a steady increase in cooling rates. The flow becomes supersonic at $\sim$170 pc (40 Myr). The transition to the supersonic regime occurs at the boundary of the cold nucleus and the hot gas at larger radii. 
This boundary coincides with the region where the gas enters the fast cooling regime $\lta10^{6}\ {\rm K}$ ($\Lambda(T)\propto T^{-1/2}$). 

The very large sonic radius, compared with that in the Bondi problem, is a key
characteristic of radiative accretion flows (especially in realistic potentials including the black hole, galaxy, and dark matter).
Even if the initial sonic point lied within $r_{\rm B}$, the mass accretion rate would eventually become so large pushing the sonic limit $r_{\rm s}>r_{\rm B}$ (cf. \citealt{Quataert:2000, Li:2012, Mathews:2012})\footnote{From another perspective, the effective $\gamma$ is lower than $5/3$ in the radiative case, which increases the sonic radius.}. Nevertheless, cool core systems typically display a minimum $t_{\rm cool}/t_{\rm ff}$ -- linked to the eventual sonic point -- above the Bondi radius (e.g. \citealt{McCourt:2012}).
As shown in the large-scale density map with the velocity field overlaid (right panel), the inflow is significantly enhanced even at the large distance of $r\sim20$ kpc, which consequently boosts $\dot{M}_\bullet$.

We note that the solutions are characterised by good numerical stability with very low oscillations both in space and in time. Achieving this is non-trivial especially in the presence of radiative cooling which tends to amplify oscillations. The key to prevent such oscillations is to properly model the sink region (see \S\ref{s:sink}). The gas must be properly evacuated when central resolution is very high. Otherwise, the gas rapidly accumulates and produces severe back-pressure on the inflowing gas at larger radii. This modifies the inner solution (e.g. $v_r>0$) and leads to artificial oscillations. This is the main reason why it is crucial to adopt the vacuum sink region, especially for the case of accretion driven by TI.

Although adding radiative cooling to hydrodynamics is more physical than resorting to the use of the standard adiabatic Bondi solution, the fact that the flow is still perfectly symmetric, steady, and unperturbed still implies that this simulated accretion does not adequately approximate that occurring in a realistic astrophysical system. In the next Sections, we increase the realism of the accretion process but note in passing that greater realism of these simulations does not affect our finding that $\dot M_\bullet$ is significantly boosted beyond the prediction from the Bondi formula.

\section[]{Accretion with cooling \& turbulence}  \label{s:stir_cool}
Observed systems are never characterised by perfect spherical symmetry. In a cooling medium, turbulence can seed new structures and nonlinear instabilities in the atmospheres provided that the seeded fluctuations are sufficiently large. Turbulent motions in the hot medium can be driven by numerous astrophysical phenomena including AGN feedback, galaxy motions, supernovae, and mergers (e.g. \citealt{Ruszkowski:2011,Vazza:2012}). We keep our model as general as possible in order to better understand basic physics of the accretion process. Therefore, continuous non-compressive turbulence is driven throughout the evolution with a low velocity dispersion reaching $150-180$ km/s when fully developed, i.e. $M\sim0.35$$\,$-$\,$$0.4$. In \S\ref{s:strong1}, we test stronger turbulence up to the transonic regime, a less general case, but still relevant when the atmosphere is heated by violent events.

\subsection[]{Dynamics}
\begin{figure} 
      \begin{center}     
      \end{center}
     \subfigure{\includegraphics[scale=0.45]{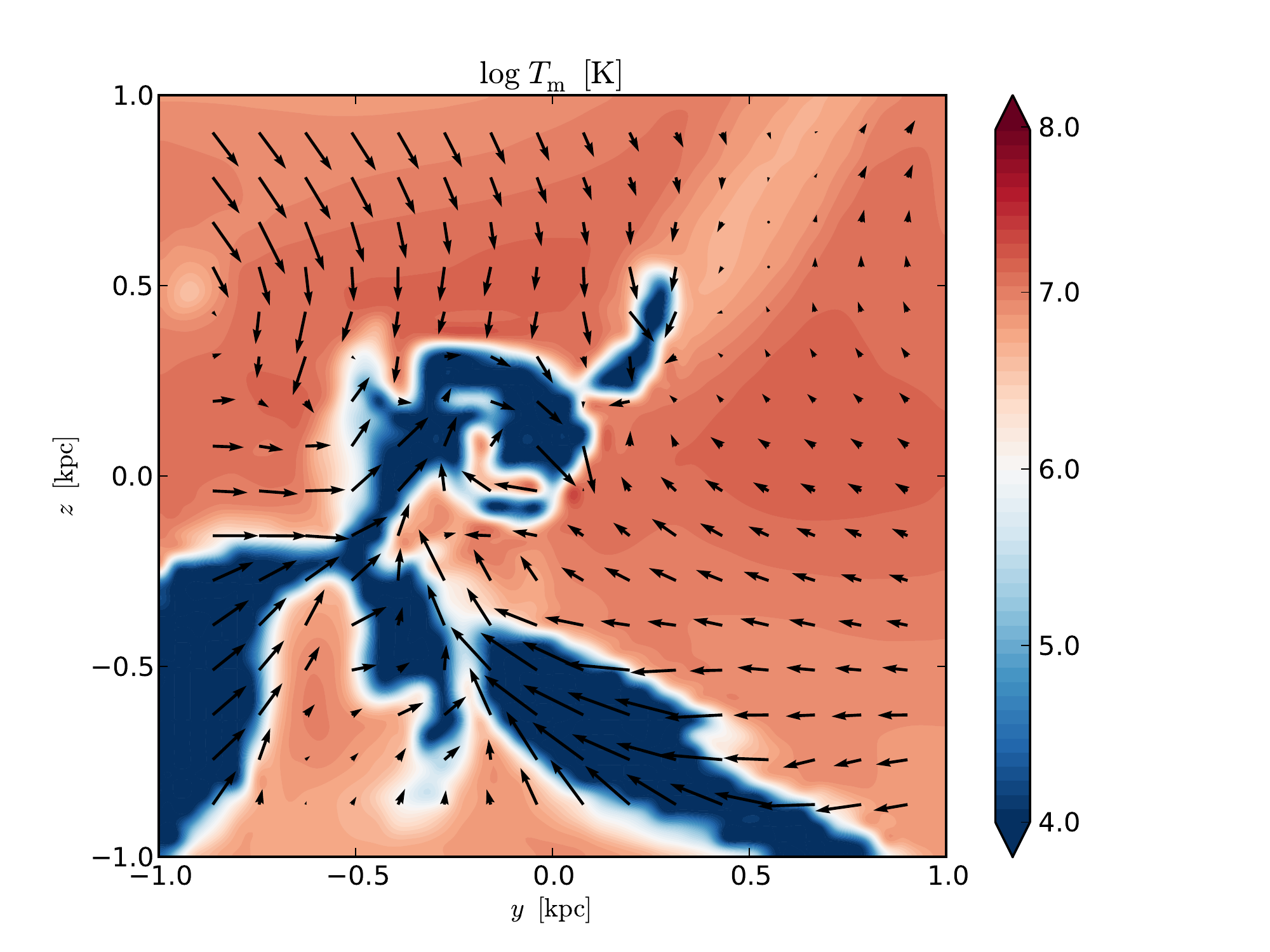}}
     \caption{Accretion with cooling \& subsonic turbulence ($\sigma_v\,$$\sim$$\,150$ km s$^{-1}$, $M\sim0.35$): mid-plane cross section through $T_{\rm m}$ ($t=40$ Myr). Including moderate turbulence generates perturbations in the hot medium. These perturbations grow nonlinearly via thermal instability and produce cold filamentary structures.
The dynamics of the gas is chaotic, driven by collisions and tidal motions, which substantially reduce the clouds angular momentum and accelerate the sinking rate.
    \label{stir_cool1}}
\end{figure}

For the present case it is more instructive to start by discussing cross sections through the temperature and velocity distributions (Figure \ref{stir_cool1}). The first key result of this more realistic model is the formation of dense cold clouds  ($T\sim10^4$ K). This phenomenon is simply explained through the growth of the nonlinear thermal instability (cf. \citealt{Field:1965, Krolik:1983, Pizzolato:2005, Sharma:2012}, or \citealt{Rees:1977} for a cosmological perspective). This occurs despite the fact that there is no distributed heating, which can promote the growth of even small fluctuations (\S\ref{s:heat} below). The moderate-amplitude stirring destroys the initial symmetric conditions and induces perturbations in density. Overdense regions then start to cool progressively faster. The process becomes nonlinear (although not exponentially as in the heated case) and cold dense filaments condense out of the hot phase after $\sim$10 Myr (Fig.~\ref{stir_cool1}). As the cold gas falls down in the potential well, compression accelerates cloud cooling, although in part delayed by adiabatic heating.

In order to become nonlinear, the amplitude of the density fluctuations should be $>10$ per cent in the cooling-only case. Otherwise perturbations are advected inwards without being able to fully develop (e.g. \citealt{Balbus:1989,Binney:2009,Li:2012}). The key point is that a very common subsonic turbulence ($M\sim0.35$) can induce sufficient overdensity amplitude to generate TI and extended multiphase gas.   

\addtocounter{figuresub}{-2}     

Another crucial result is that the mode of accretion has completely changed. Cold accretion is now fully {\it chaotic} (due to the high sensitivity of the long-term detailed dynamics on TI) and {\it stochastic} (given the random nature of the fluctuations)\footnote{In this work, we alternate the use of `chaotic' and `stochastic', since they are linked (see also \S\ref{s:heat}); note also that specialised mathematical work may present different definitions of `chaotic'.}. 
Cold clouds and filaments form first where $t_{\rm cool}/t_{\rm ff}$ has a minimum\footnote{In the absence of heating, the threshold for multiphase gas formation is also proportional to the perturbation amplitude (cf. \citealt{Sharma:2012}).} of $\sim4\,$-$\,$5 at 250 pc (see \S\ref{s:stir_cool_prof}). Over time the cold blobs and filaments start to appear at progressively larger radii, where the ratio $t_{\rm cool}/t_{\rm ff}$ is initially higher and where there is less (spherical) geometric compression. 
The accretor region is small compared with the typical cross section of the condensing clouds/filaments, thus they start to collide multiple times: in part they are accreted, in part they form a transient rotational structure (torus). 
New clouds fall down and collide with the clumpy torus, drastically altering its dynamics and shape. The presence of the torus further facilitates accretion, since it increases the cross section for capturing the clouds. This compensates for the otherwise expected decrease in the accretion rate due to the cloud finite angular momentum. 

The cold blobs should {\it not} be considered tiny bullets with insignificant cross section. As we quantify in \S\ref{s:stir_cool_mdot}, the accretion rate is now only slightly reduced compared with the unperturbed cooling case, and it is still up to two orders of magnitude higher than the Bondi prediction. Overall, accretion is significantly chaotic rather than ordered with unperturbed ballistic cloud trajectories. In passing, we note that starting from the assumption of chaotic accretion, \citet{King:2007} and \citet{Nayakshin:2007} inferred interesting 
observational implications which we review in \S\ref{s:comp}.

\subsection[]{Radial profiles}  \label{s:stir_cool_prof}

\addtocounter{figure}{-1}
\addtocounter{figuresub}{1}      

\begin{figure} 
      \begin{center}
      \subfigure{\includegraphics[scale=0.28]{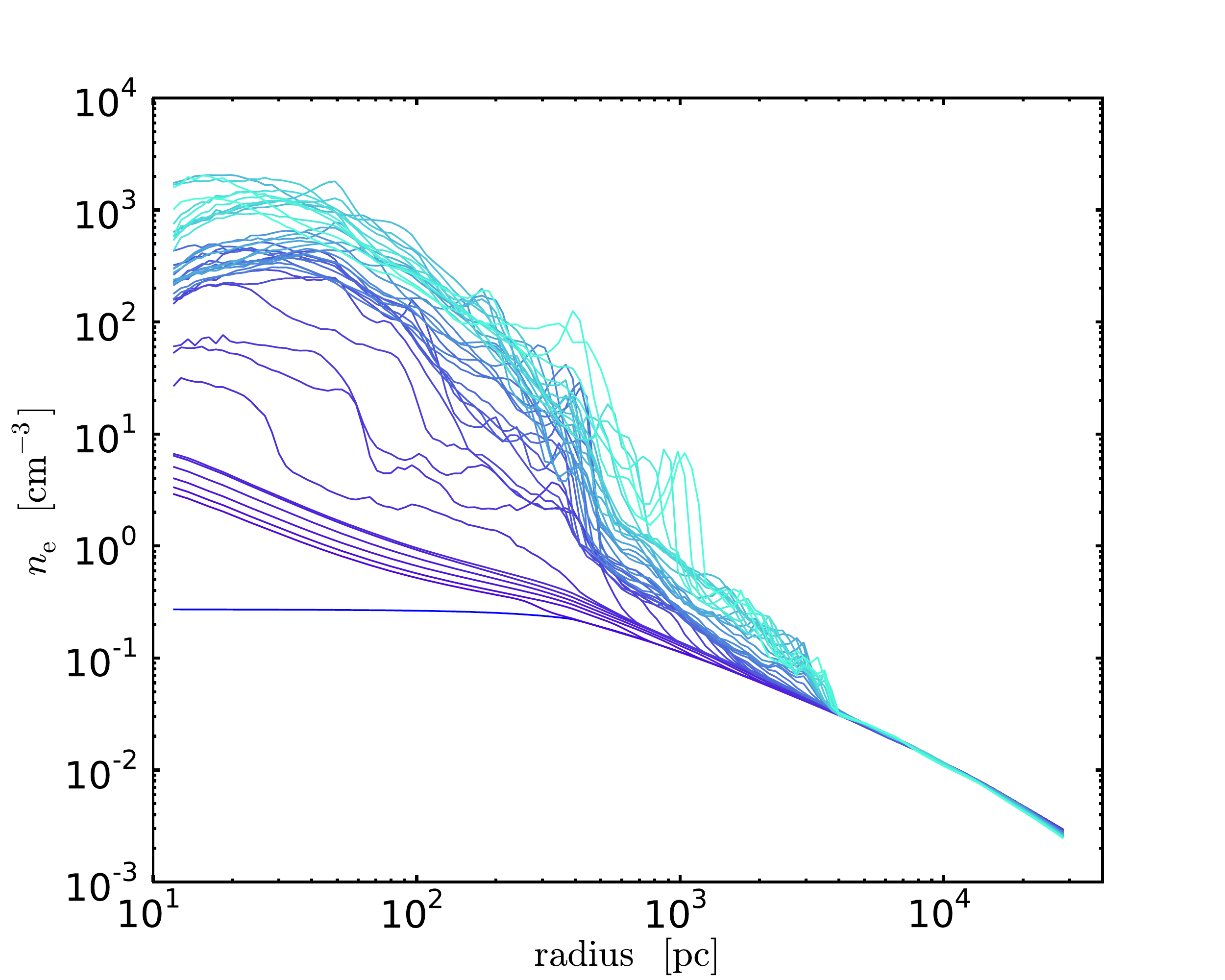}}
      \subfigure{\includegraphics[scale=0.28]{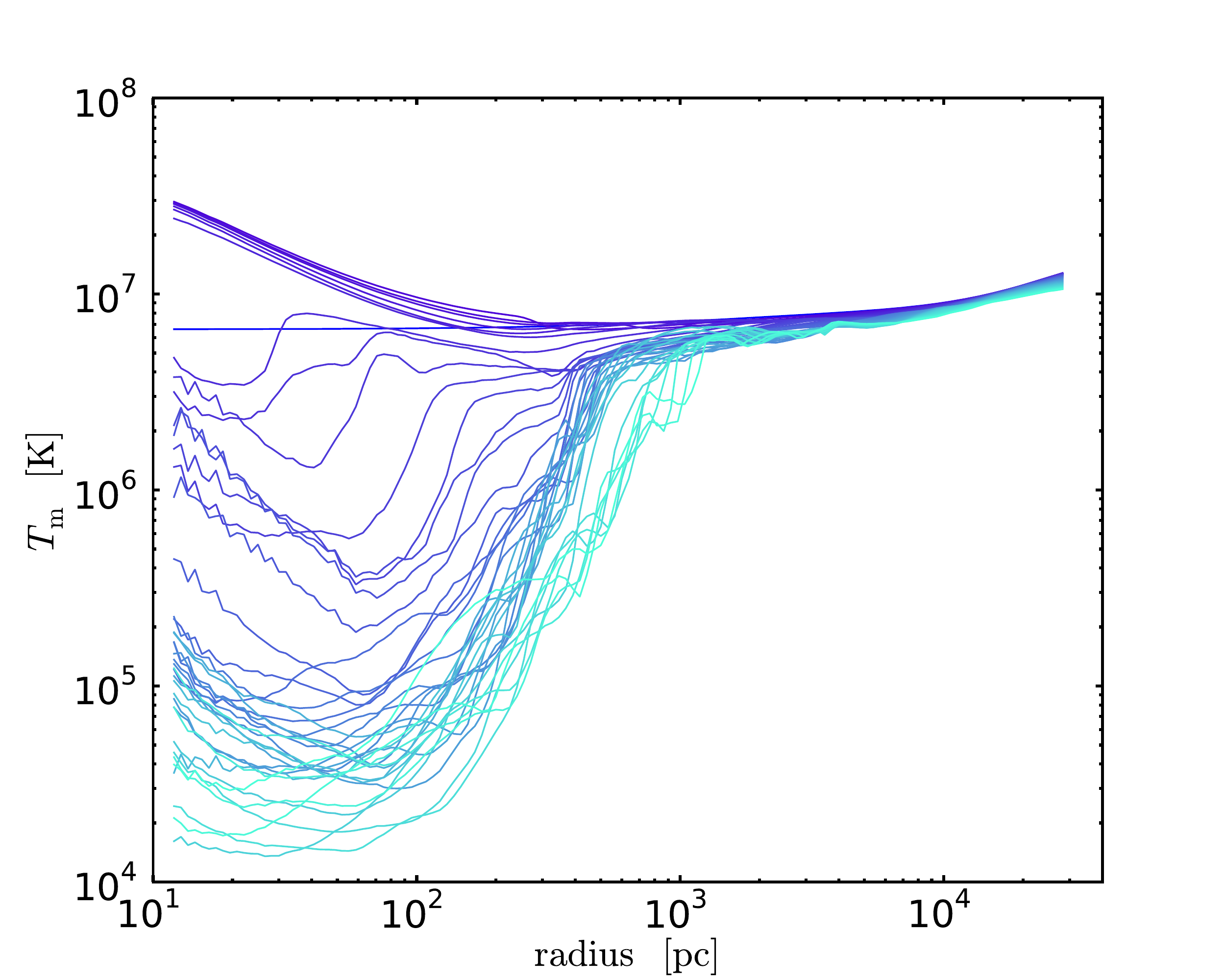}}
      \subfigure{\includegraphics[scale=0.28]{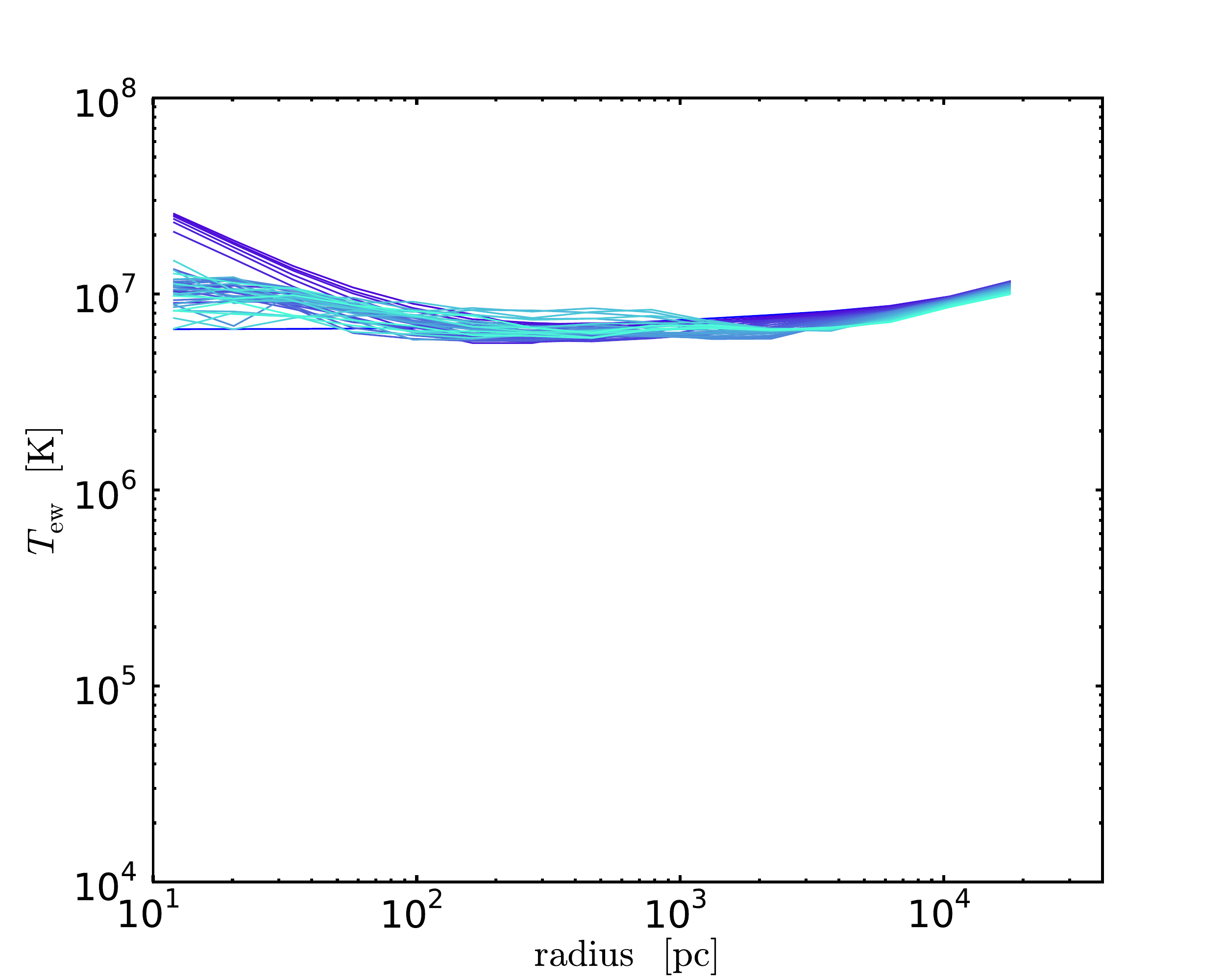}} 
      \end{center}
      \caption{Accretion with cooling \& turbulence ($M\sim0.35$): mass-- and emission--weighted radial profiles of density and temperature (cf. Fig.~\ref{cool2}). Note the condensation of warm/cold gas out of the hot phase via thermal instability, and the fluctuations imparted by turbulence. 
      \label{stir_cool2}}    
\end{figure}  
  
Unlike in the pure unperturbed cooling run (\S\ref{s:cool}), turbulence-induced overdensities lead to the formation of substantial amounts of cold gas clumps. This happens within the first $\sim$5 Myr. 
Just as in the pure cooling case the temperature still drops to $\lta10^5$ K, but the chaotic nature of the process is now evident in the mass-weighted profiles (Figure \ref{stir_cool2}). Cold blobs manifest themselves as noticeable density fluctuations superimposed on a strong increase in density within the central 3 kpc (up to $10^3$ cm$^{-3}$). The average central density is now an order of magnitude higher compared with the unperturbed cooling run.

On the other hand, if the density and temperature is weighted with X-ray emission, the picture becomes very different. The temperature distribution shows less fluctuating profiles, and the density is only slightly peaked, up to $\sim$10 cm$^{-3}$ (not shown -- see Fig.~\ref{heat2} for a similar example). Remarkably, $T_{\rm ew}$ (bottom panel) is almost flat, from few pc up to 10 kpc, and slightly hotter than the initial condition, $\sim$$10^7$ K. We note that high resolution X-ray observations could shed light on the mode of accretion by determining the temperature gradient in galactic nuclei. 
NGC 4472 and NGC 4261 might be two interesting cases of flat inner $T$ profile (\citealt{Humphrey:2009}).
Notice that the mass of the black hole can be no longer estimated via the temperature peak, as in the adiabatic Bondi-like case, due to presence of cooling and the departure from hydrostatic equilibrium. We also note that, for isolated or small galaxies ($T_{\rm vir} < 0.5$ keV), the overall $T$ gradient is usually negative (\citealt{Gaspari:2012b}), due to the lack of circumgalactic gas, and may be erroneously identified with the hot mode regime.

The minimum in the $t_{\rm cool}/t_{\rm ff}$ profile determines where the onset of the TI and cloud formation will begin. In our case, this minimum occurs around 250 pc (see Figure \ref{heat3} for a similar trend). 
Beyond $\sim$7 kpc, the cooling time becomes too long ($n_{\rm e}<10^{-2}$ cm$^{-2}$) to produce significant cooling during the evolution. 
In the other regime, within the Bondi radius ($\lta\,$80 pc), the dynamical time is too short for the gas to condense while being advected inwards. It is thus not surprising that most of the multiphase gas in the core of galaxies, groups, and clusters is observed from hundreds pc to several kpc (e.g. \citealt{Mathews:2003, Rafferty:2008, McDonald:2010, McDonald:2011a, Davis:2012, Werner:2013}). This circumnuclear region is the nursery of cold gas and star formation. In spiral galaxies, where the whole system can have $t_{\rm cool}/t_{\rm ff}<10$ up to several tens kpc, TI clouds may be even more common (e.g. the high-velocity clouds in the Milky Way and M31). 

Outside the core ($\sim$5 kpc), the profiles are very similar to the initial conditions. Only a very small amount of gas is flowing out of the box due to turbulence. Therefore, it would not be appropriate to study the evolution for several tens Myr in smaller domains ($\lta10$$\,$-$\,$15 kpc). Further, gas outflow from the computational domain would be exacerbated by too strong turbulence. The large dynamical range achieved in the present simulations is thus crucial to properly model also large scales.

Another reason to keep turbulence at low levels is that stirring motions heat the gas via turbulent dissipation. The specific heating rate due to the dissipation of turbulence can be approximated as $\dot{e}_{\rm turb} \sim \sigma_v^3/L$ ($\propto M^3$; subject to order unity uncertainty). Subsonic velocity dispersions of $\sim$150 km s$^{-1}$ ($L\gta4$ kpc) produce negligible heating compared with radiative cooling, as shown by the $T$ profiles at large radii (Fig.~\ref{stir_cool2}). On the other hand, strong transonic turbulence (\S\ref{s:strong1}) significantly increases the thermal energy of the gas and damps thermal instability.

\addtocounter{figure}{-1}
\addtocounter{figuresub}{1}      

\begin{figure} 
      \center
      \subfigure{\includegraphics[scale=0.3]{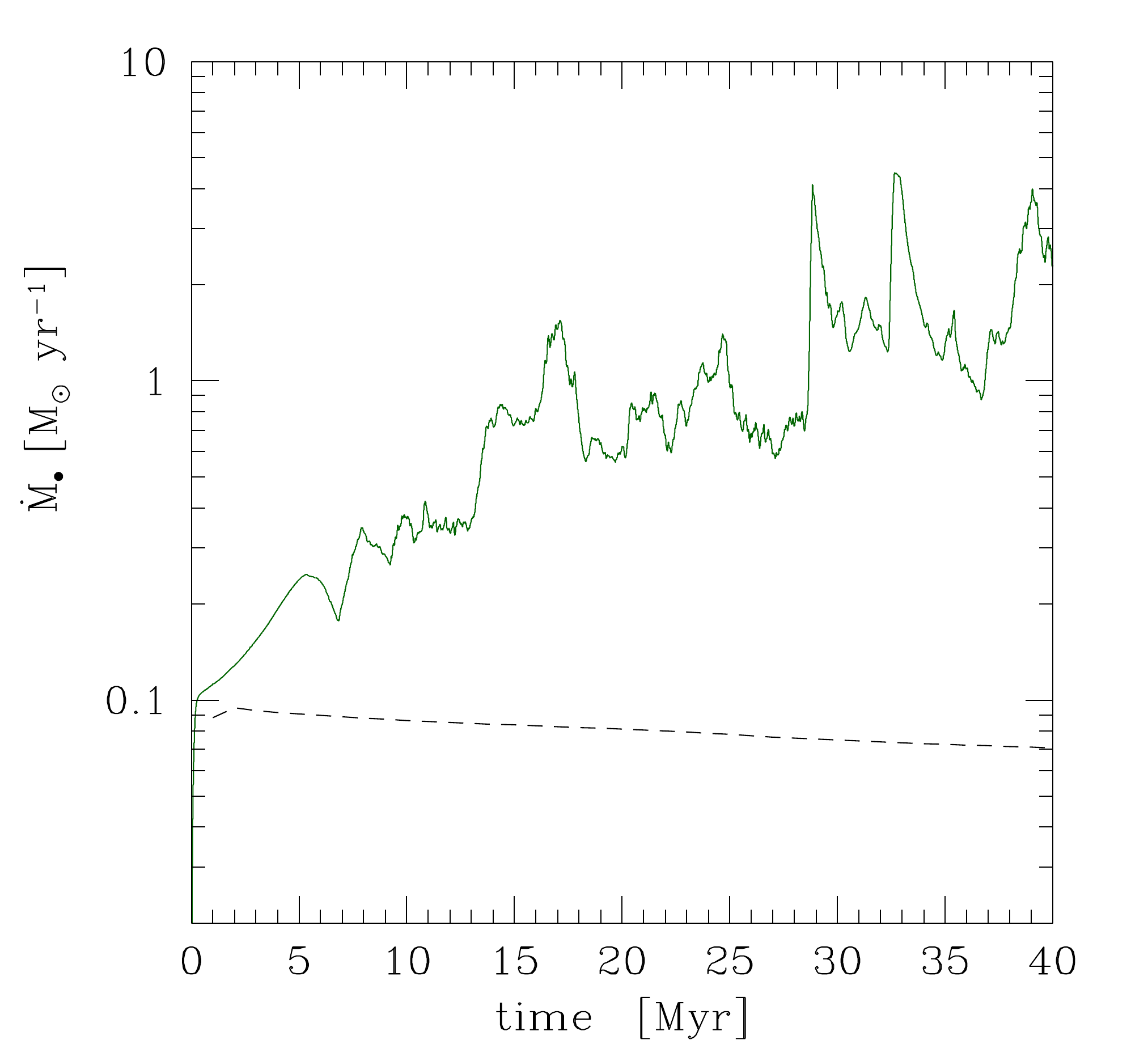}}
      \subfigure{\includegraphics[scale=0.3]{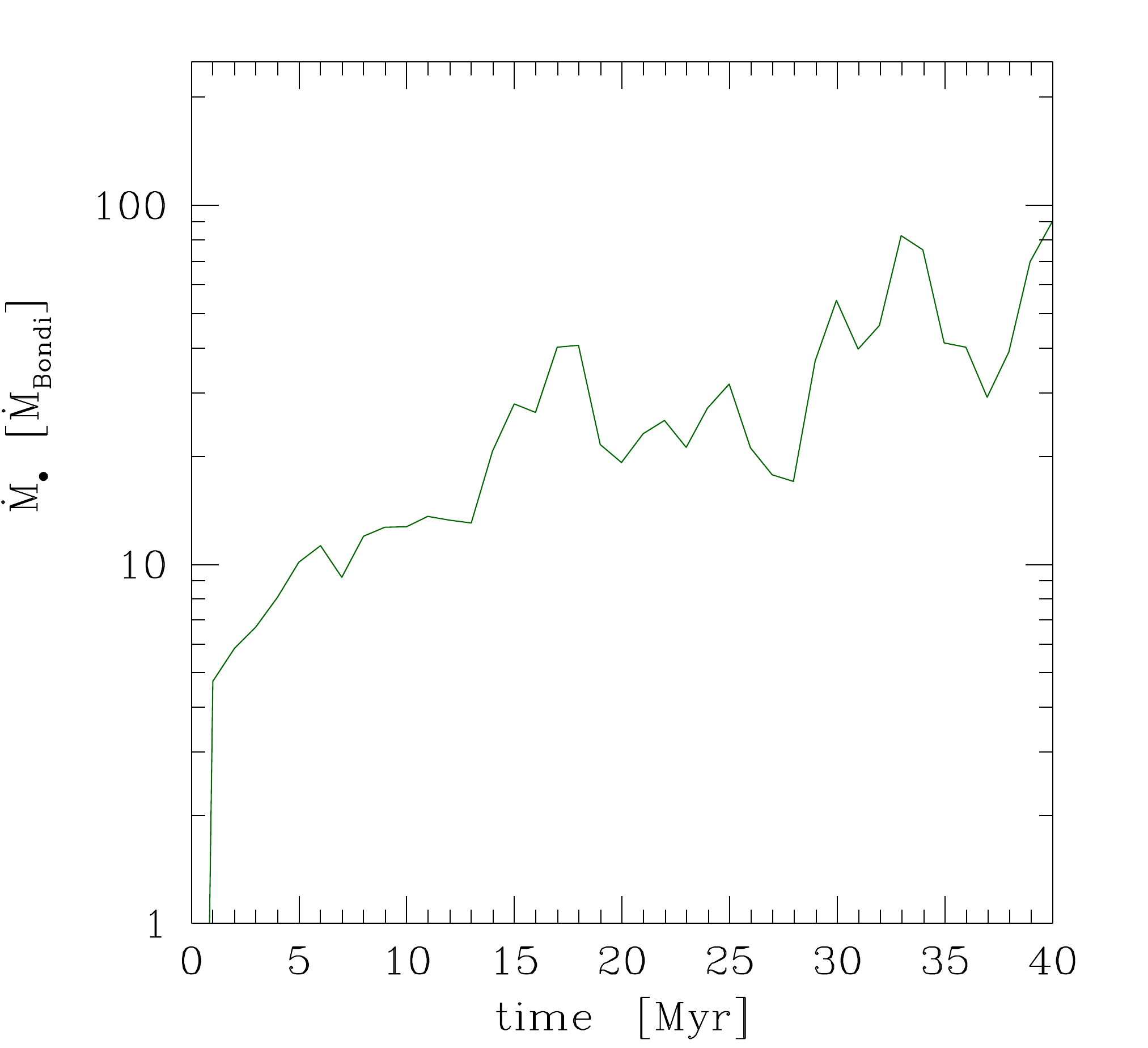}}
      \subfigure{\includegraphics[scale=0.3]{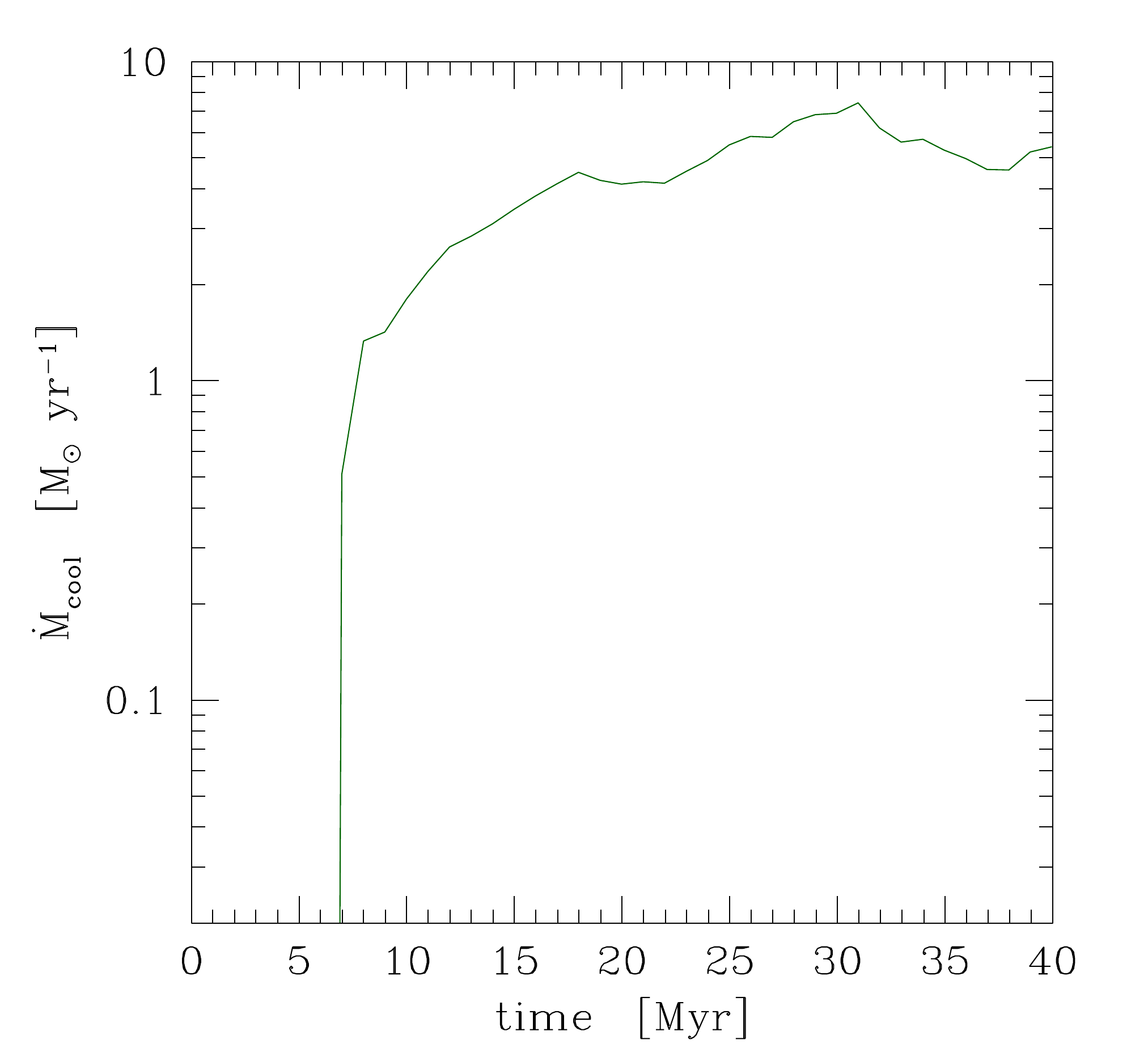}}
      \caption{Accretion with cooling \& turbulence ($M\sim0.35$): evolution of the accretion and cooling rate (cf. Fig.~\ref{cool1}; the top panel has a fast average of 0.1 Myr).
      The dashed line corresponds to the adiabatic run. Although in the presence of turbulent motions (cf. \S\ref{s:stir}), the accretion rate is again boosted by up two orders of magnitude with respect to the Bondi rate at a given time, while connected with the cooling rates. \label{stir_cool3}}  
\end{figure}  

\subsection[]{Accretion \& cooling rate}\label{s:stir_cool_mdot}

The accretion rate shows typical signs of stochastic evolution (Figure \ref{stir_cool3}). Until 6 Myr the accretion rate is similar to the one found in the unperturbed cooling run because the cooling is still feeble. At this time, 
turbulence is already well established, with $\sigma_v\simeq120$ km s$^{-1}$.
After this transient period, $\dot M_\bullet$ begins to fluctuate and the peaks in accretion rate correspond to moments when dense cold clouds cross the accretor region. The typical period of fluctuations is $\sim$5 Myr. The timescale relevant to reaching approximate steady state is the cooling time, rather than, for example, the sound-crossing time of the box (single accretion events occurs on an even shorter timescale, $t_{\rm ff}$).

The accretion rate reaches a peak value of 4$\,$-$\,$5 $\msun$ yr$^{-1}$, similar to that found in the unperturbed cooling run at $\sim$30 Myr. However, the level of accretion is now lower on average by a factor of a few. 
Compared with the Bondi formula (using gas quantities averaged on kpc scale), the average accretion rate is still up to two orders of magnitude greater than $\dot M_{\rm B}$ (middle panel).
The relevant result is that radiative losses dominate the dynamics of accretion even in the presence of moderate turbulence. In the adiabatic case, vortical and bulk motions lead instead to a decrement in the accretion rate (\S\ref{s:stir}).

The absence of heating has a strong effect on the cooling rate (bottom panel; $T\lta10^5$ K), showing values very similar to the pure cooling evolution (slightly reduced by turbulent motions), with peaks up to 7 $\msun$ yr$^{-1}$. The cooling rate also shows a moderate stochastic evolution, with a positive trend with time. $\dot{M}_\bullet$ is again linked to the cooling rate (considering in particular the peaks), within a factor of a few, since there is a delay between cooling and accretion.

\subsection[]{Strong turbulence}\label{s:strong1}
We also investigated the effect of stronger stirring on accretion. This case is perhaps less general but could be applicable to situations when more vigorous gas stirring is induced by strong AGN outbursts, stellar feedback, major mergers. Such events could result in transonic/supersonic velocity dispersions and compressible turbulence. Here we describe only key similarities and differences
between these runs and those discussed above. 

First, we considered turbulence with the velocity dispersion of $\sim$300 km s$^{-1}$ ($M\sim0.7$). This was done by increasing the energy per mode $\epsilon^\ast$. As in the previous run (\S\ref{s:stir_cool_mdot}), we see the formation of thermal instabilities and subsequent chaotic accretion. Chaotic accretion phase begins after about one cooling time since the beginning of the simulation ($\sim$7 Myr). Accretion rate 
is again highly fluctuating with slightly lower instantaneous peaks 
reaching 2$\,$-$\,$3 $\msun$ yr$^{-1}$ and a similar 5 Myr fluctuation period. We also tried 100 times longer correlation time $\tau_{\rm d}$, but found that this did not significantly alter the evolution. Overall, the results are similar to the previous run with cooling and weaker stirring. However, turbulent heating starts to become more relevant: away from the black hole sphere of influence, on kpc scales, the temperature increases by $\sim$30$\,$-$\,$35 percent and the cooling/accretion rates are reduced by a similar fraction.

When the velocity dispersions become transonic the dynamics completely changes. 
Strong turbulence leads in fact to substantial dissipational heating ($\propto \sigma_v^3$), which
easily increases $t_{\rm cool}/t_{\rm ff}$ over 30, damping thermal instability. The suppression of the TI growth may be also supported by adiabatic processes (\citealt{Sanchez-Salcedo:2002}): random compressions heat the gas and fluctuations can re-expand before having time to cool. However, in the present run with $\sigma_v\sim600$$\,$-$\,$$650$ km s$^{-1}$ ($M\sim1.5$) we see significant variations in entropy $K$, meaning that adiabatic processes play a secondary role.

Turbulent dissipation timescale can be approximated as $t_{\rm turb}\sim e_{\rm th}/ \dot{e}_{\rm turb}\sim M^{-2}\, t_{\rm eddy}$, where $e_{\rm th}$ is the specific thermal energy and $t_{\rm eddy}\simeq L/\sigma_v$ is the eddy turnover time (with a factor of order unity uncertainty). Assuming a characteristic scale $L\simeq4$ kpc and transonic Mach number, yields $t_{\rm turb}\sim6$$\,$-$\,$$8$ Myr ($\sim t_{\rm cool}$), which implies that turbulent heating becomes very important. In fact, during the 40 Myr evolution, the temperature raises by a factor of circa 3. Cooling is therefore suppressed by two orders of magnitude ($<0.01$ $\msun$ yr$^{-1}$), and so is thermal instability. The evolution is dominated only by strong stirring and only hot gas is present in the atmosphere. Consequently, $\dot{M}_\bullet$ shows a progressively strong decline, from 0.1 down to $10^{-3}$ $\msun$ yr$^{-1}$ (30 Myr),
due to the high temperature and large vorticity (see \S\ref{s:stir}). 
The minima of $\dot{M}_\bullet$ display values even one order of magnitude smaller than those computed via the Bondi formula. Strong turbulence tends also to smooth out the central density profile and to drive more gas out of the boundaries. The decrement in the central density is 1.5$\,$-$\,$2$\times$, a sign that turbulent diffusion is also quite relevant via the flattening of gradients. At these high Mach numbers, dissipation tends to dominate over diffusion (with a net increase of entropy), albeit the two contributions are not trivial to disentangle.

In summary, turbulence over 500$\,$-$\,$600 km s$^{-1}$ ($M \gta 1$), or more generally when $t_{\rm turb}\gta t_{\rm cool}$,  has the exact opposite influence on the dynamics than radiative cooling. It dramatically reduces the black hole accretion rates, while heating the gas\footnote{Such high turbulence seems to produce too much heating, compared with observations.
However, a realistically lower level of turbulence (200$\,$-$\,$400 km s$^{-1}$; \citealt{dePlaa:2012}) is an integral element of a successful heating model, as it efficiently redistributes and isotropises the feedback energy (\citealt{Gaspari:2012b}).}.
The only channel of accretion is thus a very feeble hot mode.
The whole accretion process is overall
determined by the competition of three timescales: $t_{\rm ff}$, $t_{\rm cool}$ {\it and} $t_{\rm turb}$.

\section[]{Adiabatic Accretion with turbulence} \label{s:stir}

\addtocounter{figuresub}{-2}     

\begin{figure} 
      \begin{center}
      \subfigure{\includegraphics[scale=0.3]{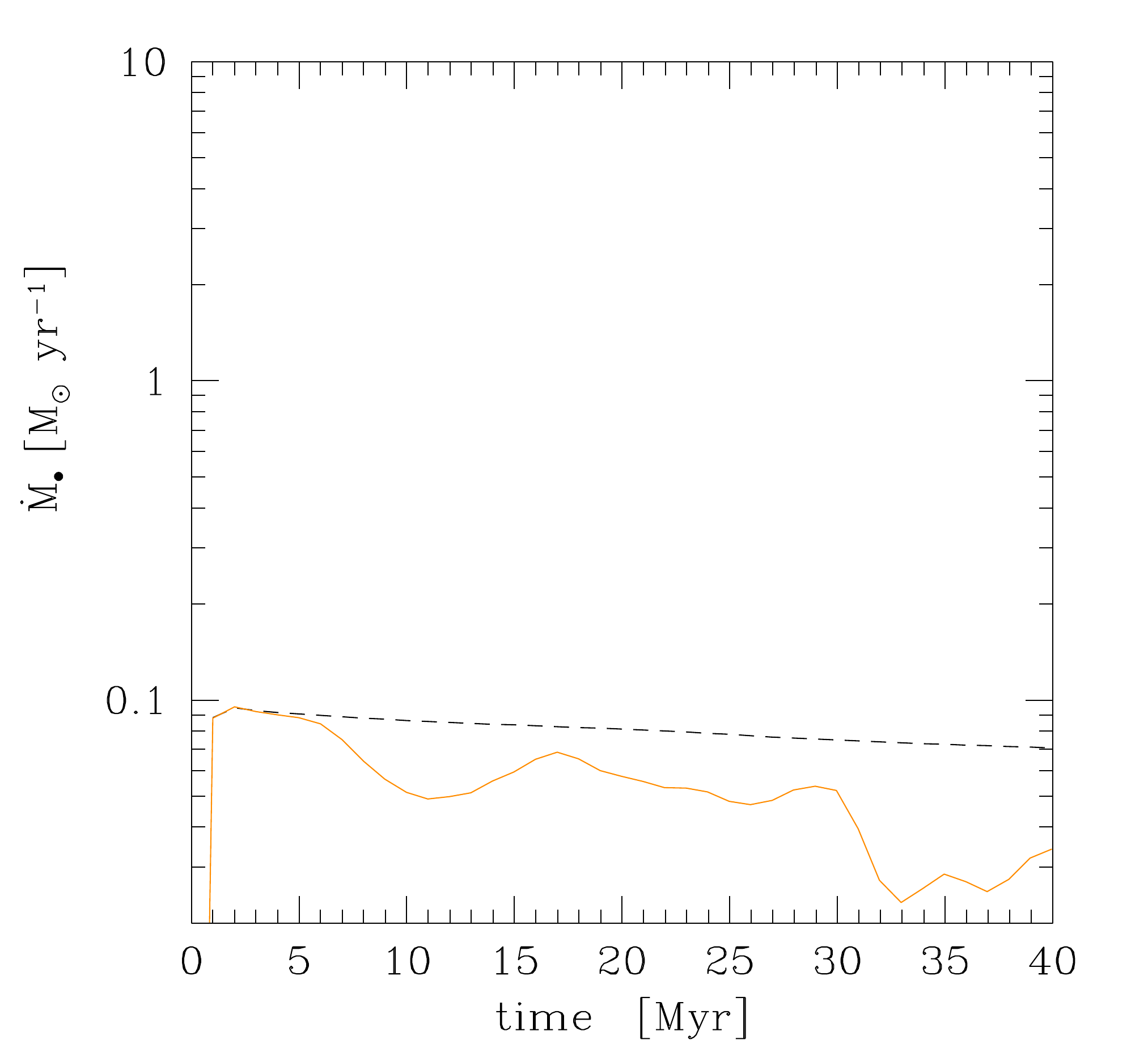}}
      \end{center}
      \caption{Adiabatic accretion with subsonic turbulence ($\sigma_v\,$$\sim$$\,150$ km s$^{-1}$, $M\sim0.35$): evolution of the accretion rate (cf. Fig.~\ref{pure1}).       
       Note the significant decline in the accretion rate compared with the Bondi-like case without turbulence. This decrease is due to the presence of turbulent eddies (vortical and bulk motions relative to the accretor).       
      \label{stir1}}
\end{figure}  

\addtocounter{figure}{-1}
\addtocounter{figuresub}{1}      

\begin{figure}
      \begin{center}
      \subfigure{\includegraphics[scale=0.28]{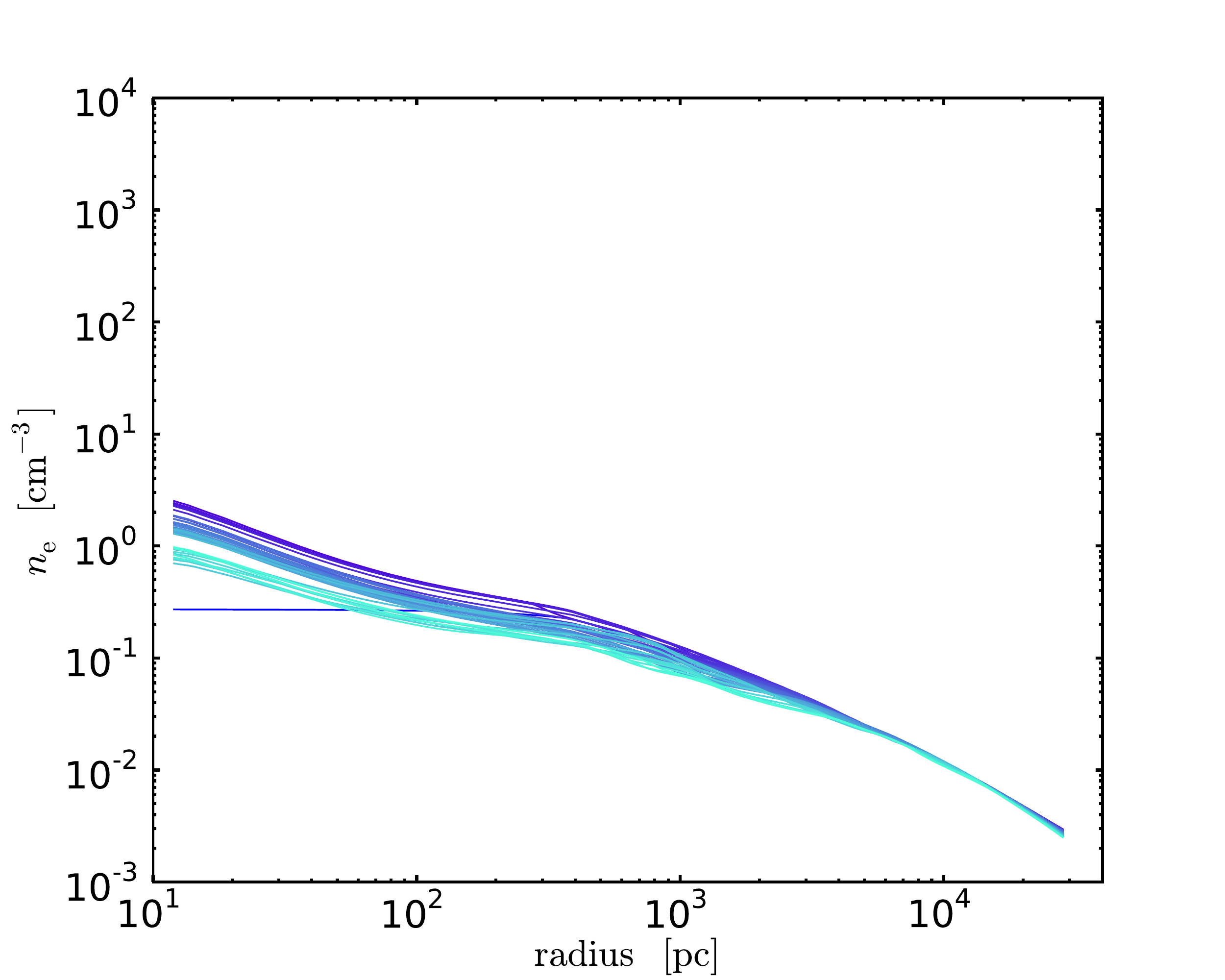}}
      \subfigure{\includegraphics[scale=0.28]{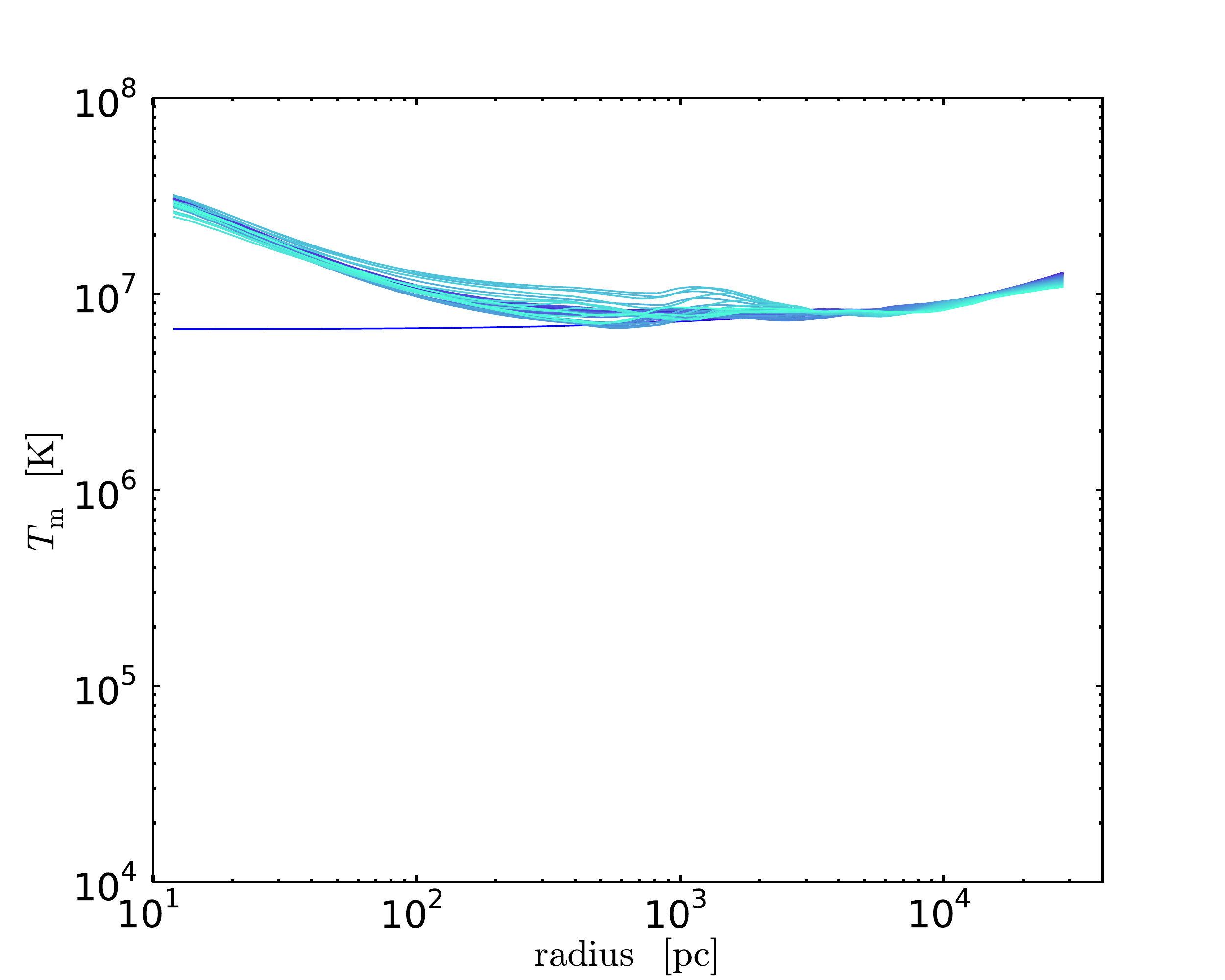}}
      \end{center}
      \caption{Adiabatic accretion with subsonic turbulence ($M\sim0.35$): mass-weighted profiles of density and temperature (cf. Fig.~\ref{pure2}). The profiles are similar to the Bondi case, though more shallow due to turbulent diffusion, and displaying fluctuations.
      \label{stir2}}    
\end{figure}  

\addtocounter{figure}{-1}
\addtocounter{figuresub}{1}   

\begin{figure} 
      \begin{center}
       \subfigure{\includegraphics[scale=0.45]{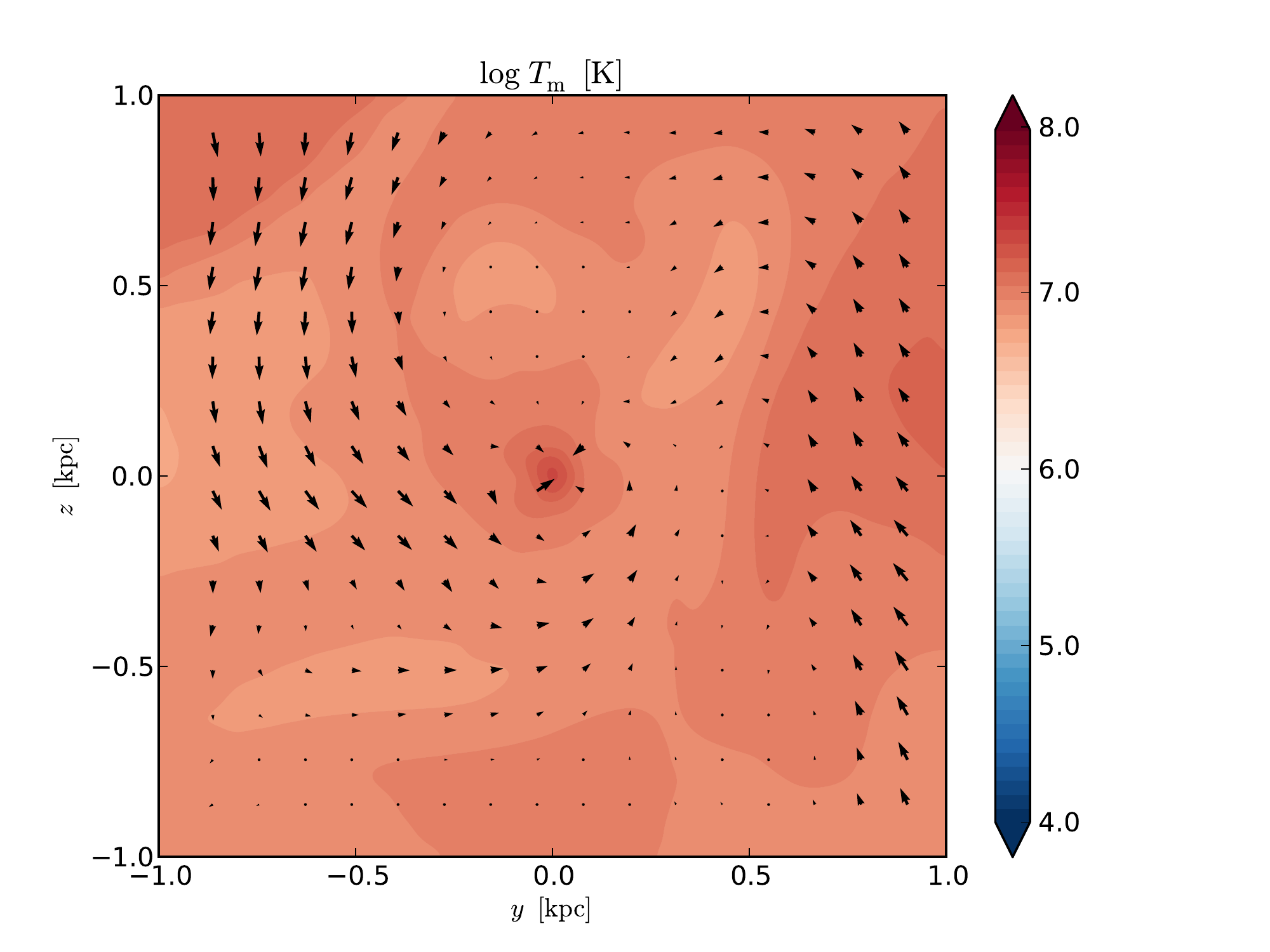}}
      \end{center}
      \caption{Adiabatic accretion with subsonic turbulence ($M\sim0.35$): mid-plane temperature cut (cf. Fig.~\ref{pure3}). The stochastic nature of turbulence and the driven turbulent eddies are evident.
      \label{stir3}}    
\end{figure}  

It is worth to briefly analyse the evolution without radiative cooling, while still in the presence of moderate levels of solenoidal turbulence, $\sigma_v\sim150$$\,$-$\,$180 km s$^{-1}$ ($M\sim0.35$$\,$-$\,$$0.4$). The turbulent energy is just $\sim$8 per cent of the thermal energy ($E_{\rm turb} \simeq 0.55\, M^2\, E_{\rm th}$). 
The following evolution could represent a phase
of slight overheating, e.g. after an AGN outburst or major merger, in which
thermal instability is temporarily suppressed ($t_{\rm cool}/t_{\rm ff}>10$) and the hot mode is thus the only channel
of accretion. In this Section we emphasise only the relevant differences between such a case and its radiatively cooling counterpart discussed in \S\ref{s:stir_cool}. 

The key result is that the accretion rate declines as opposed to the cooling run (Figure \ref{stir1}): $\dot M_\bullet$ drops by a factor of a few below 0.07 $\msun$ yr$^{-1}$, which was the minimum value reached in the unperturbed adiabatic model. At final time, the decrease is a factor of $\sim$2.5. The dominant timescale for quasi-steady state is the eddy turnover time ($t_{\rm eddy}\sim25$$\,$-$\,$30 Myr).
The accretion rate normalised to the Bondi rate (not shown) is very low: $\sim$2$\,$-$\,$3 $\dot{M}_{\rm B}$ (the small bias due to the galactic gradients;  cf. \S\ref{s:adi1}). The $\dot{M}_\bullet$ evolution still shows a stochastic behaviour, though with a reduced amplitude.

When the solenoidal turbulent motions are significant, the vorticity driven via the baroclinic instability becomes important and inhibits accretion (cf. Krumholz et al 2005, 2006; we remark that we do not fix an initial rotational velocity). This happens because the medium is stratified\footnote{The process is analogous to the formation of meteorological cyclones.}. On the other hand, when turbulence induces a stronger transient bulk motion, the reduction of $\dot M_\bullet$ is mainly associated with the non-zero relative velocity between the accretor and the flow motion (\citealt{Bondi:1944}).

The radial emission- or mass-weighted profiles (Fig.~\ref{stir2}),
are very similar to those corresponding to the Bondi-like run (\S\ref{s:adi2}). The only difference is the presence of moderate fluctuations on the order of few tens of percent, and the ability of weak turbulent diffusion to slightly flatten the central density profile (which in minor part helps reducing $\dot M_\bullet$).
We remark that the heating due to the dissipation of turbulent motions is negligible for this level of stirring, as indicated by the temperature profile at large radii. 

The temperature map (Fig.~\ref{stir3}) reveals the presence of turbulent eddies on the scales of a few kpc, i.e. the injection scale, which are cascading to smaller scales. 
The fluctuations driven in the current evolution are typically quasi isentropic, with more significant fluctuations in pressure, as opposed to the runs with cooling enabled. As shown in the previous Section, increasing $\sigma_v$ progressively magnifies the impact of turbulent dissipation, hence leading to a mixture of entropy and pressure fluctuations. Fluctuations with varying entropy prevail when turbulence becomes transonic and, of course, when the cooling/heating terms are active.

In summary, adding moderate stirring to the purely adiabatic case drives hot accretion with significantly reduced $\dot M_\bullet$. With stronger turbulent motions, the accretion rate is reduced by over two orders of magnitude (not shown).
In less massive galaxies this state should be much more frequent, since central heating has more severe consequences in less bound systems (\citealt{Gaspari:2012b}), and turbulent velocities may easily reach $M \gta1$, given the lower gas sound speed. Sgr A$^\ast$ and NGC 3115 might be two exemplary cases (\citealt{Wong:2011}).
Finally, this run demonstrates that the radiative cooling, in combination with turbulence (and heating; see below), is the main cause of the dramatic boosts of matter channelled to the very centre, compared with the Bondi expectation.

\section[]{Accretion with heating, cooling \& turbulence}  \label{s:heat}

\addtocounter{figuresub}{-2}     

\begin{figure} 
      \begin{center}
      \subfigure{\includegraphics[scale=0.305]{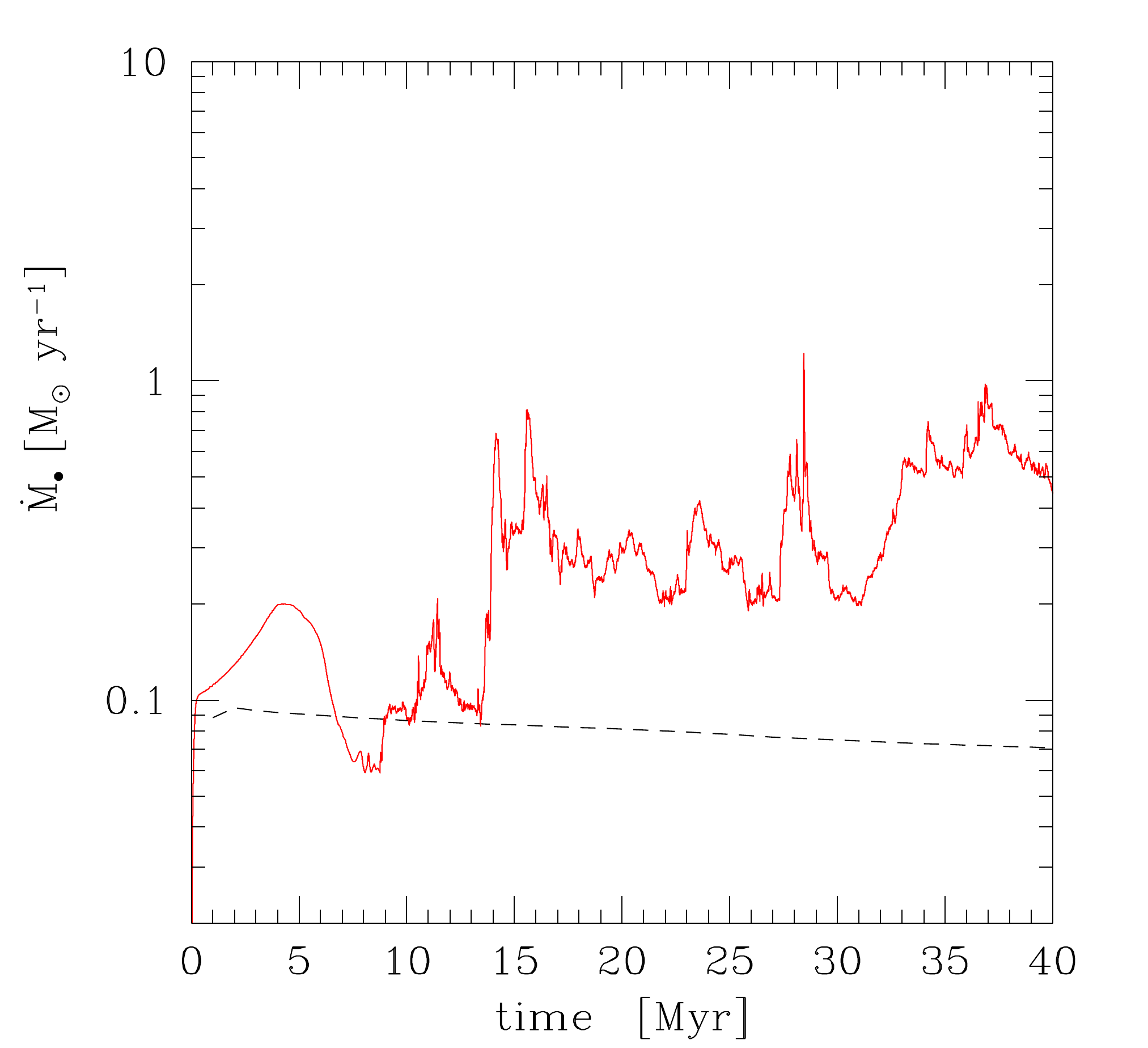}}
      \subfigure{\includegraphics[scale=0.305]{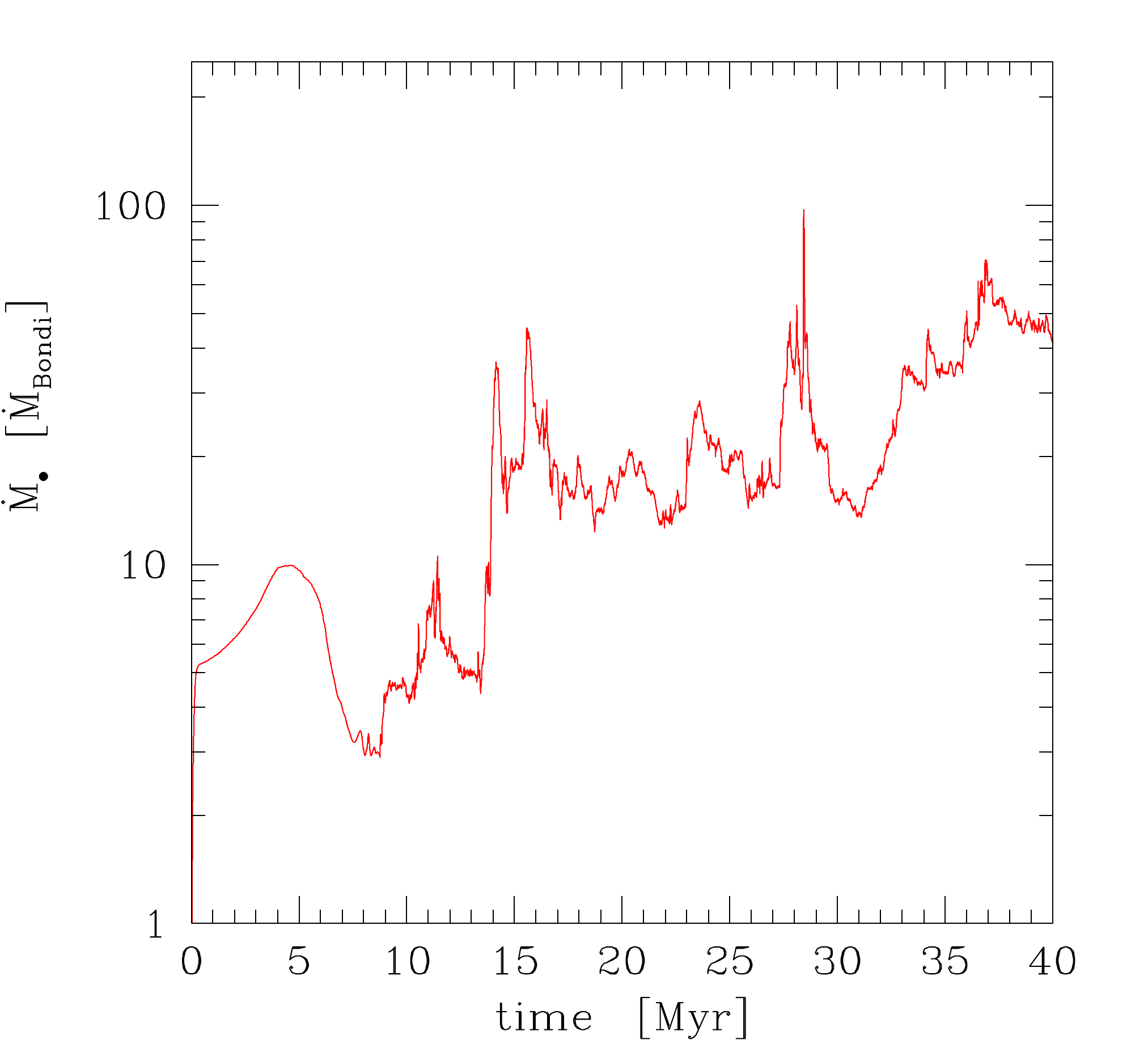}}
      \subfigure{\includegraphics[scale=0.305]{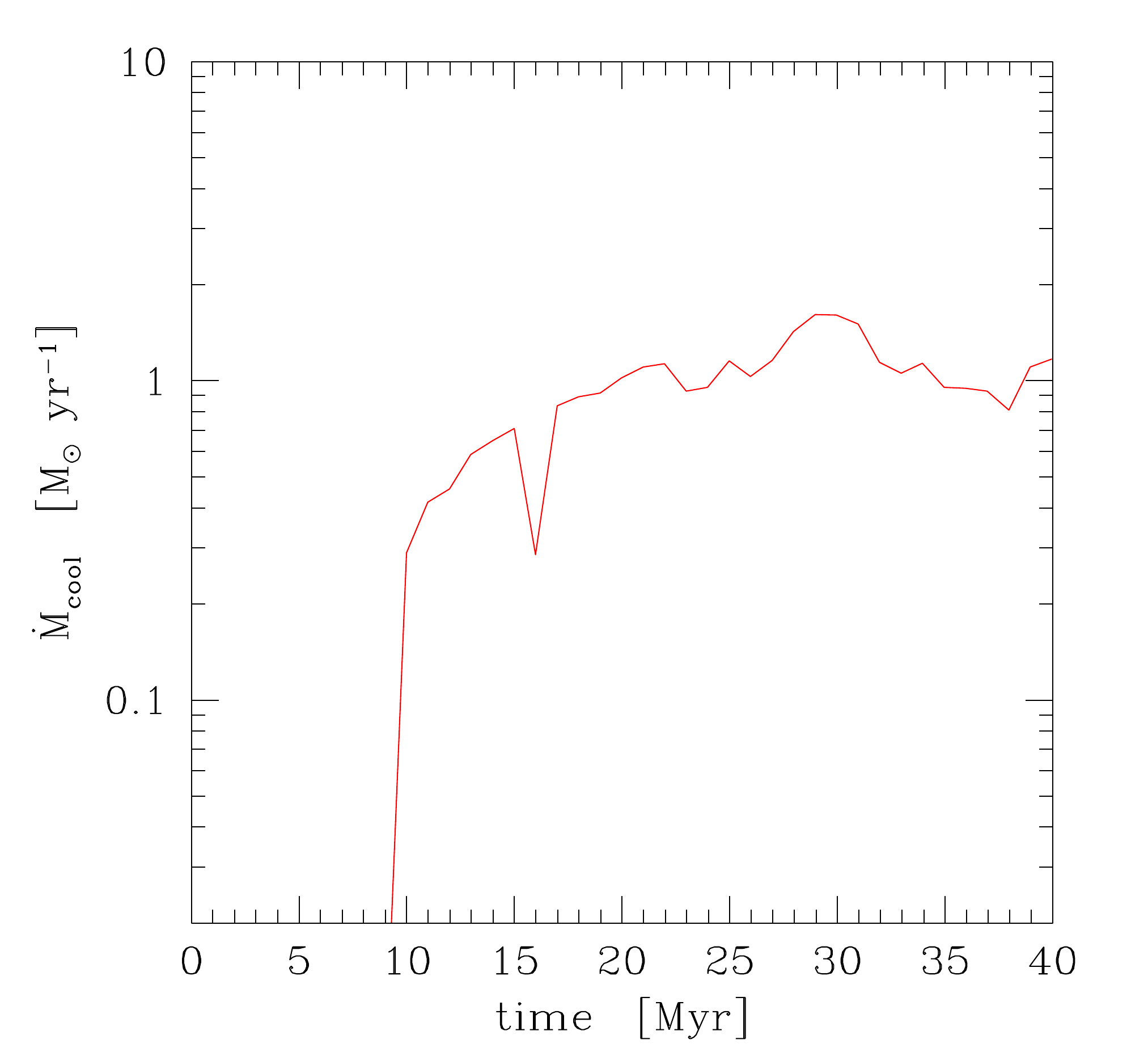}}
      \end{center}
      \caption{Accretion with heating, cooling \& turbulence ($M\sim0.35$): evolution of the accretion and cooling rate (cf. Fig.~\ref{cool1}; top two panels have a fast average of 0.1 Myr). 
       Even in the presence of heating, which reduces the average cooling rate, the accretion rate is still boosted by TI up to two orders of magnitude, compared with the Bondi prediction. 
       The accreted hot gas remains always below 5 per cent of the accreted cold phase.
       \label{heat1}}  
\end{figure}  

Hot gas in galaxies, groups, and clusters is not only cooling but it also appears to be kept in quasi thermodynamical equilibrium. The main heating mechanism that prevents catastrophic cooling is likely AGN feedback, possibly helped by mergers, conduction, and stellar evolution. The SMBH acts as a thermostat, triggering AGN outflows/jets, and subsequent buoyant bubbles, shocks, and turbulent motions (\S\ref{s:gheat}). The latter phenomena are able to efficiently thermalise the kinetic energy, and to isotropically heat the medium (\citealt{Gaspari:2012a}). The next step in increasing the level of realism of our model is thus to include spatially-distributed global heating. As our intention is to implement physics that is applicable to a range of situations, we use a simple heating prescription that, in agreement with observations, guarantees global average balance of heating and cooling. In this model the gas is globally thermally stable, which does not preclude the possibility of the development of local TI (\S\ref{s:gheat}). In addition to heating, the simulation discussed in this Section includes cooling and turbulence at the level identical to that considered in the previous Section ($\sigma_v\sim150$$\,$-$\,$180 km s$^{-1}$, $M\sim0.35$), with all other key parameters remaining unchanged. Recall that for these velocities, the energy injection due to the dissipation of turbulent motions is negligible compared with the cooling rate. In \S\ref{s:strong2} we experiment with stronger stirring.

\subsection[]{Accretion \& cooling rate} \label{s:heat_acc}
The evolution of the accretion rate reveals the multiphase structure of the flow (Figure \ref{heat1}). Before a cooling time ($\lta7$ Myr), most of the accreted gas is hot ($T>10^6$ K):
the $\dot M_\bullet$ evolution is smooth and tends to slowly decline due to the driven turbulent motions (cf. \S\ref{s:stir}), falling below the rate observed in the unperturbed adiabatic case, $\dot{M}_\bullet\sim0.08\ \msun$ yr$^{-1}$.
From 7 Myr onward, TI becomes exponentially nonlinear and cold clouds condense out of the hot phase progressively boosting the accretion rate. Consequently, the accreted hot gas remains below 5 per cent of the accreted cold phase throughout the rest of the evolution.

\addtocounter{figure}{-1}
\addtocounter{figuresub}{1}     

\begin{figure*} 
      \begin{center}
      \subfigure{\includegraphics[scale=0.28]{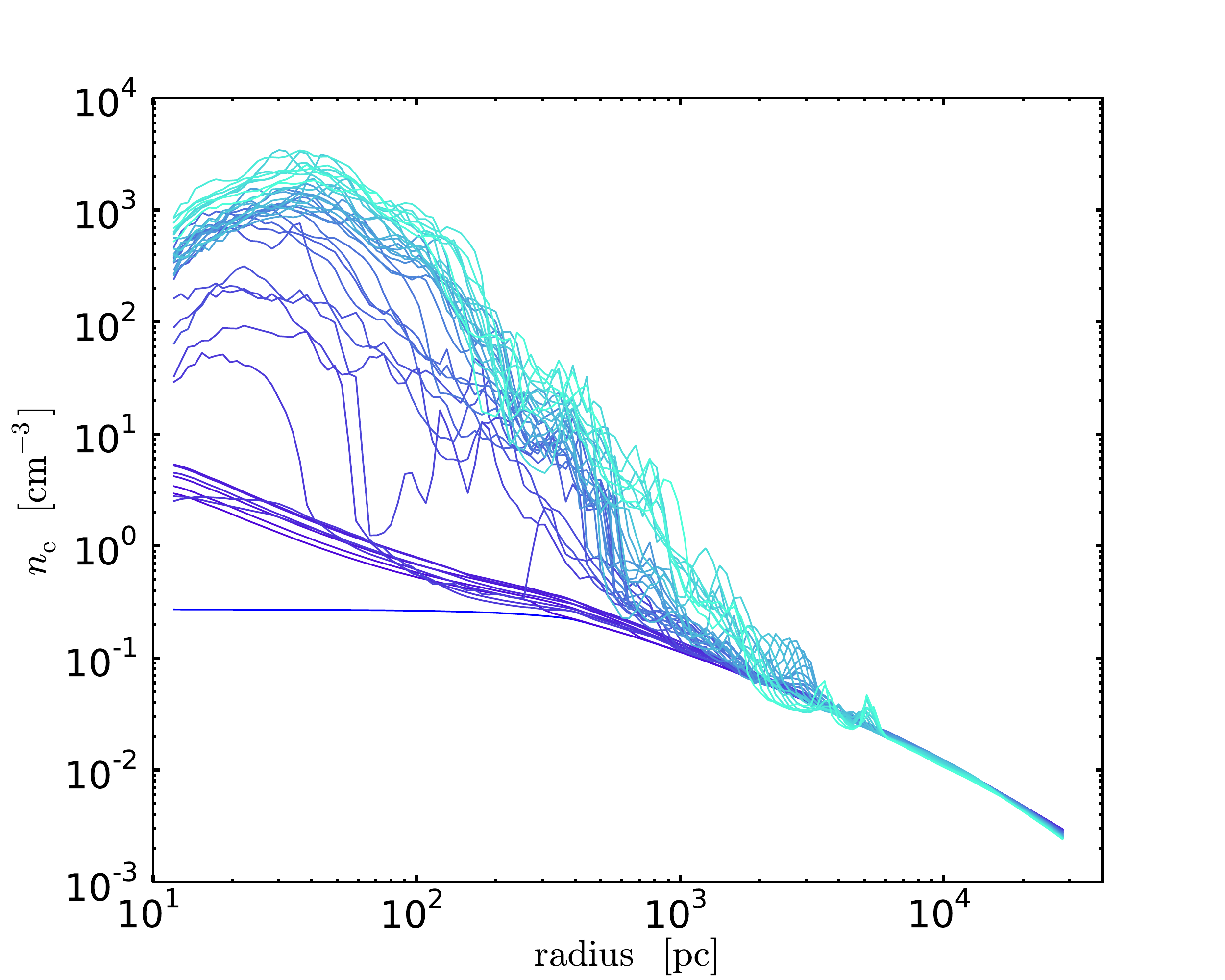}}
      \subfigure{\includegraphics[scale=0.28]{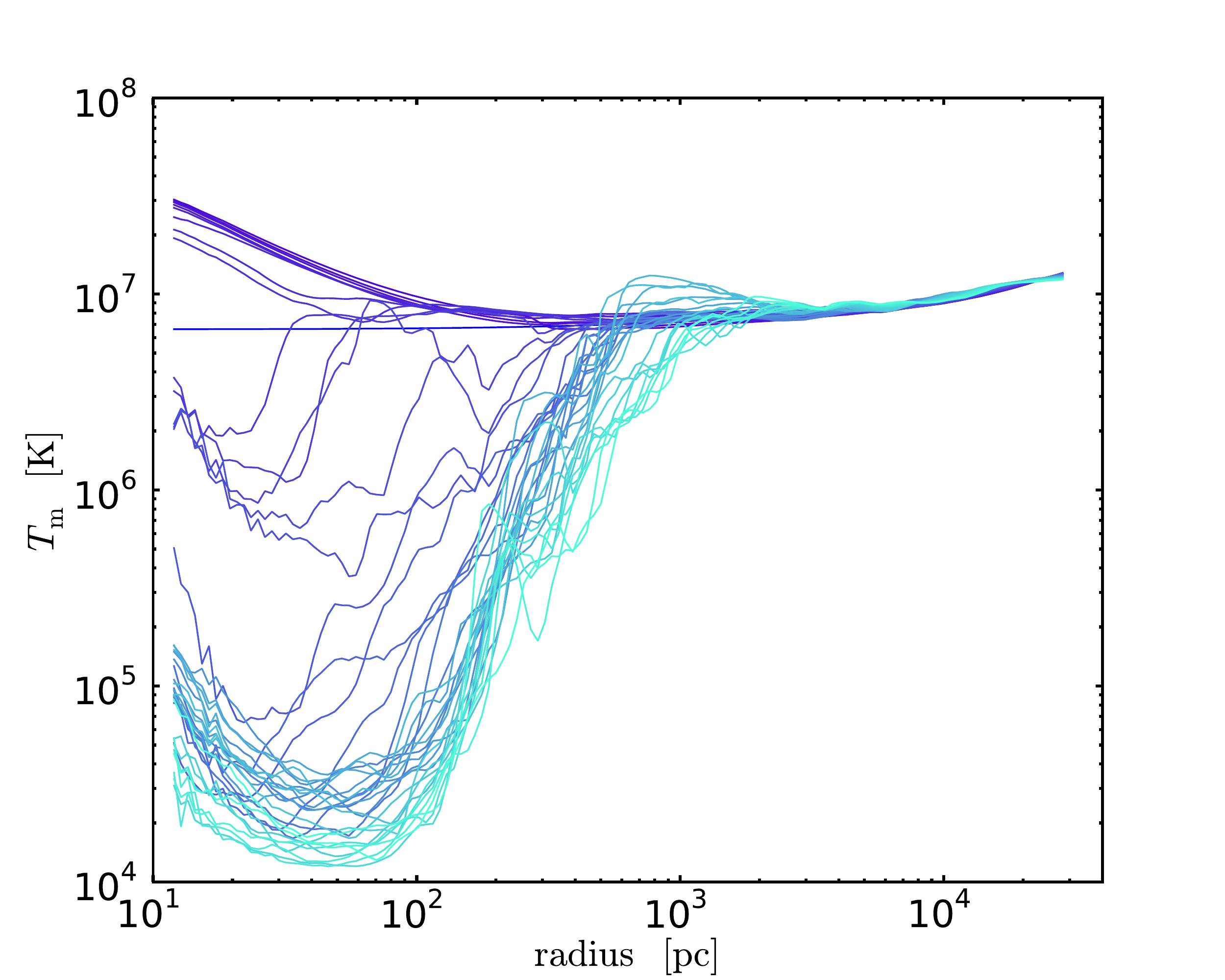}}
      \subfigure{\includegraphics[scale=0.28]{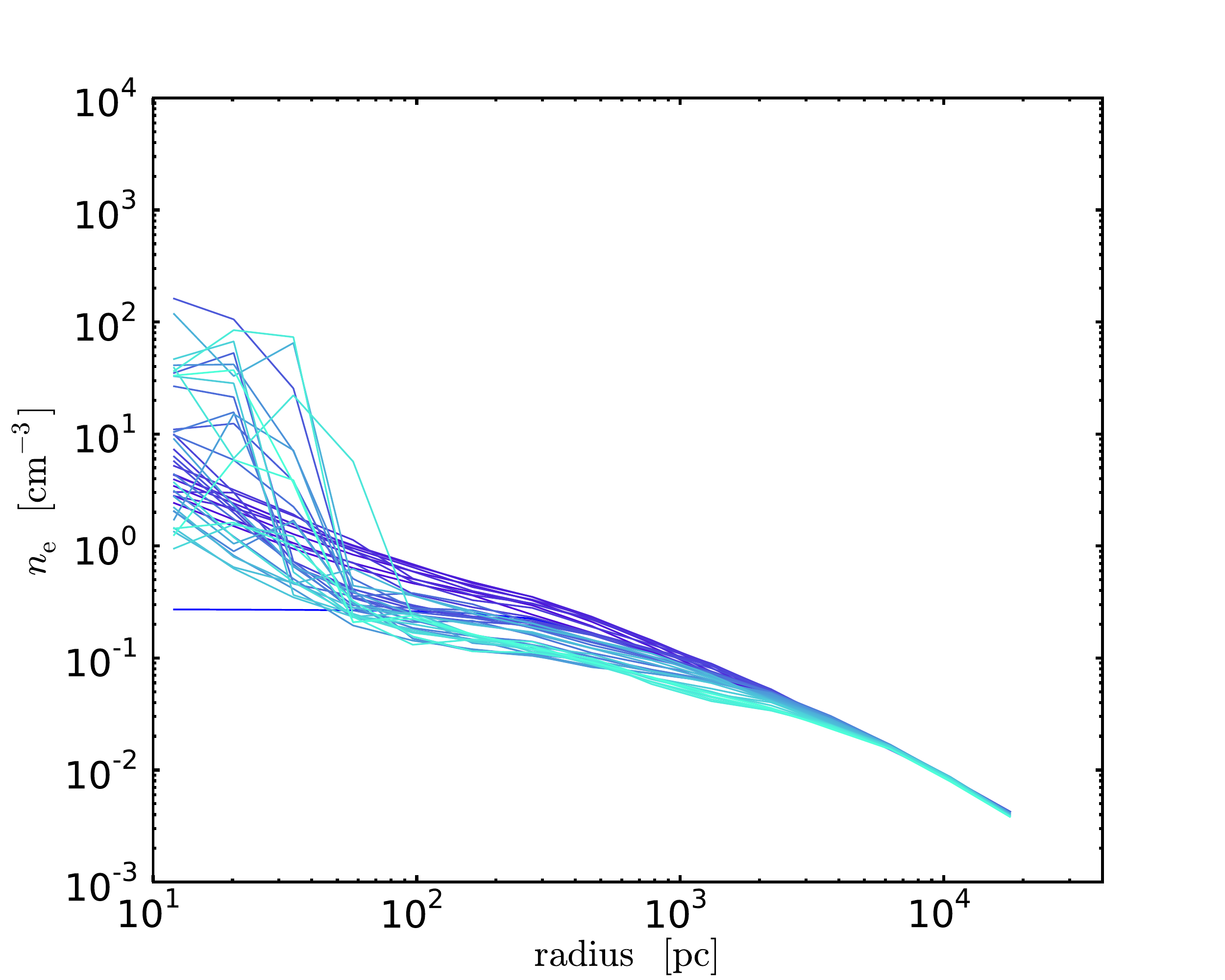}}
      \subfigure{\includegraphics[scale=0.28]{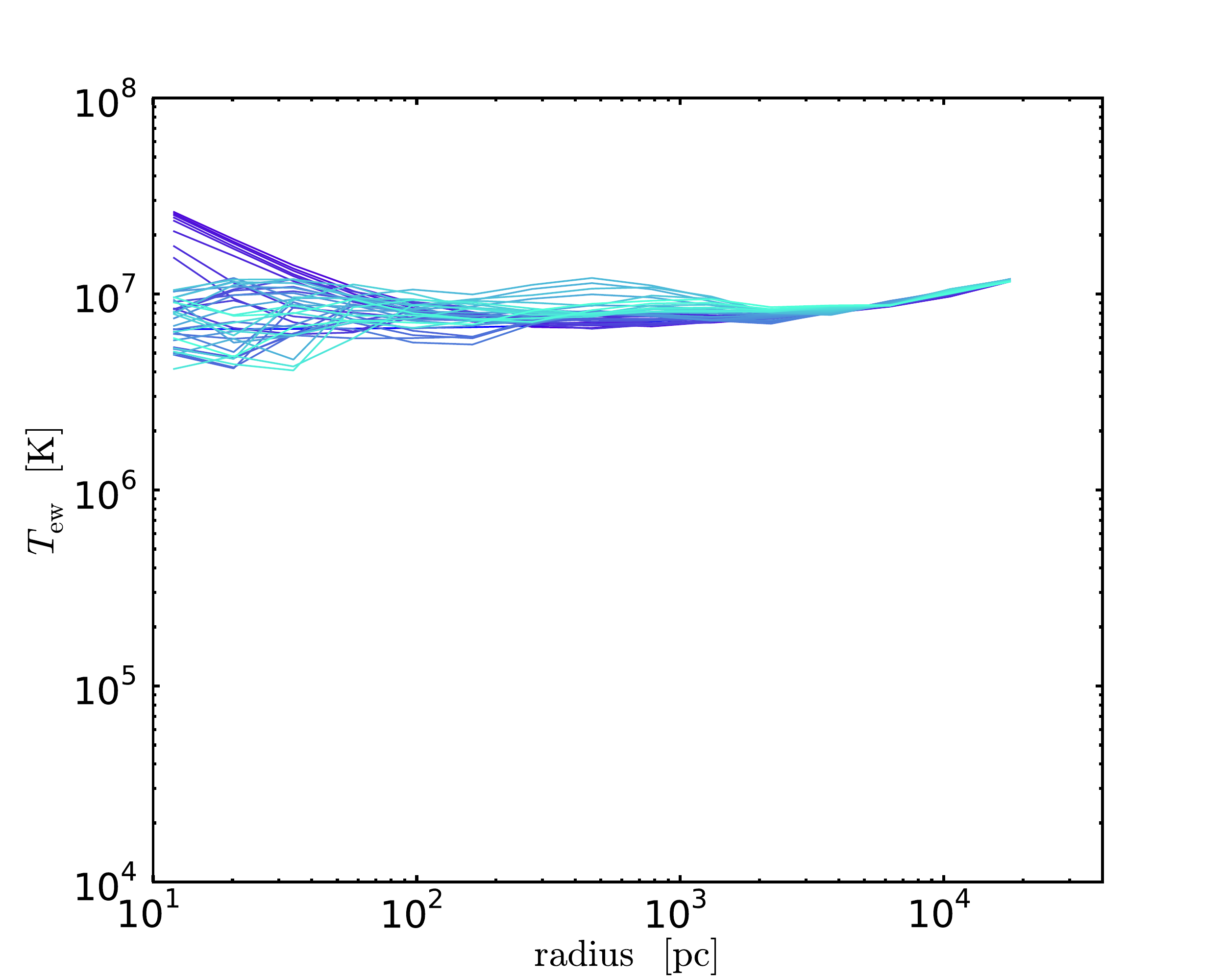}} 
      \end{center}
      \caption{Accretion with heating, cooling \& turbulence ($M\sim0.35$): 3D mass- ({\it top}) and emission-weighted ({\it bottom}) radial profiles of density and temperature (cf. Fig.~\ref{cool2}). The extended multiphase structure and stochastic nature of the accretion flow is evident within 10 kpc, but it is largely concealed by weighting with X-ray emission (projection along line of sight will aggravate this effect). The emission-weighted temperature profile is remarkably flat: this prediction could be tested with the next generation of X-ray telescopes (resolving radii $< 100$ pc) to discriminate between the hot and cold mode of accretion.
      \label{heat2}}    
\end{figure*}  

Similarity between the present and the previous stirring-plus-cooling run is evident (cf. Fig.~\ref{stir_cool3}). However, heating introduces important differences. The fluctuations in the accretion rate have considerably higher frequency and subsequent accretion peaks can appear in just 2 Myr.  
In fact, in a heated atmosphere TI grows exponentially rather than linearly, as would be the case in the cooling-plus-turbulence case (see \citealt{McCourt:2012} for the analytic dispersion relation). When $t_{\rm cool}/t_{\rm ff} \lta 10$,
buoyancy is not able to prevent the nonlinear condensation from falling. Cooling is further boosted by the geometrical compression, leading to the runaway sinking of cold blobs. 

Other than driving a more dynamic atmosphere,
the effect of global heating appears as a significant reduction in the average cooling rate, 
oscillating around 1 $\msun$ yr$^{-1}$ (bottom panel), i.e. $\lta$$\,$15 per cent of the pure cooling rate (in the late stage). 
The cooling rate, as well as the tightly linked $\dot{M}_\bullet$, approximately reaches the statistical steady state after a few $t_{\rm cool}$, 
while in the cooling-only run it tends to slightly rise up to $10\times$ higher.
We note that accretion is sub-Eddington in all cases considered in this work ($\dot M_{\rm Edd} \simeq 70$ 
$\msun$ yr$^{-1}$), but could be super-Eddington at high redshift when the black hole has considerably smaller mass.
 
A key result from this experiment is that, even in the presence of heating, using the Bondi formula (on the kpc scale, as commonly done in cosmological and large-scale simulations) leads to the accretion rates underestimated by up to two orders of magnitude, with typical values around $50\times$ near final time. This is caused by the fact that the dense cold filaments drive accretion despite the fact that their volume filling is negligible compared with the hot phase. Just as in the cooling case with turbulence, the sizes of the clouds and filaments are still sufficiently large to facilitate collisions between them. This helps to keep the accretion rate at these high levels.

We remark that our work provides a physical basis for the high value of the boost parameter assumed in cosmological simulations ($\alpha \gg 1$), which is required to reproduce the concurrent growth of black holes and their host galaxies (e.g. \citealt{Sijacki:2007}).

\subsection[]{Radial profiles}  \label{s:heat_prof}

Since thermal instabilities and filamentary extended cold gas grow in the presence of the distributed heating, the mass-weighted profiles are dominated again by the dense cold phase within few kpc from the centre (Fig.~\ref{heat2}) and show sharp increase (decrement) in the density (temperature). As for $\dot M_\bullet$, the temporal fluctuations have now higher amplitude and frequency, a signal that the multiphase gas is now more clumpy and filamentary.

In the linear regime, and for a homogeneous background, we would see thermal instabilities forming at all distances from the centre as long as $t_{\rm cool}/t_{\rm ff}$ is sufficiently small.
However, in the presence of a declining density gradient, $t_{\rm cool}/t_{\rm ff}$ displays a minimum at $\simeq250$ pc (Fig.~\ref{heat3}). Indeed, we observe the formation of the first small filamentary cold structures near this radius. In the nonlinear regime, small-scale perturbations have also the potential to reach higher amplitudes (\citealt{Burkert:2000}). After this initial phase, we observe the formation of cold clouds with a wide range of sizes, from distances of several 10 pc up to $\sim$8 kpc, where $t_{\rm cool}/t_{\rm ff}$ remains below 10 (Fig.~\ref{heat2}$\,$-$\,$\ref{heat3}; the radial average masks some of the clouds). 
This value of the cooling to free-fall time ratio is the threshold that approximately corresponds to the onset of nonlinear cold gas condensation (\citealt{Gaspari:2012a, Sharma:2012}). Instabilities form whenever this ratio falls below the threshold in a certain region, hence setting the scale of the local cloud. 

We note that turbulence acts as an effective diffusive process,  like thermal conduction, preventing most of the TI to reach the smallest resolution (and thus helping convergence; \S\ref{s:conv}). The usually tiny Field length (\citeyear{Field:1965}) -- the smallest scale of TI collapse set by conduction in the absence of stratification -- may thus not be relevant in this context.

After 40 Myr, the dropout of the cold gas from the hot phase causes the ambient medium to become progressively more tenuous, leading to $t_{\rm cool}/t_{\rm ff}\gta10$ at all radii. The growth of new cold clouds is inhibited, since in this regime  buoyancy dominates the dynamics exciting gravity waves and following convective stability.
At this point, a strong AGN feedback event will likely be triggered due to the cold accretion of the previously formed clouds, followed by a phase of slight overheating (\S\ref{s:stir}). Consequently, the feedback input will rapidly decrease, allowing cooling to lower $t_{\rm cool}/t_{\rm ff}$ and to start the loop again (\citealt{Gaspari:2012a} for a detailed analysis of the self-regulated feedback cycle over several Gyr).

The presence of heating has the effect to sustain the temperature radial profile at large radii. In fact, the $T$ profile remains adherent to the initial condition over several kpc ($\sim$$10^7$ K), while in the cooling-only unperturbed case it is steadily declining. 
Removing the contribution to the profiles from the cold gas via emission weighting results in a roughly constant temperature profile (Fig.~\ref{heat2}, bottom right panel). The X-ray emission-weighted density profile (not projected) demonstrates that the temporal fluctuations\footnote{The spatial variations are significantly smeared out by the large radial bins, commonly used also in observational analysis.} are larger compared with the fluctuations in the no-heating case, especially within the inner 100 pc (bottom left panel). 

We emphasise that current X-ray telescopes -- typically resolving radii $\gta 1$ kpc (except for few nearby galaxies) -- have severe difficulty in determining the mode of accretion via the $T_{\rm ew}$ and $n_{\rm e}$ profiles (cf. the magnitude of errors in \citealt{Wong:2011}). Figure \ref{heat2} also shows that extrapolating the observed profiles from kpc scales (e.g. \citealt{Allen:2006}) is misleading, since the X-ray profiles resemble Bondi hot accretion, while the accretion rates and dynamics could be profoundly different (the same is valid for the adiabatic case with turbulence in \S\ref{s:stir}). 

\subsection[]{Dynamics: chaotic cold accretion}\label{s:heat_maps}
\addtocounter{figure}{-1}
\addtocounter{figuresub}{1}       

\begin{figure} 
      \begin{center}     
      \subfigure{\includegraphics[scale=0.28]{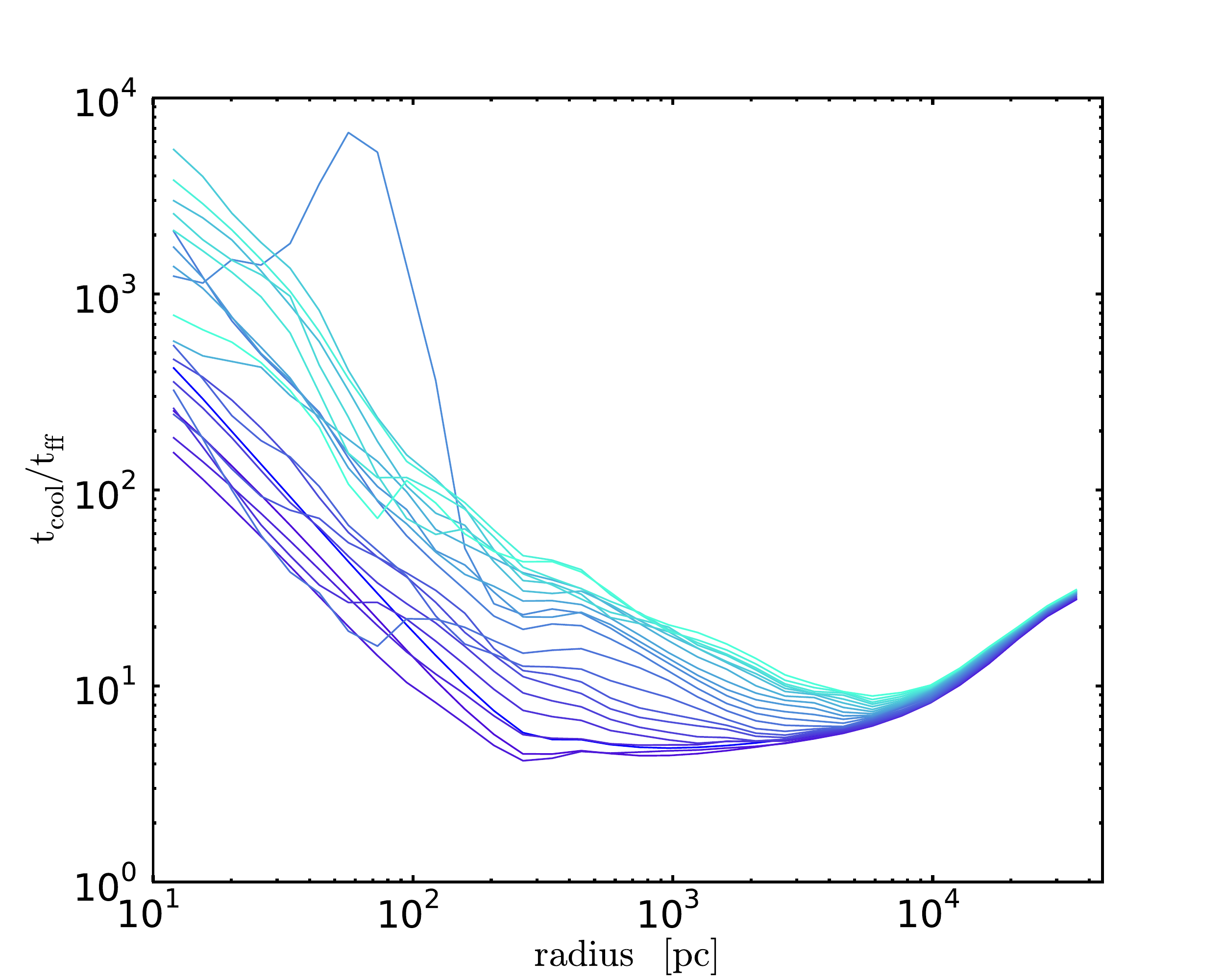}}
      \end{center}
      \caption{Accretion with heating, cooling \& turbulence ($M\sim0.35$): volume-weighted ratio of the cooling time and free-fall time for the hot phase in radial shells ($T>0.1$ keV; every 2 Myr). The threshold of $\sim$10 marks the point where the condensation of extended cold gas becomes possible (100 pc$\,$-$\,$8 kpc). The dropout of cold gas out of the hot phase results in a steady increase of this ratio, since $t_{\rm cool}$ of the more tenuous hot gas becomes longer.
      \label{heat3}}  
\end{figure}

\addtocounter{figure}{-1}
\addtocounter{figuresub}{1}   

\begin{figure} 
     \subfigure{\includegraphics[scale=0.46]{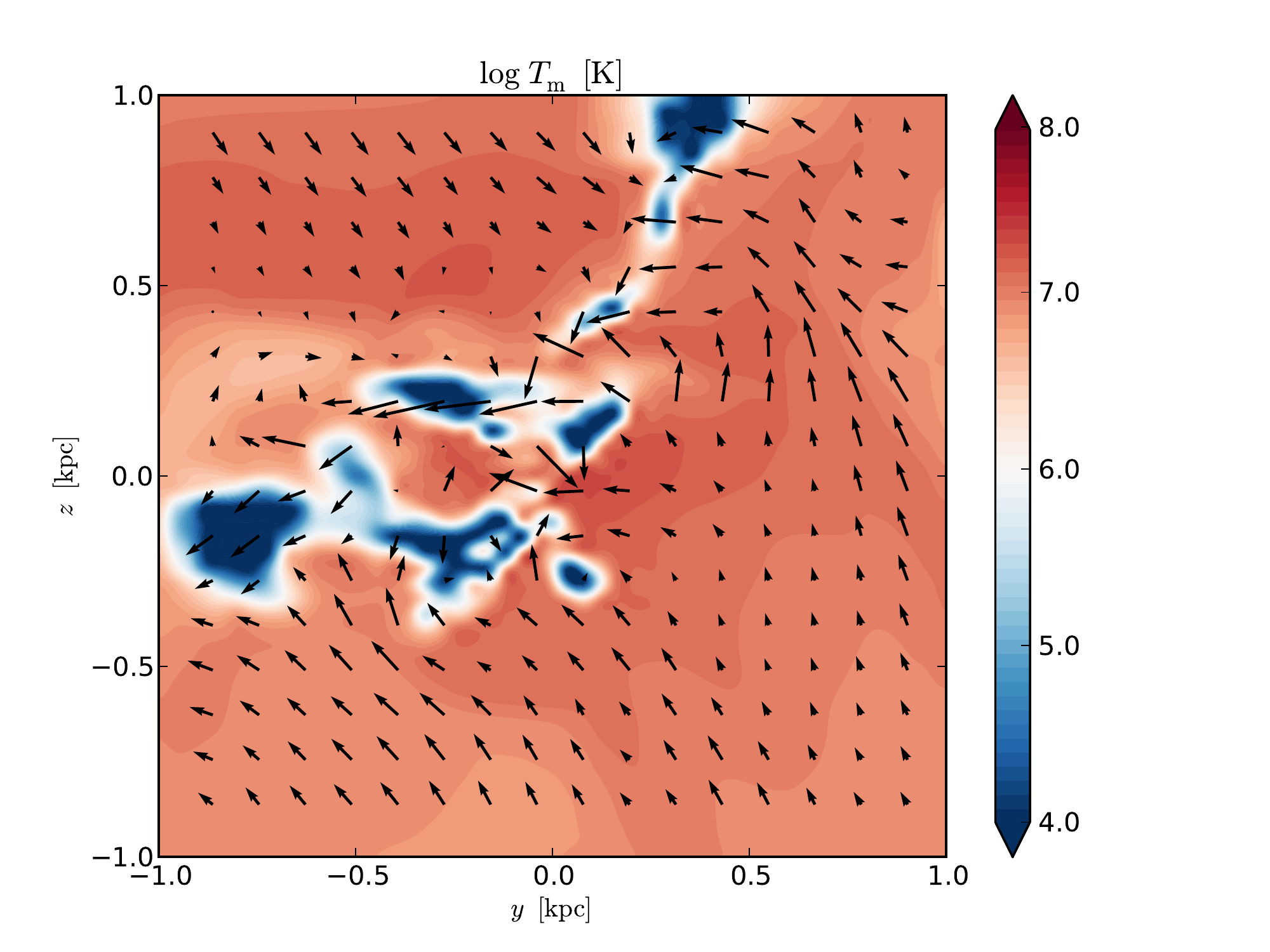}}
     \subfigure{\includegraphics[scale=0.46]{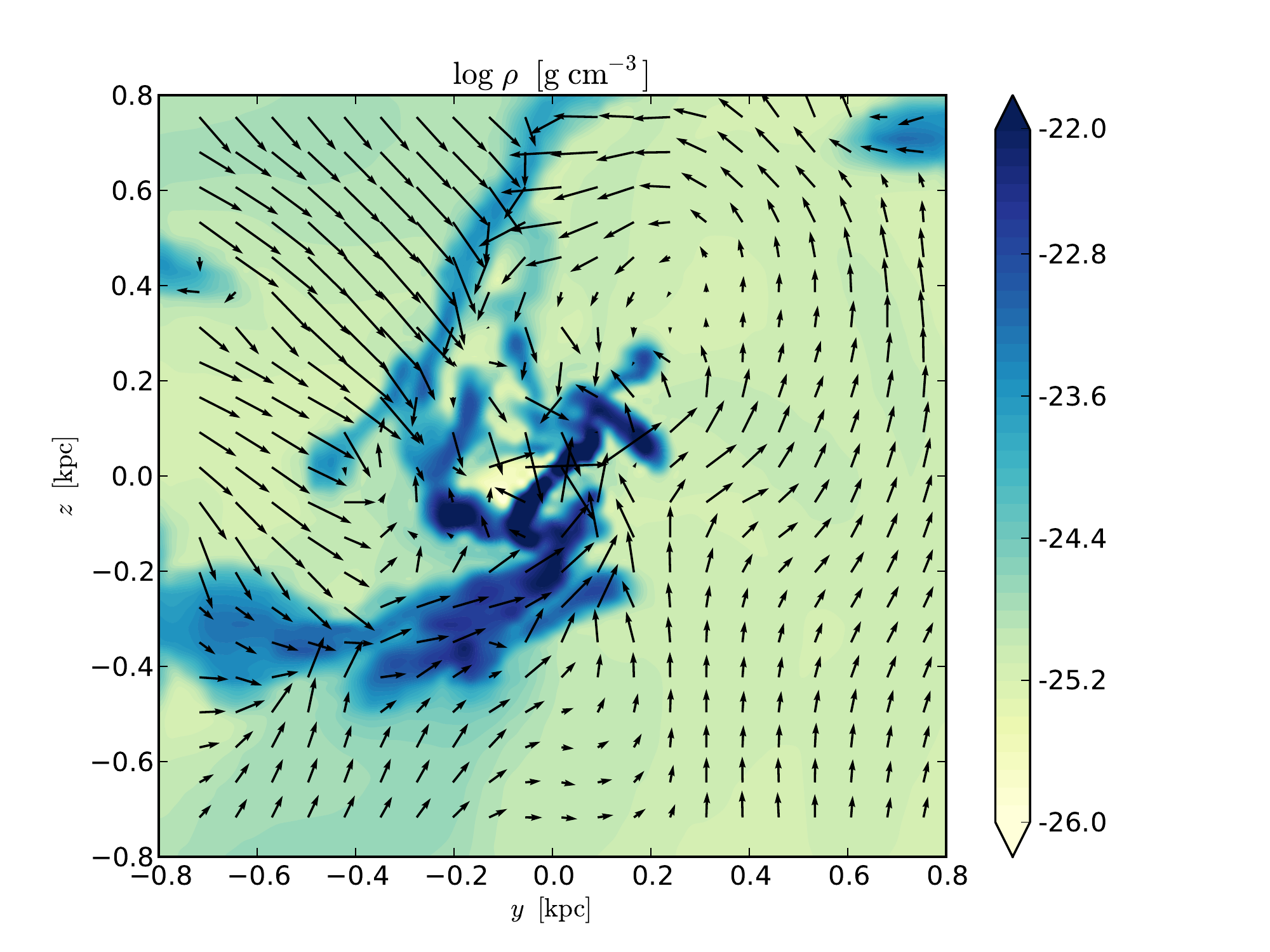}} 
    \caption{Accretion with heating, cooling \& turbulence ($M\sim0.35$). 
    Both maps show representative snapshots of the temperature (top) and density (bottom) distribution, with the velocity field overlaid.
    The mode of accretion is {\it cold} and {\it chaotic}, driven by frequent collisions and tidal motions between the cold clouds, extended filaments and the central intermittent torus. 
    These strong interactions reduce the angular momentum of the cold gas, leading to the boost of $\dot M_\bullet$.
      \label{heat4}}  
\end{figure}

The temperature and density maps clearly reveal that accretion in a realistic astrophysical situation is cold and chaotic. This is a crucial result with important implications for the evolution of galaxies, groups, and clusters (\S\ref{s:disc}).

Figure \ref{heat4} shows a representative example of the temperature (top) and density (bottom) distribution. After few cooling times, the evolution reaches equilibrium in the statistical sense. Thermal instabilities quickly enter the nonlinear regime and extended cold filaments drop out of the hot phase within $\lta8$ kpc from the centre. In this region $t_{\rm cool}/t_{\rm ff}\lta10$.

The denser gas at $10^4$ K forms the nucleus of the extended cold filaments. 
The massive nucleus 
is often surrounded by layers of cooling gas in which $t_{\rm cool}/t_{\rm ff}\sim4$$\,$-$\,$6 and temperatures are approximately a few $10^6$ K. This regime is highly thermally unstable due to strong line emission. The scenario is strikingly similar to what is seen in the multi-wavelength observations of cool cores. These observations reveal co-spatial presence of filamentary cold/warm gas in H$\alpha$, FUV, and soft X-ray band (\citealt{McDonald:2009, McDonald:2010, McDonald:2011a}). A small offset can arise in regions of strong stirring or when AGN outflows are active. 
Nevertheless, the overall correspondence between X-ray and H$\alpha$ emission suggests that the filaments developed through thermal instability, like in our models.

Figure \ref{heat4}  shows that multiphase gas forms on several scales with various topologies and complex dynamics -- a trademark of chaotic systems. The cold gas is always dense, in rough pressure equilibrium with the hot ambient.
The combined action of tidal forces and turbulent motions can easily stretch smaller spherical clouds (top panel) into longer and thinner filaments, extending even 2 kpc in length (bottom). 
Magnetic forces could further contribute to the development of filamentary structures in the flow, due to the action of anisotropic thermal conduction (\citealt{Sharma:2010}). 

As revealed by Figure \ref{heat5}, the temperature fluctuations are mainly due to entropy variations (non isentropic),
while the pressure remains rather smooth (quasi isobaric), in contrast to the evolution of the adiabatic case with stirring (\S\ref{s:stir}).
Adiabatic heating or cooling due to turbulent motions has therefore a minor relevance in chaotic cold accretion (cf. \S\ref{s:cool_prof}), 
apart from its effect on seeding TI.
The final product is a strongly biphase medium, with the low entropy phase emerging from the high entropy background.

Collisions between cold clouds are very common and may occur more frequently than once per Myr. The collision frequency increases as the clouds get closer to the accretor. Consequently, larger clouds are formed due to merging of smaller clouds. Tangled filaments increase their cross section over time (even up to $\sim$1 kpc$^2$) and experience even more encounters and shear with other clouds. The morphology of the cold phase continuously changes and, as the clouds grow in size, they may become more spherical. However, larger clouds can again become filamentary or may even get dissociated by violent motions and tidal forces. The evolution is fully chaotic.
In fact, the accretion rate and orbits of the clouds at a given time and position can not be exactly predicted (in a way similar to meteorological weather).
This scenario could also explain the frequent variability of the AGN luminosity.

We emphasise that approximating clouds as compact and dense clouds on ballistic trajectories (e.g. in semi-analytic treatments or in numerical simulations of isolated clouds) would not capture the essential physical processes. In idealised atmospheres, a moderate amount of angular momentum could prevent accretion. However, cold clouds are entities with finite cross section, which can often form extended threads. This leads to numerous inelastic collisions with a significant cancellation of the gas angular momentum. Even if the gas does not possess net non-zero angular momentum, turbulence seeds non-zero local angular momentum in the forming clouds, which is further magnified by the baroclinc instability toward inner radii.
The friction between the cold clouds with the hot phase, although minor, also helps to reduce the angular momentum of the clouds. These processes substantially boost the accretion rate (cf. the turbulence-only model in \S\ref{s:stir}).
Notice that, during the secular evolution, the cold gas will not always display an inflowing kinematics, since it can be uplifted by violent feedback events (outflows, jets, bubbles; \citealt{Gaspari:2012a}), sustained by the turbulent weather.

An interesting intermittent structure present in the flow is the nuclear torus (Fig.~\ref{heat2} and \ref{heat4}), i.e. cold gas possessing higher angular momentum within the central 100 pc (via baroclinic instability). Under idealised conditions this structure would develop into a classic perfectly rotating disc (\S\ref{s:init}; \citealt{Shakura:1973}), and some angular momentum transport process would be needed to facilitate accretion. 
In chaotic cold accretion, the momentum of the intermittent torus is instead removed by dissipative collisions with new infalling clouds that continuously bombard the torus (especially when they have opposite momentum). The presence of a dense obscuring torus within 100 pc from the centre, which is supported by an extensive AGN literature (\citealt{Bianchi:2012} for a review), is thus a natural consequence of chaotic cold accretion. The dense TI clouds could also represent the narrow and broad line region of AGN, situated on scales greater or less than the torus, respectively.
 
Overall, the key process is the angular momentum cancellation through frequent collisions (cf. \citealt{Nayakshin:2007, Pizzolato:2010}), not only between the cloud-torus but also between cloud-cloud. Even if the bulge had a significant initial rotational velocity (not modelled here, since generally low for ellipticals; \S\ref{s:init}), as long as $\sigma_v > v_{\rm rot, gas}$, rotational support and a dramatic decline in $\dot M_\bullet$ do not occur
(as shown by \citealt{Hobbs:2011}). Our simulated turbulence always resides in this regime. 
With no turbulence, the thin disc induced by gas circularisation on roughly 50-100 pc, would instead be a serious hindrance for accretion: the typical viscous timescale appears $\sim$1 Gyr at 1 pc (\citealt{King:2007}).
Other works (\citealt{Hopkins:2010, Mayer:2010}) suggest that instabilities within the disc-like structures can nevertheless alleviate the angular momentum problem.

\addtocounter{figure}{-1}
\addtocounter{figuresub}{1} 

\begin{figure} 
     \subfigure{\includegraphics[scale=0.46]{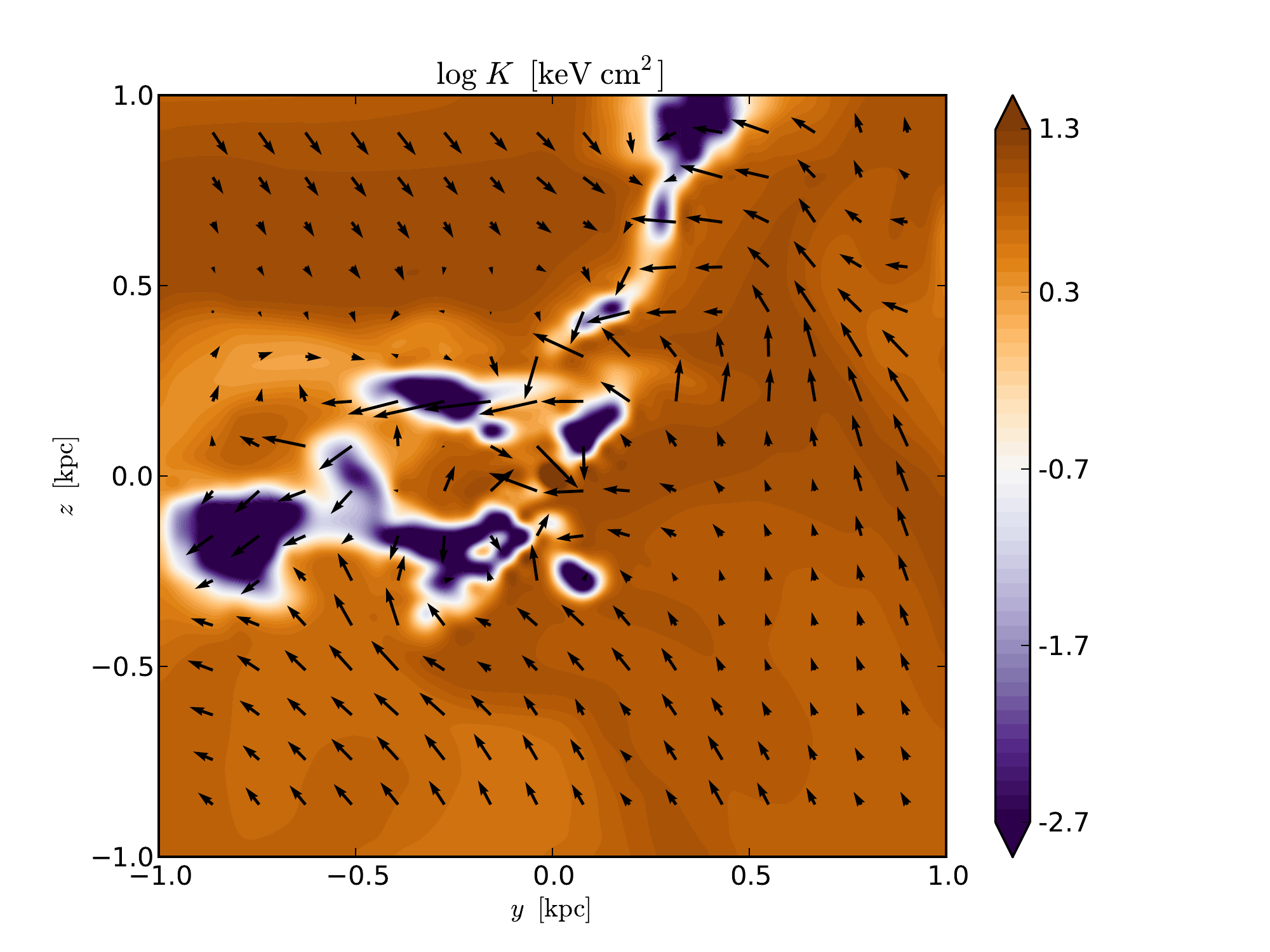}}     
    \caption{Accretion with heating, cooling \& turbulence ($M\sim0.35$). 
    Entropy distribution of the same map presented in the top panel of Fig.~\ref{heat4} ($K\equiv k_{\rm B}T/n^{2/3}_{\rm e}$). Most perturbations induced by thermal instability are not isentropic, but quasi isobaric. Adiabatic processes have therefore a minor relevance in chaotic cold accretion. The medium is strongly biphase, with the low entropy phase emerging from the high entropy background. \label{heat5}}  
\end{figure}

Self-gravity is not expected to dramatically change the rate of accretion or momentum extraction from the torus. For self-gravity to be important Toomre Q parameter must be $<1$, or $M_{\rm disc}\gta(H/R)M_\bullet$, where $M_{\rm disc}$ is the critical disc mass and $H/R$ is the aspect ratio. From the simulations, a common aspect ratio is roughly 0.1, leading to $M_{\rm disc}\gta 3\times 10^{8}$ $\msun$. 
We estimate that our typical torus masses are $\lta 10^7$ $\msun$, hence self-gravity may be secondary. 
On the other hand, we warn that the torus
is a continuously warped and dismantled structure, which can not develop in a classic thin disc (e.g. $H/R\sim10^{-3}$), due to the continuous action of turbulence and heating. Toomre Q parameter might thus not be crucial for this clumpy strucure.
From another perspective, \citet{Nayakshin:2007} argued that the torus could never grow as massive in the first place since
the stochastic accretion timescale is much shorter than the viscous timescale ($\propto R^2$), hence preventing 
star formation.
Finally, the cold clouds and filaments are not massive enough to overcome the external potential (BH, stars, matter; cf. \citealt{Li:2012}). Considering just the BH influence, the tidal disruption radius is
$r_{\rm t}\sim(M_\bullet/\rho_{\rm cloud})^{1/3}\gta 1$ kpc, for $\rho_{\rm cloud} \lta 10^{-22}$ g cm$^{-3}$,
much greater than that of any cloud ($\rho_{\rm cloud}$ declines with $r$). Only a small fraction of the cloud cores, where gas is considerably denser, can be Jeans unstable and self-gravitating. The core could then form a molecular cloud ($T<10^4$ K), triggering star formation. Subsequent stellar feedback can reheat the gas over $10^4$ K, leading to more turbulence and collisions. We defer these effects to future high-resolution MHD studies.

\subsection{Strong turbulence}\label{s:strong2}
We briefly discuss models with cooling and heating, but consider stronger turbulence by increasing the energy per mode. 

The first run reaches steady turbulent velocities of $\sim$300 km s$^{-1}$ ($M\sim0.7$), i.e. about two times larger velocity dispersion than in the previous Section. The overall evolution results to be similar. The accretion rate initially drops to lower levels (0.02 $\msun$ yr$^{-1}$) and for a longer time.
However, after 12 Myr, $\dot M_\bullet$ is again boosted by the sinking of cold clouds that condense out of the hot phase via TI. The accretion rate settles to a quasi-stable stochastic regime, but exhibits slightly higher peaks up to 1$\,$-$\,$2 $\msun$ yr$^{-1}$, showing shorter period (less than 2 Myr).
Similar to the weaker stirring case, the normalised accretion rate stays in the range of 50$\,$-$\,$100 $\dot{M}_{\rm B}$, and the cooling rate oscillates around 1 $\msun$ yr$^{-1}$.
The maps show the same stochastic dynamics as found in the previous Section, with several collisions, merging, and tidal motions. In contrast to the reference case, the radial profiles now reveal that turbulent heating starts to be effective, increasing the temperature by $\sim40$ per cent. 

We also tested the transonic stirring regime, driving turbulence up to $\sim$550 km s$^{-1}$ ($M\sim1.3$), this time only for the initial 5 Myr (continuous turbulence suppresses TI; \S\ref{s:strong1}).
This level of turbulence mimics the dynamics of a very strong AGN feedback event ($\sim$10$^{45}$ erg s$^{-1}$). 
Half the evolution (15$\,$-$\,$20 Myr) is dominated by turbulent dissipation, showing a drastic decline of $\dot M_\bullet$ (as in \S\ref{s:strong1}) and cooling rates reduced well below $1\ \msun$ yr$^{-1}$. 
The density and temperature fluctuations in the mass- and emission-weighted profiles are substantial, 
showing a $T$ increase by a factor of $\sim$2.
However, since the stirring is not continuous, after this hot accretion period,
we observe again the formation of extended multiphase gas, with
$\dot{M}_\bullet$ boosted up to $\sim1\ \msun$ yr$^{-1}$ or 100 times $\dot{M}_{\rm B}$. 

We conclude that continuous stirring, in a heated atmosphere, and for the common Mach numbers $M\sim0.2$$\,$-$\,$0.7, leads to the formation of TI and a strong boost in the accretion rate, while not violating the observational constraints on the large-scale profiles of the galaxy/group.
On the other hand, for transonic stirring, turbulent dissipation becomes significant and can stifle TI by increasing the $t_{\rm cool}/t_{\rm ff}$ unless the driving vanishes after few Myr (as in the self-regulated feedback loop). 
The secular accretion history is thus driven by the competition of $t_{\rm ff}$ versus $t_{\rm cool}$ versus $t_{\rm turb}$,
alternating between the cold and hot accretion mode (with the latter significantly less efficient).

\subsection{Convergence \& the sub-parsec region} \label{s:conv}
As a last step, we test convergence of the fiducial heated simulation with moderate stirring (\S\ref{s:heat_acc}) by performing three dedicated runs. We focus on the convergence of the accretion rate rather than the detailed cloud morphology.

In the first test, we start the evolution from the same initial conditions as before, but we increase the central resolution by 4 times to $\sim$0.2 pc. The system is integrated for half the usual evolution time, i.e. until $\simeq20$ Myr. As shown in Figure \ref{conv1}, the higher resolution run (magenta) is consistent with the previous reference accretion model (black).\footnote{Convergence is meant here in the statistical sense as the set of random numbers for the Fourier stirring modes is different for different resolutions ($\sigma_v$ remains the same).} 
The cooling rate is also convergent throughout the evolution, and radial profiles and maps of density/temperature show identical values as in Figure \ref{heat2}.

\renewcommand{\thefigure}{\arabic{figure}}

\begin{figure} 
      \begin{center}
      \subfigure{\includegraphics[scale=0.32]{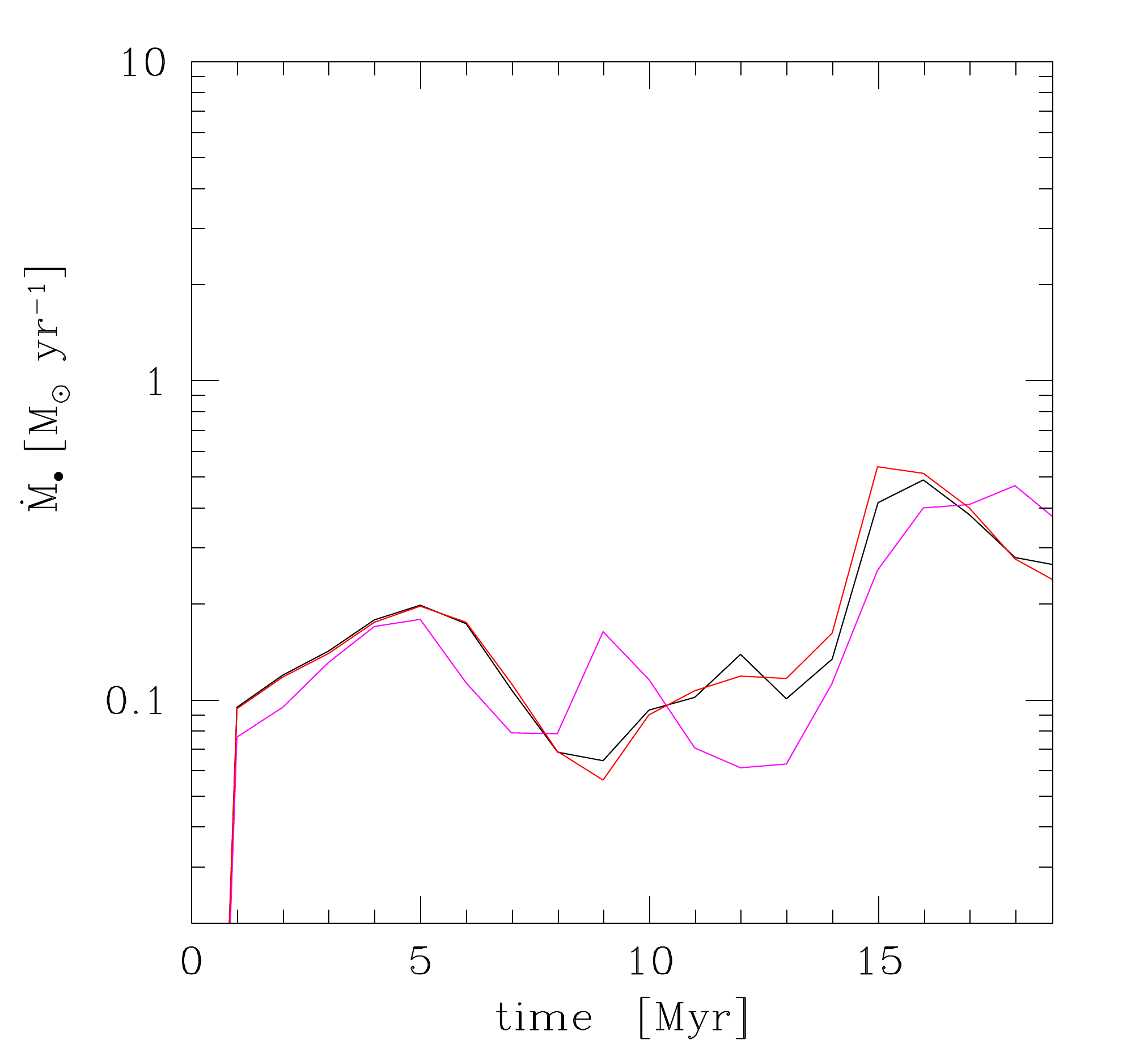}}
      \end{center}
    \caption{Convergence test with $4\times$ higher (magenta) and $2\times$ lower resolution (red) than the fiducial model with heating, cooling and moderate turbulence (black -- \S\ref{s:heat_acc}). The accretion rate shows that the fiducial run with resolution of $\sim$0.8 pc is well in the convergence limit (for the magenta line in the statistical sense, since the random noise is not identical). Increasing or reducing the resolution leads to a very similar evolution and dynamics.}
    \label{conv1}
\end{figure}

In the second test, the radial extent of the fixed AMR shells is halved, thus reducing the resolution by a factor of two at each radius. Since the maximum resolution is the same (as in \S\ref{s:heat_acc}), we are able to drive stirring with the same random numbers in real/phase space. The key result is that convergence is clearly achieved (Fig.~\ref{conv1}, red line), since the accretion rate matches that for the higher resolution case (black line). Occasionally, the accretion rate in the lower resolution case is higher by a few percent due to the formation of cold blobs over few grid zones.
However, most of the cold blobs are typically described by a thousand or more cells (in volume). 
Turbulent diffusion acts also as an effective diffusion, preventing the cold clumps to collapse to the minimum resolution
(the details of the cold phase will be studied in a following work).
We note that while the clouds emerge in the entire radial range where $t_{\rm cool}/t_{\rm ff}\lta10$, the majority of the them form near the inner better resolved radius where this condition holds (Fig.~\ref{heat3}). The evolution of the cooling rate is also convergent, smoothly increasing to 1 $\msun$ yr$^{-1}$ at 20 Myr. Finally, over 40 Myr, profiles and maps show identical  behaviour described in \S\ref{s:heat}. 
We conclude that the pc resolution of the fiducial models ensures that the results are robust.

\begin{figure} 
      \begin{center}
      \subfigure{\includegraphics[scale=0.42]{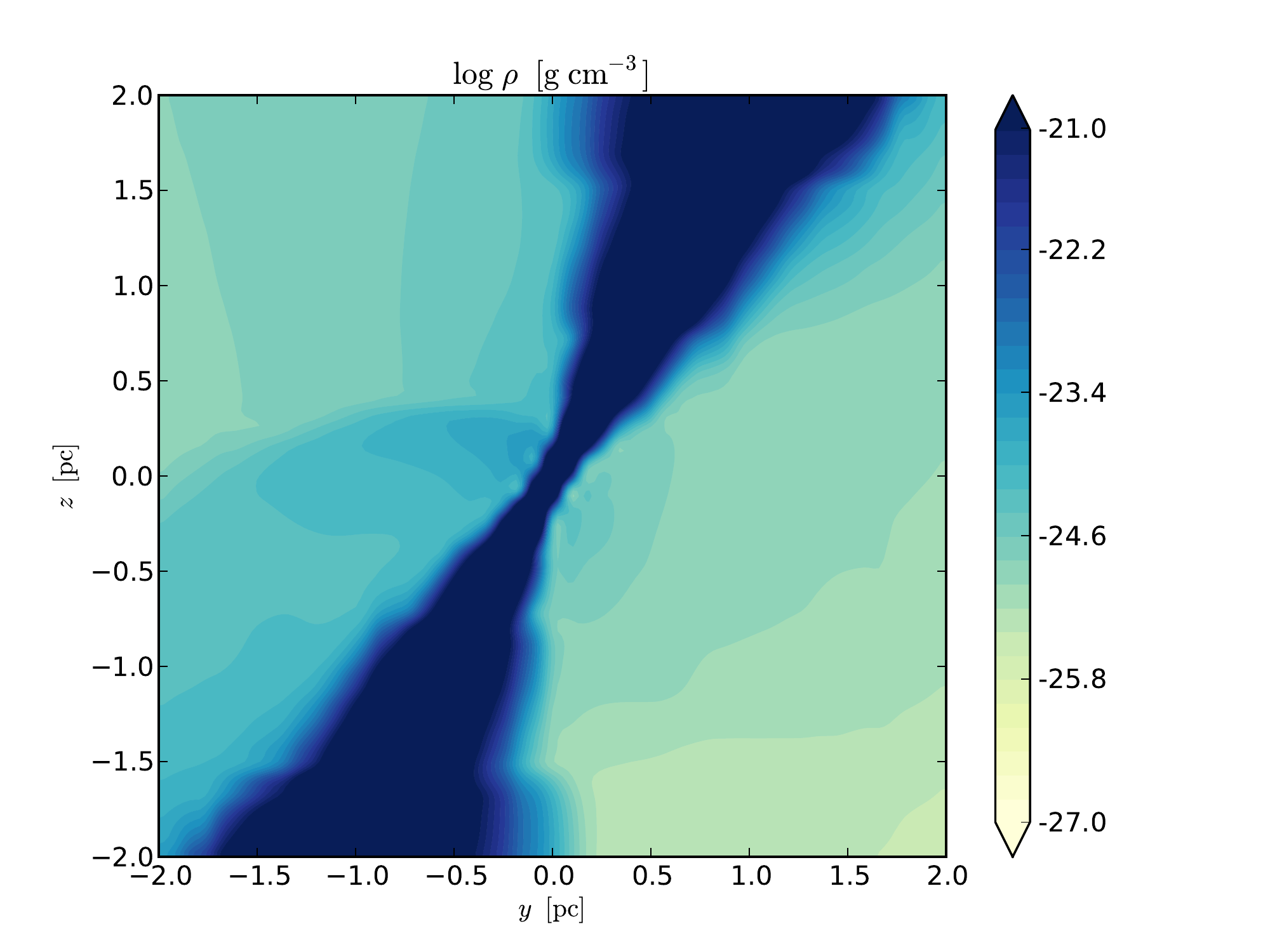}}
      \end{center}
    \caption{Cross section of the gas density distribution from the deep zoom-in test with resolution of $\sim$20 $R_{\rm S}$. The accretion rate is still maintained at high levels, since the cold clouds are stretched and channelled via a funnel toward the BH by its strong gravity. 
    \label{conv2}}
\end{figure}

The last test is the most expensive, pushing the AMR capability of the code and reaching the resolution of $\sim20\ R_{\rm S}\sim 6\times10^{-3}$ pc (the dynamical range is almost 10 million). The time evolution needs to be drastically decreased down to 1 Myr. We restarted the fiducial run from the final time (40 Myr), keeping the same conditions in the 52 kpc box but zooming in on the central region. 
The previous sink region is filled with hot diffuse gas at $\sim$$10^7$ K. After the initial transient phase due to the restarting\footnote{The system experiences a brief phase of purely hot accretion (few 10 kyr). Refilling the older sink region with dense cold gas, induces instead an initially higher $\dot M_\bullet$, but the later evolution is then identical to the previous case with hot gas refilling.}, the multiphase gas steadily accretes reaching an average value slightly less than that of the fiducial run in 1 Myr, $\sim$0.4 $\msun$ yr$^{-1}$.
The accretion rate remains high even on sub-pc scales, since the strong gravitational attraction (\S\ref{s:sink}) channels the cold gas toward the black hole via a narrow funnel (Figure \ref{conv2}). 
In a few dynamical times, most of the cold clouds experiencing chaotic collisions can be potentially captured and channelled toward the sub-pc region. 
Below a few tens of $R_{\rm S}$, the accretion process is expected to be more complex. We will perform dedicated simulations to study this region including additional physical processes discussed in $\S\ref{s:disc}$.

\section{Comparison with other works} \label{s:comp}
In this Section, we review and compare previous work in the accretion literature related to our simulations, which can shed further insights into chaotic accretion. 
We do not attempt to provide an exhaustive picture of previous accretion physics models. 

Regarding analytic calculations, \citet{King:2006,King:2007} and \citet{Nayakshin:2007} postulated that black holes may grow through a series of randomly oriented accretion events, via a succession of minor mergers. 
Their main hypothesis 
was that the driver of accretion is the cancellation of angular momentum via collisions between the clouds and torus. This avoids the problem of accreting the gas through viscous dissipation in a classic thin disc, which has a large inflow time ($\sim$1 Gyr near 1 pc). As pointed out by the previous authors, important observational implications could arise from chaotic accretion. The self-gravity catastrophe of the disc can be avoided at radii up to tens of parsecs, 
preventing global star formation and inefficient gas accretion. Nuclear star formation may be though possible in rings near $\sim$0.01 pc (the self-gravity radius), especially at low accretion rates (e.g. Sgr A$^\ast$ or dwarf galaxies). Further, the spin of the black hole could be significantly lowered, leading to faster BH growth, due to the low radiative efficiency, and random jet orientation disconnected from large-scale discs or bars. \citet{King:2007} also estimated the BH luminosity function, emerging from a chaotic evolution, which is encouragingly similar to the {\it SDSS} data in \citet{Heckman:2004}. 

Our simulations show that a turbulent, cooling and heated system experiences chaotic dynamics at radii up to almost 10 kpc due to thermal instabilities, where $t_{\rm cool}/t_{\rm ff}\lta 10$, not only below 10 pc and without the requirement of merging satellites. 
We found the torus a very volatile and clumpy structure, which does not develop into a steady thin disc, since it is frequently dismantled ($<1$ Myr) by the turbulent and heated environment (feedback events would accentuate this behaviour). The accretion is driven by angular momentum cancellation via collisions between both the cloud and torus {\it and} the clouds themselves.

\citet{Pizzolato:2005} and \citet{Soker:2009} highlighted a major flaw of pure Bondi accretion as a source of feedback heating, i.e. the slow communication time ($> 1$ Gyr) between the kpc-scale gas and the BH.  
They also emphasised the importance of cooling. Using analytic models, they focused on the condensation of a cold blob, suggesting that the cold phase is a key element for BH accretion (they named the model `cold feedback mechanism'). As shown by our numerical simulations, the angular momentum barrier is avoided by frequent collisions, with typical timescale shorter than the inflow time, especially in the inner 100 pc (cf. \citealt{Pizzolato:2010}), in minor part helped by the drag of the ICM.

\citet{Hobbs:2012_subgrid} emphasised the failure of classic Bondi-Hoyle prescription in cosmological simulations, 
since on large scale the potential is dominated by the galaxy and, combined with cooling, the flow might enter the free-fall regime (see also \citealt{Rees:1977,Quataert:2000}). They suggested a subgrid formula interpolating between Bondi-Hoyle and $\dot{M_\bullet}\sim M_{\rm gas} (< r)/t_{\rm ff}$. However, their calculations assume isothermal gas and no heating, and thus can not resolve TI or a multiphase medium.
In the presence of a heated atmosphere (\S\ref{s:heat}), the cooling flow is moderate and the free-fall regime is attained only by single TI clouds, not the entire gas mass enclosed within the halo. We find the cooling rate (in the galactic core) a simpler and effective approximation, which avoids the limitations of the Bondi formula (see \S\ref{s:cool_mdot}). If $t_{\rm cool}\ll t_{\rm ff}$, then the accretion rate should be limited by the free-fall time, as argued by the previous authors. This could happen for gas-rich high-$z$ galaxies, especially at large radii. We remark that the cold mode of black hole accretion is not identical to the massive cold flow associated with galaxy formation and occurring at $r\gta r_{vir}$ (\citealt{Keres:2005}), since they work in different density and length-scale regimes.

Using smoothed particle hydrodynamics (SPH), \citet{Hobbs:2011} studied the evolution of a rotating (60 km s$^{-1}$) and gaseous shell (30$\,$-$\,$100 pc) surrounding the BH, for a few dynamical times. The gas is assumed isothermal.
In the unperturbed case, $\dot M_\bullet$ is strongly suppressed. On the other hand, adding turbulence with $\sigma_v\gta v_{\rm rot}$ boosts the accretion rate by up to three orders of magnitude.
When turbulence is supersonic, dense filaments are compressed inducing `ballistic' accretion.
However, our computed long-term evolution 
of an extended non-isothermal atmosphere 
experiencing cooling plus heating shows a different behaviour, i.e. multiple collisions between condensed cold clouds,
deviating from bullet-like orbits, even in the presence of subsonic turbulence. 

\citet{Barai:2012} employed SPH simulations to study the strong X-ray radiation from a quasar affecting a dense absorbing shell (0.1$\,$-$\,$200 pc) 
for $\lta 1$ Myr. When $L \sim0.01 \ L_{\rm Edd}$ and ionisation $\xi\sim500$, the shell fragments in dense long filaments, due to the exponential growth of thermal instability (in the range $r\sim1$$\,$-$\,$30 pc). 
This photo-ionised TI (cf. \citealt{Krolik:1983}) occurs in a very narrow regime ($0.01-0.02\ L_{\rm Edd}$), below which radiation becomes irrelevant and above which a spherical outflow develops, suppressing accretion.

The same authors pointed out that the SPH method has relevant limitations when studying instabilities, stating that inherent fluctuations of SPH can lead to a spurious TI growth. As shown by comparison studies (e.g. \citealt{Agertz:2007, Tasker:2008, Mitchell:2009, Bauer:2012}) fluid instabilities, discontinuities, mixing, and turbulence are poorly resolved or even suppressed by SPH techniques. The finite number of total SPH particles which can be sinked is also a limitation (the boundary problem), allowing only a brief time integration (\citealt{Barai:2011}). 
We do not claim that grid codes are unaffected by numerical diffusion and deficiencies,
nevertheless we believe they are more suited to study the turbulent and stochastic dynamics, developing key instabilities (cf. figure 4 in \citealt{Bauer:2012}). 

Finally, we note that, in contrast to the previous numerical works, we have studied the mode of accretion for several tens Myr, in a realistic galactic atmosphere,
linking the 100 kpc to the subpc scales, under the action of cooling, heating, and turbulence.\\

\section{Summary \& discussion}  \label{s:disc}

The simplicity of classic accretion theory makes it extremely appealing to apply the Bondi formula (Eq.~\ref{MdotB}) to the interpretation of observations, build analytical models, and parameterise the black hole accretion rate in large-scale simulations. However, the Bondi model requires a number of strong assumptions that can be grossly violated in a realistic astrophysical environment. The Bondi flow must be adiabatic, steady, unperturbed, with no vorticity or heating,  set by constant boundary conditions at large radii, among the most notable.
In an attempt to increase the realism of the accretion, we presented a systematic investigation of the effects of radiative cooling, global heating, and turbulence on the black hole mass accretion rates in extreme dynamical range simulations (up to 10 million), and compared the simulated evolution with the prediction of the classic Bondi model. Our 3D simulations also allow us to isolate the effect of non-vanishing gradients in density and temperature on the accretion rate. Our model is fairly general and applicable to a range of systems, such as hot halos in galaxies, groups, and clusters. Below we present the summary of our key findings, discuss crucial implications and limitations of the model, as well as future directions.\\

1. {\it Adiabatic \& unperturbed accretion}. \\
First, we studied adiabatic accretion in the presence of a realistic declining density and slightly increasing temperature profiles. That is, we relaxed the assumption of the flat distribution of thermodynamic variables assumed in the Bondi model.
When evaluated near $r_{\rm B}$, the Bondi formula provides an excellent estimate for the actual accretion rate (within a few percent), even if the accretion rate slowly declines over time. Computing instead the Bondi rate on the kpc scales (as often done in literature), leads to the underestimate in $\dot M_\bullet$ by a factor of 2$\,$-$\,$4. This is due to the presence of non-vanishing density and temperature gradients; this bias can not reach a factor of $\gta100$. We emphasise that the relativistic potential induces a finite sonic point (near the pc scale), even in the adiabatic flow. \\

2. {\it Cooling \& unperturbed accretion}. \\
The standard Bondi prescription can result in unrealistic accretion rates. The main reason for this is the violation of adiabaticity: astrophysical hot plasmas are radiating, losing thermal energy and pressure support via condensation.
In a hot radiative atmosphere, the accretion rate is boosted by two orders of magnitude, compared with the pure Bondi values. After one cooling time, the conditions near the Bondi radius are completely altered by cooling, i.e. from that point on, accretion is dominated by the thermally unstable cold gas rather than the spatially uniform hot phase. Within a few 100 pc, the mass-weighted temperature profiles drop to $10^4$ K and the densities reach over 100 cm$^{-3}$. On the kpc scale, the $T$ profile is only slightly decreasing, partially sustained by compressional heating.
However, after 40 Myr, a strong cooling flow progressively alters also these regions (if no feedback is present).\\

3. {\it Cooling \& turbulent accretion}.\\
Adding stochastic stirring motions characterised by $\sigma_v\sim100$$\,$-$\,$300 km s$^{-1}$ ($M\sim0.25-0.7$), further increases the realism of the simulation. These motions can be produced by AGN outflows, bubbles, supernovae, mergers or cosmic flows. The key result is the nonlinear growth of thermal instabilities and the formation of multiphase gas even for  moderate turbulence. Accretion is no longer symmetric and is instead completely chaotic. The cooling rates ($T\lta10^5$ K) are still high ($\gta8$ $\msun$ yr$^{-1}$), and the profiles differ from the previous case only for the continuous fluctuations.
At transonic velocities $M\gta1$, turbulent heating becomes significant and starts inhibiting TI and cooling: the key threshold is $t_{\rm turb}/t_{\rm cool}\lta 1$, where $t_{\rm turb}\sim M^{-2}\, L/\sigma_{\rm v}$.\\

4. {\it Adiabatic \& turbulent accretion}.\\
Without cooling, the effect of the reference subsonic turbulence ($M\sim0.35$) is instead to reduce the accretion rate by a factor of $\sim$3. Accretion is reduced by orders of magnitude when $M>1$. The profiles resemble the Bondi hot mode solution with flatter density and temperature profiles, due to turbulent diffusion. In the subsonic regime and in a stratified atmosphere, the generation of vorticity via the baroclinic instability plays a key role in the reduction of $\dot M_\bullet$. When turbulence leads to a transient strong bulk motion, the non-zero relative velocity between the accretor and the flow further reduces the accretion rate. \\

5. {\it Heating, cooling \& turbulent accretion}.\\
The last and most realistic scenario includes spatially-distributed heating. This is motivated by the fact that atmospheres appear to remain in global thermal equilibrium due to self-regulated AGN feedback, possibly helped by conduction, mergers and stellar evolution. The heating is broadly applicable to a range of situations. The heating offsets cooling globally, but not locally, leading to the growth of nonlinear thermal instability.
In the heated atmosphere, the total cold mass and cooling rate are reduced by an order of magnitude. 
On the other hand, (quasi isobaric) fluctuations are amplified and TI now grows exponentially. The result is a very frequent condensation of cold clouds and filaments, in the radial range of $\sim$100 pc$\,$-$\,$10 kpc, where $t_{\rm cool}/t_{\rm ff}\lta10$. The key point is that in a common astrophysical scenario, {\it accretion is mainly cold and chaotic}. The dynamics is driven by stochastic dissipative {\it collisions} which cause a significant reduction of angular momentum in the cold phase and, hence, a significant boost of $\dot M_\bullet$. The accretion rate is fluctuating with peaks over 1 $\msun$ yr$^{-1}$.  Utilising the Bondi formula with the density and temperature evaluated on kpc scales would underestimate the true accretion rate by up to a factor of 50$\,$-$\,$100.

It is remarkable that subsonic turbulence (100$\,$-$\,$300 km s$^{-1}$) can induce thermal instabilities, multiphase gas, and chaotic cold accretion. In fact, such a level of velocity fluctuations seems to be very common in gaseous halos (e.g. \citealt{Schuecker:2004, Rebusco:2006, Churazov:2012, dePlaa:2012, Sanders:2013}; and even higher in the interstellar medium: \citealt{Vazquez-Semadeni:2012}) and ca be confirmed by the upcoming {\it Astro-H} mission 
(e.g. \citealt{Zhuravleva:2012,Shang:2012a,Shang:2012b}).

The cold phase shows a plethora of morphologies on different scales: from clouds to a central torus to thin long filaments, stretched by the turbulent and tidal forces. 
The cold phase, although almost in free-fall, can not be considered collisionless. A significant cross section of clouds is a crucial factor facilitating frequent interactions (especially within 5$\,$-$\,$10 $r_{\rm B}$), and in boosting $\dot M_\bullet$ beyond the Bondi prediction. 
Chaotic clouds seem excellent candidates to 
explain the common variability in the AGN luminosity, as well as the deflection or mass-loading of jets via entrainment.
The ubiquitous presence of cold and molecular gas observed in the cores of galaxies, groups, and clusters (e.g. \citealt{Edge:2001, Mathews:2003, Salome:2004, Rafferty:2008, McDonald:2010, McDonald:2011b, Davis:2012, Russell:2012, Werner:2013}) seems to corroborate the cold accretion scenario. The dense TI clouds could also represent the narrow and broad line region of AGN, situated on scales greater or less than the torus, respectively. 

The angular momentum problem is a crucial issue in transporting gas from sub-kpc to much smaller scale. For instance, local viscous stresses, which are thought to dominate the angular momentum transport in classic discs, become ineffective at $r \gta 0.01-0.1$ pc \citep{Goodman:2003}.
Chaotic cold accretion drives gas accretion by angular momentum cancellation via collisions between the cloud-torus and cloud-cloud over several kpc. The torus is indeed a volatile and clumpy structure, continuously dismantled by the turbulent and heated environment, which prevents 
the development of a strong angular momentum barrier (locally induced by turbulence and enhanced by the baroclinc instability; \S\ref{s:stir}).

The stochastic nature of the accretion process is also revealed via fluctuations in the radial mass-weighted profiles: below 1 kpc the density can exceed $10^3$ cm$^{-3}$ (X-ray brightness increase would be visible only within $r_{\rm B}$). The X-ray temperature is instead remarkably flat, even down to tens of pc. The next generation of deeper observations could clearly determine the current mode of accretion through the observations of the temperature profiles. The adiabatic Bondi-like model predicts in fact a rising X-ray temperature below $0.5\ r_{\rm B}$, although this could be also achieved when the central feedback becomes very active. Note that extrapolating the (emission-weighted) profiles from the kpc scale is instead misleading, since cold and hot accretion display the same behaviour. NGC 4472 and NGC 4261 may be two cases of inner flat $T$ profile, while NGC 4649 and NGC 1332 may represent the hot mode regime (\citealt{Humphrey:2008, Humphrey:2009}). 

We performed convergence tests and obtained consistent results. Lowering or increasing the resolution, does not change the mass accretion rates: $\dot{M}_\bullet$ stays on high levels, with similar cooling rates and chaotic dynamics. On sub-pc scales the cold phase is stretched and channelled toward the accretor via a funnel, while the accretion rate remains on a level comparable to that on larger scales.

\subsection{Improvements \& additional physics} \label{s:disc_phys}

In future, we intend to study in more detail the extended multiphase gas. This phase not only provides the fuel for accretion, but it is also associated with star formation and is co-spatial with dust and molecular clouds visible in the optical, infrared, and radio bands.
Limited resolution may favour the condensation of the less resolved clouds and filaments, which could be also achieved by magnetic draping. On the other hand, numerical diffusion tends to 
prevent the initial development of instabilities on the smallest scales, an effect that could be mimicked by thermal conduction. 
All these effects can affect the distribution of the detailed emission measure as a function of $T$, which we will study next.

The magnetic field could play a role in shaping the cold phase, while its average low strength of few $\mu$G might be insufficient to modify the global accretion dynamics.
The cold filaments could be though characterised by enhanced magnetic fields with vectors aligned along the filaments. Magnetic fields will also play a role in determining the degree of mixing between the hot and cold phase, with  anisotropic thermal conduction likely equalising $T$ along the filaments (\citealt{Sharma:2010}). On the other hand, turbulence could act as the effective (isotropic) diffusive process, making conduction and the typically tiny Field length secondary.

Strong X-ray radiation could also influence the evolution (\citealt{Buff:1974,Ostriker:1976,Krolik:1983, Barai:2012}). In particular, Compton heating could have the major impact, especially near a few $R_{\rm S}$, where the cold clouds reach extreme column densities and opacities. The coupling with the hot plasma on large scales appears instead inefficient due to its high ionisation and low density. A (rare) quasar-like Eddington regime may be necessary to produce sufficient radiation pressure to alter the global dynamics.

In this study, we did not include the strong initial phase of BH feedback, focusing more on the fuelling stage. 
Bipolar kinetic outflows or jets can entrain and uplift a fraction of the infalling clouds along the direction perpendicular to the torus (\citealt{Gaspari:2012a}), and induce higher turbulence and new cold phase interactions. Some analytic theories (e.g. \citealt{Blandford:1999}) also predict that the gas reaching the very centre may become unbound. However, the fate of the infalling gas depends on the mode of accretion. The presence of cold gas condensation near $r_{\rm B}$ means that the outer boundary conditions for the sub-pc region would be different than normally assumed, and analytic accretion theories (as ADAF or thin disc) might lead to different results.

\subsection{Key implications \& applications}

Considering the chaotic nature of cold accretion, it is not trivial to predict the exact accretion rate. However, we find that the cooling rates (in the galaxy core) are tightly linked to the BH accretion rates, $\dot{M}_\bullet \approx \dot{M}_{\rm cool,\, core}$, both in the heated and pure cooling cases. Due to low resolution, cosmological simulations often assume substantially boosted Bondi rates by a factor of $\sim$100 (e.g. \citealt{Springel:2005, DiMatteo:2005, Sijacki:2007, Booth:2009}), or high feedback efficiencies, in order to explain the growth of BHs and their host galaxy. 
Interestingly, we find from first principles, that the boost factor is about two orders of magnitude just as required by the cosmological models (the reference $\dot{M}_{\rm Bondi}$ was computed on kpc scales, as is commonly done in large-scale simulations).
We suggest thus to apply the subgrid prescription $\dot{M}_\bullet \approx \dot{M}_{\rm cool,\, core}$, which can physically reproduce both the increased magnitude and the realistic fluctuating evolution of the accretion rate, avoiding ad-hoc factors dependent on numerics\footnote{Different boost parameters lead to profoundly different star formation history and black hole growth (\citealt{Booth:2009}).}. Overall, it is the cold mode that drives the BH growth/feedback, explaining why large-scale simulations are forced to strongly modify Bondi `hot' accretion in order to match observations.

The TI scenario provides a direct link between the large scales of the hot galactic halo and the relatively tiny SMBH. The TI criterion determines how much material can decouple from the hot halo and rain down onto the black hole leading to its growth. The presence of the instability and the rate of its development are on the other hand controlled by the feedback from the BH. In the cold mode, the response time of the black hole to the increased rate of accretion (a few $t_{\rm ff}$) is much faster than in the case of the hot mode. Therefore, a very efficient feedback can be established, leading to the proper thermodynamical self-regulation of the galaxy, group or cluster, with a duty cycle on the order of the cooling time.

In this symbiotic mechanism, the Magorrian relation (\citeyear{Magorrian:1998}) could appear naturally, since the accretion rate is potentially linked to the gas mass of the {\it entire} galaxy/bulge (up to several kpc, where $t_{\rm cool}/t_{\rm ff}\lta10$), 
while in Bondi hot accretion the supply is severely limited by the region $\lta r_{\rm B}$ ($< 100$ pc).
Chaotic cold accretion could thus be the key feeding mechanism able to establish the linear scaling
$M_\bullet \propto M_{\rm \ast, bulge}$ (\citealt{McConnell:2012}). 
The proper normalisation, $\sim$$10^{-3}$, is certainly shaped by the evolution of halos, recurrent chaotic TI events, and varying star formation rates regulated by AGN feedback, which may also explain the significant scatter in the relation\footnote{Momentum- or energy-driven feedback may also determine the conversion of $M_\bullet - M_{\rm \ast}$ to the scaling with the stellar velocity dispersion, $M_\bullet \propto \sigma_\ast^4-\sigma^5$ (\citealt{King:2005,Fabian:2012, Nayakshin:2012}).}.
It is interesting to note that, assuming a high-redshift BH growth during few TI events,
we can crudely estimate $M_\bullet = \dot M_\bullet \,t_{\rm active} \sim \dot M_{\rm cool, net}\, t_{\rm ff}$.
Since cooling and heating must preserve an equilibrium of roughly $10$\% (\S\ref{s:gheat}), this implies $\dot M_{\rm cool, net}\sim0.1\, \dot M_{\rm cool} \sim 0.1\, M_{\rm gas}/t_{\rm cool}$. Substituting in our estimate,  $M_\bullet \sim \dot M_{\rm cool, net}\, t_{\rm ff} \sim0.1 \, M_{\rm gas}\, (t_{\rm ff} / t_{\rm cool}) \sim 10^{-2}\, M_{\rm gas,core}$, using the TI-ratio threshold of 10.
The conversion to stars is not trivial, since star formation rates strongly vary with the cosmological epoch.
For example, during starbursts much of the gas can form stars, which can not be longer accreted, accumulating in the bulge.
We plan therefore to perform dedicated cosmological simulations to exactly quantify the normalisation factor and the relation between gas, stars and black hole.

Based on the symbiotic scenario, more massive black holes may reside in more massive galaxies, due to the larger available fuel up to few 10 kpc ($\dot M_\bullet\approx\dot M_{\rm cool}\propto M_{\rm gas, core}$), 
rather than due to the larger radius of influence (as in Bondi theory where $\dot M_\bullet \propto r_{\rm B}^2\propto M^2_\bullet$). The symbiotic link between the tiny black hole and the {\it extended} environment should find its strongest manifestation in the cores of clusters, where black holes seem to grow up to several $10^{10}$ $\msun$ (\citealt{McConnell:2011, Fabian:2013}). 

Cold accretion should play a crucial role in high-$z$ galaxies, since the intergalactic medium appears strongly multiphase, as evidenced by the Lyman-$\alpha$ forest, and the rich gas supply has not yet massively converted into (collisionless) stars. The massive TI condensation could quickly drive fast accretion beyond the Eddington limit, activating the quasar regime and potentially providing an explanation for the presence of SMBHs at high-$z$ (\citealt{Haiman:2013}). Simulations with radiative transfer are required to understand this further.
We note that if $t_{\rm cool} \ll t_{\rm ff}$ then the accretion rate formula should be limited by the free-fall time (see \S\ref{s:comp}).
The reference cooling rate for $\dot M_\bullet$ must be nevertheless computed in the galactic core, in order to distinguish it from the {\it different} massive cold flow linked to galaxy formation at $r\gta r_{\rm vir}$.
At lower redshift, when hot halos are already well formed and more tenuous, the feedback driven by thermal instability should become less violent and more recurrent, gently preserving the core of the system in quasi-stable thermal equilibrium for several Gyr (e.g. \citealt{Gaspari:2012c}). 

In closing, we remark that the hot mode is not absent from the secular evolution. After strong feedback or dynamic events, the system typically experiences a phase of overheating and substantial turbulence, which is more prolonged in less massive/more diffuse halos (isolated ellipticals, spiral galaxies, etc.), due to the lower binding energy
and gas sound speed. 
During this stage, thermal instability is suppressed. The only channel of accretion is hot gas, with accretion rates severely reduced by turbulent motions and heating (\S\ref{s:strong1}-\ref{s:stir}). NGC 3115 and Sgr A$^\ast$ might be two exemplary cases. Overall, the secular accretion history results to be substantially polarised, showing deviations of orders of magnitude either above or below the Bondi rate, driven by the competition between three key timescales: $t_{\rm ff}$ versus $t_{\rm cool}$ versus $t_{\rm turb}$ (stratification vs. net cooling vs. turbulence).

\section*{Acknowledgments}
The FLASH code was in part developed by the DOE NNSA-ASC OASCR Flash centre at the University of Chicago. MR acknowledges NSF grant AST 1008454 and NASA grant 12-ATP12-0017.
SPO acknowledges NASA grant NNX12AG73G for support.  
We acknowledge the CLS centre and University of Michigan for the availability of high-performance computing resources.
MG thanks in particular F. Brighenti, along with E. Churazov, N. Soker, A. Tchekhovskoy, and K. Yang for interesting discussions. The post-processing analysis was in part performed with YT (\citealt{Turk:2011}).

\bibliographystyle{mn2e}
\bibliography{biblio}

\providecommand{\SortNoop}[1]{}
\begin{thebibliography}{122}
\expandafter\ifx\csname natexlab\endcsname\relax\def\natexlab#1{#1}\fi

\bibitem[{{Agertz} {et~al}\mbox{.}(2007){Agertz}, {Moore}, {Stadel}, {Potter},
  {Miniati}, {Read}, {Mayer}, {Gawryszczak}, {Kravtsov}, {Nordlund}, {Pearce},
  {Quilis}, {Rudd}, {Springel}, {Stone}, {Tasker}, {Teyssier}, {Wadsley}, \&
  {Walder}}]{Agertz:2007}
{Agertz} O. {et~al.}, 2007, \mnras, 380, 963

\bibitem[{{Allen} {et~al}\mbox{.}(2006){Allen}, {Dunn}, {Fabian}, {Taylor}, \&
  {Reynolds}}]{Allen:2006}
{Allen} S.~W., {Dunn} R.~J.~H., {Fabian} A.~C., {Taylor} G.~B., {Reynolds}
  C.~S., 2006, \mnras, 372, 21

\bibitem[{{Baganoff} {et~al}\mbox{.}(2003){Baganoff}, {Maeda}, {Morris},
  {Bautz}, {Brandt}, {Cui}, {Doty}, {Feigelson}, {Garmire}, {Pravdo}, {Ricker},
  \& {Townsley}}]{Baganoff:2003}
{Baganoff} F.~K. {et~al.}, 2003, \apj, 591, 891

\bibitem[{{Balbus} \& {Soker}(1989)}]{Balbus:1989}
{Balbus} S.~A., {Soker} N., 1989, \apj, 341, 611

\bibitem[{{Barai} {et~al}\mbox{.}(2011){Barai}, {Proga}, \&
  {Nagamine}}]{Barai:2011}
{Barai} P., {Proga} D., {Nagamine} K., 2011, \mnras, 418, 591

\bibitem[{{Barai} {et~al}\mbox{.}(2012){Barai}, {Proga}, \&
  {Nagamine}}]{Barai:2012}
{Barai} P., {Proga} D., {Nagamine} K., 2012, \mnras, 424, 728

\bibitem[{{Bauer} \& {Springel}(2012)}]{Bauer:2012}
{Bauer} A., {Springel} V., 2012, \mnras, 423, 2558

\bibitem[{{Bianchi} {et~al}\mbox{.}(2012){Bianchi}, {Maiolino}, \&
  {Risaliti}}]{Bianchi:2012}
{Bianchi} S., {Maiolino} R., {Risaliti} G., 2012, ADA\&A, 2012

\bibitem[{{Binney} {et~al}\mbox{.}(2009){Binney}, {Nipoti}, \&
  {Fraternali}}]{Binney:2009}
{Binney} J., {Nipoti} C., {Fraternali} F., 2009, \mnras, 397, 1804

\bibitem[{{Blandford} \& {Begelman}(1999)}]{Blandford:1999}
{Blandford} R.~D., {Begelman} M.~C., 1999, \mnras, 303, L1

\bibitem[{{Bondi}(1952)}]{Bondi:1952}
{Bondi} H., 1952, \mnras, 112, 195

\bibitem[{{Bondi} \& {Hoyle}(1944)}]{Bondi:1944}
{Bondi} H., {Hoyle} F., 1944, \mnras, 104, 273

\bibitem[{{Booth} \& {Schaye}(2009)}]{Booth:2009}
{Booth} C.~M., {Schaye} J., 2009, \mnras, 398, 53

\bibitem[{{Brighenti} \& {Mathews}(2003)}]{Brighenti:2003}
{Brighenti} F., {Mathews} W.~G., 2003, \apj, 587, 580

\bibitem[{{Buff} \& {McCray}(1974)}]{Buff:1974}
{Buff} J., {McCray} R., 1974, \apj, 189, 147

\bibitem[{{Burkert} \& {Lin}(2000)}]{Burkert:2000}
{Burkert} A., {Lin} D.~N.~C., 2000, \apj, 537, 270

\bibitem[{{Caon} {et~al}\mbox{.}(2000){Caon}, {Macchetto}, \&
  {Pastoriza}}]{Caon:2000}
{Caon} N., {Macchetto} D., {Pastoriza} M., 2000, \apjs, 127, 39

\bibitem[{{Cattaneo} {et~al}\mbox{.}(2009){Cattaneo}, {Faber}, {Binney},
  {Dekel}, {Kormendy}, {Mushotzky}, {Babul}, {Best}, {Br{\"u}ggen}, {Fabian},
  {Frenk}, {Khalatyan}, {Netzer}, {Mahdavi}, {Silk}, {Steinmetz}, \&
  {Wisotzki}}]{Cattaneo:2009}
{Cattaneo} A. {et~al.}, 2009, \nat, 460, 213

\bibitem[{{Cattaneo} \& {Teyssier}(2007)}]{Cattaneo:2007}
{Cattaneo} A., {Teyssier} R., 2007, \mnras, 376, 1547

\bibitem[{{Cavagnolo} {et~al}\mbox{.}(2008){Cavagnolo}, {Donahue}, {Voit}, \&
  {Sun}}]{Cavagnolo:2008}
{Cavagnolo} K.~W., {Donahue} M., {Voit} G.~M., {Sun} M., 2008, \apjl, 683, L107

\bibitem[{{Churazov} {et~al}\mbox{.}(2002){Churazov}, {Sunyaev}, {Forman}, \&
  {B{\"o}hringer}}]{Churazov:2002}
{Churazov} E., {Sunyaev} R., {Forman} W., {B{\"o}hringer} H., 2002, \mnras,
  332, 729

\bibitem[{{Churazov} {et~al}\mbox{.}(2012){Churazov}, {Vikhlinin},
  {Zhuravleva}, {Schekochihin}, {Parrish}, {Sunyaev}, {Forman},
  {B{\"o}hringer}, \& {Randall}}]{Churazov:2012}
{Churazov} E. {et~al.}, 2012, \mnras, 421, 1123

\bibitem[{{Croton} {et~al}\mbox{.}(2006){Croton}, {Springel}, {White}, {De
  Lucia}, {Frenk}, {Gao}, {Jenkins}, {Kauffmann}, {Navarro}, \&
  {Yoshida}}]{Croton:2006}
{Croton} D.~J. {et~al.}, 2006, \mnras, 365, 11

\bibitem[{{Davis} {et~al}\mbox{.}(2012){Davis}, {Alatalo}, {Bureau},
  {Cappellari}, {Scott}, {Young}, {Blitz}, {Crocker}, {Bayet}, {Bois},
  {Bournaud}, {Davies}, {de Zeeuw}, {Duc}, {Emsellem}, {Khochfar},
  {Krajnovi{\'c}}, {Kuntschner}, {Lablanche}, {McDermid}, {Morganti}, {Naab},
  {Oosterloo}, {Sarzi}, {Serra}, \& {Weijmans}}]{Davis:2012}
{Davis} T.~A. {et~al.}, 2012, \mnras, 286

\bibitem[{{de Plaa} {et~al}\mbox{.}(2012){de Plaa}, {Zhuravleva}, {Werner},
  {Kaastra}, {Churazov}, {Smith}, {Raassen}, \& {Grange}}]{dePlaa:2012}
{de Plaa} J., {Zhuravleva} I., {Werner} N., {Kaastra} J.~S., {Churazov} E.,
  {Smith} R.~K., {Raassen} A.~J.~J., {Grange} Y.~G., 2012, \aap, 539, A34

\bibitem[{{Di Matteo} {et~al}\mbox{.}(2003){Di Matteo}, {Allen}, {Fabian},
  {Wilson}, \& {Young}}]{DiMatteo:2003}
{Di Matteo} T., {Allen} S.~W., {Fabian} A.~C., {Wilson} A.~S., {Young} A.~J.,
  2003, \apj, 582, 133

\bibitem[{{Di Matteo} {et~al}\mbox{.}(2005){Di Matteo}, {Springel}, \&
  {Hernquist}}]{DiMatteo:2005}
{Di Matteo} T., {Springel} V., {Hernquist} L., 2005, \nat, 433, 604

\bibitem[{{Diehl} \& {Statler}(2008)}]{Diehl:2008b}
{Diehl} S., {Statler} T.~S., 2008, \apj, 687, 986

\bibitem[{{Edge}(2001)}]{Edge:2001}
{Edge} A.~C., 2001, \mnras, 328, 762

\bibitem[{{Eswaran} \& {Pope}(1988)}]{Eswaran:1988}
{Eswaran} V., {Pope} S.~B., 1988, Computers and Fluids, 16, 257

\bibitem[{{Fabian}(2012)}]{Fabian:2012}
{Fabian} A.~C., 2012, \araa, 50, 455

\bibitem[{{Fabian} {et~al}\mbox{.}(2013){Fabian}, {Sanders}, {Haehnelt},
  {Rees}, \& {Miller}}]{Fabian:2013}
{Fabian} A.~C., {Sanders} J.~S., {Haehnelt} M., {Rees} M.~J., {Miller} J.~M.,
  2013, arXiv:astro-ph/1301.1800

\bibitem[{{Field}(1965)}]{Field:1965}
{Field} G.~B., 1965, \apj, 142, 531

\bibitem[{{Fisher} {et~al}\mbox{.}(2008){Fisher}, {Kadanoff}, {Lamb}, {Dubey},
  {Plewa}, {Calder}, {Cattaneo}, {Constantin}, \& {Foster}}]{Fisher:2008}
{Fisher} R.~T. {et~al.}, 2008, IBM J. Res. \& Dev., 52, 127

\bibitem[{{Fryxell} {et~al}\mbox{.}(2000){Fryxell}, {Olson}, {Ricker},
  {Timmes}, {Zingale}, {Lamb}, {MacNeice}, {Rosner}, {Truran}, \&
  {Tufo}}]{Fryxell:2000}
{Fryxell} B. {et~al.}, 2000, \apjs, 131, 273

\bibitem[{{Gaspari} {et~al}\mbox{.}(2012{\natexlab{a}}){Gaspari},
  {\SortNoop{a}}{Ruszkowski}, \& {Sharma}}]{Gaspari:2012a}
{Gaspari} M., {\SortNoop{a}}{Ruszkowski} M., {Sharma} P., 2012{\natexlab{a}},
  \apj, 746, 94

\bibitem[{{Gaspari} {et~al}\mbox{.}(2012{\natexlab{b}}){Gaspari},
  {\SortNoop{b}}{Brighenti}, \& {Temi}}]{Gaspari:2012b}
{Gaspari} M., {\SortNoop{b}}{Brighenti} F., {Temi} P., 2012{\natexlab{b}},
  \mnras, 424, 190

\bibitem[{{Gaspari} {et~al}\mbox{.}(2012{\natexlab{c}}){Gaspari},
  {\SortNoop{c}}{Brighenti}, \& {Ruszkowski}}]{Gaspari:2012c}
{Gaspari} M., {\SortNoop{c}}{Brighenti} F., {Ruszkowski} M.,
  2012{\natexlab{c}}, in Galaxy Clusters as Giant Cosmic Laboratories, Ness
  J.-U., ed., 14

\bibitem[{{Gaspari} {et~al}\mbox{.}(2011{\natexlab{a}}){Gaspari},
  {{\SortNoop{m2011a}}Melioli}, {Brighenti}, \& {D'Ercole}}]{Gaspari:2011a}
{Gaspari} M., {{\SortNoop{m2011a}}Melioli} C., {Brighenti} F., {D'Ercole} A.,
  2011{\natexlab{a}}, \mnras, 411, 349

\bibitem[{{Gaspari} {et~al}\mbox{.}(2011{\natexlab{b}}){Gaspari},
  {{\SortNoop{m2011b}}Brighenti}, {D'Ercole}, \& {Melioli}}]{Gaspari:2011b}
{Gaspari} M., {{\SortNoop{m2011b}}Brighenti} F., {D'Ercole} A., {Melioli} C.,
  2011{\natexlab{b}}, \mnras, 415, 1549

\bibitem[{{Goodman}(2003)}]{Goodman:2003}
{Goodman} J., 2003, \mnras, 339, 937

\bibitem[{{Haiman}(2013)}]{Haiman:2013}
{Haiman} Z., 2013, in Astrophysics and Space Science Library, Vol. 396,
  Astrophysics and Space Science Library, {Wiklind} T., {Mobasher} B., {Bromm}
  V., eds., p. 293

\bibitem[{{Hardcastle} {et~al}\mbox{.}(2007){Hardcastle}, {Evans}, \&
  {Croston}}]{Hardcastle:2007}
{Hardcastle} M.~J., {Evans} D.~A., {Croston} J.~H., 2007, \mnras, 376, 1849

\bibitem[{{Heckman} {et~al}\mbox{.}(2004){Heckman}, {Kauffmann}, {Brinchmann},
  {Charlot}, {Tremonti}, \& {White}}]{Heckman:2004}
{Heckman} T.~M., {Kauffmann} G., {Brinchmann} J., {Charlot} S., {Tremonti} C.,
  {White} S.~D.~M., 2004, \apj, 613, 109

\bibitem[{{Hobbs} {et~al}\mbox{.}(2011){Hobbs}, {Nayakshin}, {Power}, \&
  {King}}]{Hobbs:2011}
{Hobbs} A., {Nayakshin} S., {Power} C., {King} A., 2011, \mnras, 413, 2633

\bibitem[{{Hobbs} {et~al}\mbox{.}(2012){Hobbs}, {Power}, {Nayakshin}, \&
  {King}}]{Hobbs:2012_subgrid}
{Hobbs} A., {Power} C., {Nayakshin} S., {King} A.~R., 2012, \mnras, 421, 3443

\bibitem[{{Hopkins} {et~al}\mbox{.}(2006){Hopkins}, {Narayan}, \&
  {Hernquist}}]{Hopkins:2006}
{Hopkins} P.~F., {Narayan} R., {Hernquist} L., 2006, \apj, 643, 641

\bibitem[{{Hopkins} \& {Quataert}(2010)}]{Hopkins:2010}
{Hopkins} P.~F., {Quataert} E., 2010, \mnras, 407, 1529

\bibitem[{{Humphrey} {et~al}\mbox{.}(2008){Humphrey}, {Buote}, {Brighenti},
  {Gebhardt}, \& {Mathews}}]{Humphrey:2008}
{Humphrey} P.~J., {Buote} D.~A., {Brighenti} F., {Gebhardt} K., {Mathews}
  W.~G., 2008, \apj, 683, 161

\bibitem[{{Humphrey} {et~al}\mbox{.}(2009){Humphrey}, {Buote}, {Brighenti},
  {Gebhardt}, \& {Mathews}}]{Humphrey:2009}
{Humphrey} P.~J., {Buote} D.~A., {Brighenti} F., {Gebhardt} K., {Mathews}
  W.~G., 2009, \apj, 703, 1257

\bibitem[{{Igumenshchev}(2006)}]{Igumenshchev:2006}
{Igumenshchev} I.~V., 2006, \apj, 649, 361

\bibitem[{{Kere{\v s}} {et~al}\mbox{.}(2005){Kere{\v s}}, {Katz}, {Weinberg},
  \& {Dav{\'e}}}]{Keres:2005}
{Kere{\v s}} D., {Katz} N., {Weinberg} D.~H., {Dav{\'e}} R., 2005, \mnras, 363,
  2

\bibitem[{{King}(2005)}]{King:2005}
{King} A., 2005, \apjl, 635, L121

\bibitem[{{King} \& {Pringle}(2006)}]{King:2006}
{King} A.~R., {Pringle} J.~E., 2006, \mnras, 373, L90

\bibitem[{{King} \& {Pringle}(2007)}]{King:2007}
{King} A.~R., {Pringle} J.~E., 2007, \mnras, 377, L25

\bibitem[{{Krolik} \& {London}(1983)}]{Krolik:1983}
{Krolik} J.~H., {London} R.~A., 1983, \apj, 267, 18

\bibitem[{{Krumholz} {et~al}\mbox{.}(2005){Krumholz}, {McKee}, \&
  {Klein}}]{Krumholz:2005}
{Krumholz} M.~R., {McKee} C.~F., {Klein} R.~I., 2005, \apj, 618, 757

\bibitem[{{Krumholz} {et~al}\mbox{.}(2006){Krumholz}, {McKee}, \&
  {Klein}}]{Krumholz:2006}
{Krumholz} M.~R., {McKee} C.~F., {Klein} R.~I., 2006, \apj, 638, 369

\bibitem[{{Lee} \& {Deane}(2009)}]{Lee:2009}
{Lee} D., {Deane} A.~E., 2009, Journal of Computational Physics, 228, 952

\bibitem[{{Levine} {et~al}\mbox{.}(2008){Levine}, {Gnedin}, {Hamilton}, \&
  {Kravtsov}}]{Levine:2008}
{Levine} R., {Gnedin} N.~Y., {Hamilton} A.~J.~S., {Kravtsov} A.~V., 2008, \apj,
  678, 154

\bibitem[{{Li} \& {Bryan}(2012)}]{Li:2012}
{Li} Y., {Bryan} G.~L., 2012, \apj, 747, 26

\bibitem[{{Loewenstein} {et~al}\mbox{.}(2001){Loewenstein}, {Mushotzky},
  {Angelini}, {Arnaud}, \& {Quataert}}]{Loewenstein:2001}
{Loewenstein} M., {Mushotzky} R.~F., {Angelini} L., {Arnaud} K.~A., {Quataert}
  E., 2001, \apjl, 555, L21

\bibitem[{{Magorrian} {et~al}\mbox{.}(1998){Magorrian}, {Tremaine},
  {Richstone}, {Bender}, {Bower}, {Dressler}, {Faber}, {Gebhardt}, {Green},
  {Grillmair}, {Kormendy}, \& {Lauer}}]{Magorrian:1998}
{Magorrian} J. {et~al.}, 1998, \aj, 115, 2285

\bibitem[{{Mathews} \& {Brighenti}(2003)}]{Mathews:2003}
{Mathews} W.~G., {Brighenti} F., 2003, \araa, 41, 191

\bibitem[{{Mathews} \& {Guo}(2012)}]{Mathews:2012}
{Mathews} W.~G., {Guo} F., 2012, \apj, 754, 154

\bibitem[{{Mayer} {et~al}\mbox{.}(2010){Mayer}, {Kazantzidis}, {Escala}, \&
  {Callegari}}]{Mayer:2010}
{Mayer} L., {Kazantzidis} S., {Escala} A., {Callegari} S., 2010, \nat, 466,
  1082

\bibitem[{{McCarthy} {et~al}\mbox{.}(2008){McCarthy}, {Babul}, {Bower}, \&
  {Balogh}}]{McCarthy:2008}
{McCarthy} I.~G., {Babul} A., {Bower} R.~G., {Balogh} M.~L., 2008, \mnras, 386,
  1309

\bibitem[{{McConnell} \& {Ma}(2012)}]{McConnell:2012}
{McConnell} N.~J., {Ma} C.-P., 2012, arXiv:astro-ph/1211.2816

\bibitem[{{McConnell} {et~al}\mbox{.}(2011){McConnell}, {Ma}, {Gebhardt},
  {Wright}, {Murphy}, {Lauer}, {Graham}, \& {Richstone}}]{McConnell:2011}
{McConnell} N.~J., {Ma} C.-P., {Gebhardt} K., {Wright} S.~A., {Murphy} J.~D.,
  {Lauer} T.~R., {Graham} J.~R., {Richstone} D.~O., 2011, \nat, 480, 215

\bibitem[{{McCourt} {et~al}\mbox{.}(2012){McCourt}, {Sharma}, {Quataert}, \&
  {Parrish}}]{McCourt:2012}
{McCourt} M., {Sharma} P., {Quataert} E., {Parrish} I.~J., 2012, \mnras, 419,
  3319

\bibitem[{{McDonald} {et~al}\mbox{.}(2012){McDonald}, {Bayliss}, {Benson},
  {Foley}, {Ruel}, {Sullivan}, {Veilleux}, {Aird}, {Ashby}, {Bautz}, {Bazin},
  {Bleem}, {Brodwin}, {Carlstrom}, {Chang}, {Cho}, {Clocchiatti}, {Crawford},
  {Crites}, {de Haan}, {Desai}, {Dobbs}, {Dudley}, {Egami}, {Forman},
  {Garmire}, {George}, {Gladders}, {Gonzalez}, {Halverson}, {Harrington},
  {High}, {Holder}, {Holzapfel}, {Hoover}, {Hrubes}, {Jones}, {Joy}, {Keisler},
  {Knox}, {Lee}, {Leitch}, {Liu}, {Lueker}, {Luong-van}, {Mantz}, {Marrone},
  {McMahon}, {Mehl}, {Meyer}, {Miller}, {Mocanu}, {Mohr}, {Montroy}, {Murray},
  {Natoli}, {Padin}, {Plagge}, {Pryke}, {Rawle}, {Reichardt}, {Rest}, {Rex},
  {Ruhl}, {Saliwanchik}, {Saro}, {Sayre}, {Schaffer}, {Shaw}, {Shirokoff},
  {Simcoe}, {Song}, {Spieler}, {Stalder}, {Staniszewski}, {Stark}, {Story},
  {Stubbs}, {{\v S}uhada}, {van Engelen}, {Vanderlinde}, {Vieira}, {Vikhlinin},
  {Williamson}, {Zahn}, \& {Zenteno}}]{McDonald_Phoenix_Nat}
{McDonald} M. {et~al.}, 2012, \nat, 488, 349

\bibitem[{{McDonald} \& {Veilleux}(2009)}]{McDonald:2009}
{McDonald} M., {Veilleux} S., 2009, \apjl, 703, L172

\bibitem[{{McDonald} {et~al}\mbox{.}(2011{\natexlab{a}}){McDonald}, {Veilleux},
  \& {Mushotzky}}]{McDonald:2011a}
{McDonald} M., {Veilleux} S., {Mushotzky} R., 2011{\natexlab{a}}, \apj, 731, 33

\bibitem[{{McDonald} {et~al}\mbox{.}(2010){McDonald}, {Veilleux}, {Rupke}, \&
  {Mushotzky}}]{McDonald:2010}
{McDonald} M., {Veilleux} S., {Rupke} D.~S.~N., {Mushotzky} R., 2010, \apj,
  721, 1262

\bibitem[{{McDonald} {et~al}\mbox{.}(2011{\natexlab{b}}){McDonald}, {Veilleux},
  {Rupke}, {Mushotzky}, \& {Reynolds}}]{McDonald:2011b}
{McDonald} M., {Veilleux} S., {Rupke} D.~S.~N., {Mushotzky} R., {Reynolds} C.,
  2011{\natexlab{b}}, \apj, 734, 95

\bibitem[{{McNamara} \& {Nulsen}(2012)}]{McNamara:2012}
{McNamara} B.~R., {Nulsen} P.~E.~J., 2012, New Journal of Physics, 14, 055023

\bibitem[{{Mitchell} {et~al}\mbox{.}(2009){Mitchell}, {McCarthy}, {Bower},
  {Theuns}, \& {Crain}}]{Mitchell:2009}
{Mitchell} N.~L., {McCarthy} I.~G., {Bower} R.~G., {Theuns} T., {Crain} R.~A.,
  2009, \mnras, 395, 180

\bibitem[{{Narayan} \& {Fabian}(2011)}]{Narayan:2011}
{Narayan} R., {Fabian} A.~C., 2011, \mnras, 415, 3721

\bibitem[{{Nayakshin} \& {King}(2007)}]{Nayakshin:2007}
{Nayakshin} S., {King} A., 2007, arXiv:astro-ph/0705.1686

\bibitem[{{Nayakshin} {et~al}\mbox{.}(2012){Nayakshin}, {Power}, \&
  {King}}]{Nayakshin:2012}
{Nayakshin} S., {Power} C., {King} A.~R., 2012, \apj, 753, 15

\bibitem[{{Ostriker} {et~al}\mbox{.}(1976){Ostriker}, {Weaver}, {Yahil}, \&
  {McCray}}]{Ostriker:1976}
{Ostriker} J.~P., {Weaver} R., {Yahil} A., {McCray} R., 1976, \apjl, 208, L61

\bibitem[{{Paczy{\'n}ski} \& {Wiita}(1980)}]{Paczynski:1980}
{Paczy{\'n}ski} B., {Wiita} P.~J., 1980, \aap, 88, 23

\bibitem[{{Pen} {et~al}\mbox{.}(2003){Pen}, {Matzner}, \& {Wong}}]{Pen:2003}
{Pen} U.-L., {Matzner} C.~D., {Wong} S., 2003, \apjl, 596, L207

\bibitem[{{Peterson} \& {Fabian}(2006)}]{Peterson:2006}
{Peterson} J.~R., {Fabian} A.~C., 2006, \physrep, 427, 1

\bibitem[{{Pizzolato} \& {Soker}(2005)}]{Pizzolato:2005}
{Pizzolato} F., {Soker} N., 2005, \apj, 632, 821

\bibitem[{{Pizzolato} \& {Soker}(2010)}]{Pizzolato:2010}
{Pizzolato} F., {Soker} N., 2010, \mnras, 408, 961

\bibitem[{{Proga} \& {Begelman}(2003)}]{Proga:2003}
{Proga} D., {Begelman} M.~C., 2003, \apj, 582, 69

\bibitem[{{Quataert} \& {Narayan}(2000)}]{Quataert:2000}
{Quataert} E., {Narayan} R., 2000, \apj, 528, 236

\bibitem[{{Rafferty} {et~al}\mbox{.}(2008){Rafferty}, {McNamara}, \&
  {Nulsen}}]{Rafferty:2008}
{Rafferty} D.~A., {McNamara} B.~R., {Nulsen} P.~E.~J., 2008, \apj, 687, 899

\bibitem[{{Rafferty} {et~al}\mbox{.}(2006){Rafferty}, {McNamara}, {Nulsen}, \&
  {Wise}}]{Rafferty:2006}
{Rafferty} D.~A., {McNamara} B.~R., {Nulsen} P.~E.~J., {Wise} M.~W., 2006,
  \apj, 652, 216

\bibitem[{{Rasmussen} \& {Ponman}(2009)}]{Rasmussen:2009}
{Rasmussen} J., {Ponman} T.~J., 2009, \mnras, 399, 239

\bibitem[{{Rebusco} {et~al}\mbox{.}(2006){Rebusco}, {Churazov},
  {B{\"o}hringer}, \& {Forman}}]{Rebusco:2006}
{Rebusco} P., {Churazov} E., {B{\"o}hringer} H., {Forman} W., 2006, \mnras,
  372, 1840

\bibitem[{{Rees} \& {Ostriker}(1977)}]{Rees:1977}
{Rees} M.~J., {Ostriker} J.~P., 1977, \mnras, 179, 541

\bibitem[{{Reynolds} {et~al}\mbox{.}(1996){Reynolds}, {Di Matteo}, {Fabian},
  {Hwang}, \& {Canizares}}]{Reynolds:1996}
{Reynolds} C.~S., {Di Matteo} T., {Fabian} A.~C., {Hwang} U., {Canizares}
  C.~R., 1996, \mnras, 283, L111

\bibitem[{{Ruffert}(1994)}]{Ruffert:1994}
{Ruffert} M., 1994, \apj, 427, 342

\bibitem[{{Russell} {et~al}\mbox{.}(2012){Russell}, {McNamara}, {Edge},
  {Hogan}, {Main}, \& {Vantyghem}}]{Russell:2012}
{Russell} H.~R., {McNamara} B.~R., {Edge} A.~C., {Hogan} M.~T., {Main} R.~A.,
  {Vantyghem} A.~N., 2012, arXiv:astro-ph/1211.5604

\bibitem[{{Ruszkowski} \& {Oh}(2010)}]{Ruszkowski:2010}
{Ruszkowski} M., {Oh} S.~P., 2010, \apj, 713, 1332

\bibitem[{{Ruszkowski} \& {Oh}(2011)}]{Ruszkowski:2011}
{Ruszkowski} M., {Oh} S.~P., 2011, \mnras, 414, 1493

\bibitem[{{Salom{\'e}} \& {Combes}(2004)}]{Salome:2004}
{Salom{\'e}} P., {Combes} F., 2004, \aap, 415, L1

\bibitem[{{S{\'a}nchez-Salcedo} {et~al}\mbox{.}(2002){S{\'a}nchez-Salcedo},
  {V{\'a}zquez-Semadeni}, \& {Gazol}}]{Sanchez-Salcedo:2002}
{S{\'a}nchez-Salcedo} F.~J., {V{\'a}zquez-Semadeni} E., {Gazol} A., 2002, \apj,
  577, 768

\bibitem[{{Sanders} \& {Fabian}(2013)}]{Sanders:2013}
{Sanders} J.~S., {Fabian} A.~C., 2013, \mnras, 429, 2727

\bibitem[{{Schuecker} {et~al}\mbox{.}(2004){Schuecker}, {Finoguenov},
  {Miniati}, {B{\"o}hringer}, \& {Briel}}]{Schuecker:2004}
{Schuecker} P., {Finoguenov} A., {Miniati} F., {B{\"o}hringer} H., {Briel}
  U.~G., 2004, \aap, 426, 387

\bibitem[{{Shakura} \& {Sunyaev}(1973)}]{Shakura:1973}
{Shakura} N.~I., {Sunyaev} R.~A., 1973, \aap, 24, 337

\bibitem[{{Shang} \& {Oh}(2012{\natexlab{a}})}]{Shang:2012a}
{Shang} C., {Oh} S.~P., 2012{\natexlab{a}}, arXiv:astro-ph/1211.2375

\bibitem[{{Shang} \& {Oh}(2012{\natexlab{b}})}]{Shang:2012b}
{Shang} C., {Oh} S.~P., 2012{\natexlab{b}}, \mnras, 426, 3435

\bibitem[{{Sharma} {et~al}\mbox{.}(2012){Sharma}, {McCourt}, {Quataert}, \&
  {Parrish}}]{Sharma:2012}
{Sharma} P., {McCourt} M., {Quataert} E., {Parrish} I.~J., 2012, \mnras, 420,
  3174

\bibitem[{{Sharma} {et~al}\mbox{.}(2010){Sharma}, {Parrish}, \&
  {Quataert}}]{Sharma:2010}
{Sharma} P., {Parrish} I.~J., {Quataert} E., 2010, \apj, 720, 652

\bibitem[{{Sijacki} {et~al}\mbox{.}(2007){Sijacki}, {Springel}, {Di Matteo}, \&
  {Hernquist}}]{Sijacki:2007}
{Sijacki} D., {Springel} V., {Di Matteo} T., {Hernquist} L., 2007, \mnras, 380,
  877

\bibitem[{{Soker}(2006)}]{Soker:2006}
{Soker} N., 2006, NewA, 12, 38

\bibitem[{{Soker} {et~al}\mbox{.}(2009){Soker}, {Sternberg}, \&
  {Pizzolato}}]{Soker:2009}
{Soker} N., {Sternberg} A., {Pizzolato} F., 2009, in American Institute of
  Physics Conference Series, Vol. 1201, American Institute of Physics
  Conference Series, {Heinz} S., {Wilcots} E., eds., pp. 321--325

\bibitem[{{Springel} {et~al}\mbox{.}(2005){Springel}, {Di Matteo}, \&
  {Hernquist}}]{Springel:2005}
{Springel} V., {Di Matteo} T., {Hernquist} L., 2005, \mnras, 361, 776

\bibitem[{{Sun} {et~al}\mbox{.}(2009){Sun}, {Voit}, {Donahue}, {Jones},
  {Forman}, \& {Vikhlinin}}]{Sun:2009a}
{Sun} M., {Voit} G.~M., {Donahue} M., {Jones} C., {Forman} W., {Vikhlinin} A.,
  2009, \apj, 693, 1142

\bibitem[{{Sutherland} \& {Dopita}(1993)}]{Sutherland:1993}
{Sutherland} R.~S., {Dopita} M.~A., 1993, \apjs, 88, 253

\bibitem[{{Tasker} {et~al}\mbox{.}(2008){Tasker}, {Brunino}, {Mitchell},
  {Michielsen}, {Hopton}, {Pearce}, {Bryan}, \& {Theuns}}]{Tasker:2008}
{Tasker} E.~J., {Brunino} R., {Mitchell} N.~L., {Michielsen} D., {Hopton} S.,
  {Pearce} F.~R., {Bryan} G.~L., {Theuns} T., 2008, \mnras, 390, 1267

\bibitem[{{Turk} {et~al}\mbox{.}(2011){Turk}, {Smith}, {Oishi}, {Skory},
  {Skillman}, {Abel}, \& {Norman}}]{Turk:2011}
{Turk} M.~J., {Smith} B.~D., {Oishi} J.~S., {Skory} S., {Skillman} S.~W.,
  {Abel} T., {Norman} M.~L., 2011, \apjs, 192, 9

\bibitem[{{Vazquez-Semadeni}(2012)}]{Vazquez-Semadeni:2012}
{Vazquez-Semadeni} E., 2012, arXiv:astro-ph/1202.4498

\bibitem[{{Vazza} {et~al}\mbox{.}(2012){Vazza}, {Roediger}, \&
  {Br{\"u}ggen}}]{Vazza:2012}
{Vazza} F., {Roediger} E., {Br{\"u}ggen} M., 2012, \aap, 544, A103

\bibitem[{{Vikhlinin} {et~al}\mbox{.}(2006){Vikhlinin}, {Kravtsov}, {Forman},
  {Jones}, {Markevitch}, {Murray}, \& {Van Speybroeck}}]{Vikhlinin:2006}
{Vikhlinin} A., {Kravtsov} A., {Forman} W., {Jones} C., {Markevitch} M.,
  {Murray} S.~S., {Van Speybroeck} L., 2006, \apj, 640, 691

\bibitem[{{Werner} {et~al}\mbox{.}(2013){Werner}, {Oonk}, {Canning}, {Allen},
  {Simionescu}, {Kos}, {van Weeren}, {Edge}, {Fabian}, {von der Linden},
  {Nulsen}, {Reynolds}, \& {Ruszkowski}}]{Werner:2013}
{Werner} N. {et~al.}, 2013, \apj, 767, 153

\bibitem[{{Wong} {et~al}\mbox{.}(2011){Wong}, {Irwin}, {Yukita}, {Million},
  {Mathews}, \& {Bregman}}]{Wong:2011}
{Wong} K.-W., {Irwin} J.~A., {Yukita} M., {Million} E.~T., {Mathews} W.~G.,
  {Bregman} J.~N., 2011, \apjl, 736, L23

\bibitem[{{Yang} {et~al}\mbox{.}(2012){Yang}, {Sutter}, \&
  {Ricker}}]{Yang:2012}
{Yang} H.-Y.~K., {Sutter} P.~M., {Ricker} P.~M., 2012, \mnras, 427, 1614

\bibitem[{{Zhuravleva} {et~al}\mbox{.}(2012){Zhuravleva}, {Churazov},
  {Kravtsov}, \& {Sunyaev}}]{Zhuravleva:2012}
{Zhuravleva} I., {Churazov} E., {Kravtsov} A., {Sunyaev} R., 2012, \mnras, 422,
  2712

\end{thebibliography}

\label{lastpage}

\end{document}